\documentclass[lettersize,journal]{IEEEtran}
\usepackage{amsmath,amsfonts}
\usepackage{algorithmic}
\usepackage{array}
\usepackage[caption=false,font=normalsize,labelfont=sf,textfont=sf]{subfig}
\usepackage{textcomp}
\usepackage{stfloats}
\usepackage{xcolor}
\usepackage{color, colortbl}
\definecolor{aliceblue}{rgb}{0.94, 0.97, 1.0}
\definecolor{paleaqua}{rgb}{0.74, 0.83, 0.9}
\definecolor{palecyan}{rgb}{0.6, 0.8, 0.85}

\usepackage{tikz}

\newcommand{\tikzxmark}{\tikz[scale=0.23] {
    \draw[line width=0.7,line cap=round] (0,0) to [bend left=6] (1,1);
    \draw[line width=0.7,line cap=round] (0.2,0.95) to [bend right=3] (0.8,0.05);
}}

\usepackage[nohyperlinks]{acronym}

\usepackage{url}
\usepackage{makecell}
\usepackage{verbatim}
\usepackage{graphicx}
\usepackage{cite}
\def\BibTeX{{\rm B\kern-.05em{\sc i\kern-.025em b}\kern-.08em
    T\kern-.1667em\lower.7ex\hbox{E}\kern-.125emX}}
\usepackage{balance}

\usepackage{enumitem}

\usepackage{multirow}
\usepackage{tabularx}

\usepackage{color}
\usepackage{xcolor}
\usepackage{bbding}

\usepackage[ruled,linesnumbered]{algorithm2e}
\SetAlCapNameFnt{\footnotesize}
\SetAlCapFnt{\footnotesize}

\begin{document}

\title{Towards Zero Touch Networks: Cross-Layer Automated Security Solutions for 6G Wireless Networks}

\author{Li Yang, \IEEEmembership{Member, IEEE}, Shimaa Naser, \IEEEmembership{Member, IEEE}, Abdallah Shami, \IEEEmembership{Fellow, IEEE}, \\Sami Muhaidat, \IEEEmembership{Senior Member, IEEE}, Lyndon Ong, \IEEEmembership{Member, IEEE}, and M\'{e}rouane Debbah, \IEEEmembership{Fellow, IEEE}
\thanks{Li Yang is with the Faculty of Business and Information Technology, Ontario Tech University, Oshawa, ON L1G 0C5, Canada (e-mail: li.yang@ontariotechu.ca).}
\thanks{Shimaa Naser, Sami Muhaidat, and M\'{e}rouane Debbah are with the KU 6G Research Center, Department of Computer and Information Engineering, Khalifa University, Abu Dhabi 127788, UAE (e-mails: shimaa.naser@ku.ac.ae, sami.muhaidat@ku.ac.ae, merouane.debbah@ku.ac.ae).}
\thanks{Abdallah Shami is with the Department of Electrical and Computer Engineering, Western University, London, ON N6A 3K7, Canada (e-mail: abdallah.shami@uwo.ca).}
\thanks{Lyndon Ong was with Ciena, Hanover, MD 21076, USA (e-mail: lyndon.ong@gmail.com).}
\thanks{*Corresponding Author: Sami Muhaidat (sami.muhaidat@ku.ac.ae)}%
}

\markboth{Accepted and To Appear In IEEE Transactions on Communications}{}

\IEEEspecialpapernotice{(Invited Paper)}
\maketitle

\begin{abstract}
The transition from fifth-generation (5G) to sixth-generation (6G) mobile networks necessitates network automation to meet the escalating demands for high data rates, ultra-low latency, and integrated technology. 
Recently, Zero-Touch Networks (ZTNs), driven by Artificial Intelligence (AI) and Machine Learning (ML), are designed to automate the entire lifecycle of network operations with minimal human intervention, presenting a promising solution for enhancing automation in 5G/6G networks. However, the implementation of ZTNs brings forth the need for autonomous and robust cybersecurity solutions, as ZTNs rely heavily on automation. AI/ML algorithms are widely used to develop cybersecurity mechanisms, but require substantial specialized expertise and encounter model drift issues, posing significant challenges in developing autonomous cybersecurity measures. Therefore, this paper proposes an automated security framework targeting Physical Layer Authentication (PLA) and Cross-Layer Intrusion Detection Systems (CLIDS) to address security concerns at multiple Internet protocol layers. The proposed framework employs drift-adaptive online learning techniques and a novel enhanced Successive Halving (SH)-based Automated ML (AutoML) method to automatically generate optimized ML models for dynamic networking environments. Experimental results illustrate that the proposed framework achieves high performance on the public Radio Frequency (RF) fingerprinting and the Canadian Institute for Cybersecurity Intrusion Detection System 2017 (CICIDS2017) datasets, showcasing its effectiveness in addressing PLA and CLIDS tasks within dynamic and complex networking environments. Furthermore, the paper explores open challenges and research directions in the 5G/6G cybersecurity domain. This framework represents a significant advancement towards fully autonomous and secure 6G networks, paving the way for future innovations in network automation and cybersecurity.
\end{abstract}



\maketitle

\section{Introduction}

The evolution of wireless communication technologies has progressed from the introduction of first-generation (1G) mobile networks in the 1980s to the current fifth-generation (5G) networks \cite{6gsec1}. The generations of mobile devices have brought significant advancements, with 1G introducing mobile voice calls, second-generation mobile networks (2G) enabling text messaging, third-generation mobile networks (3G) bringing mobile data, fourth-generation mobile networks (4G) LTE ushering in the era of mobile Internet, and 5G further revolutionizing communication with high bandwidth, low latency, and support for a wide range of applications, such as virtual reality to industrial automation \cite{6gsec1}. Key enablers of 5G networks include several cutting-edge technologies, such as network slicing, Software-Defined Networking (SDN), and Network Function Virtualization (NFV) \cite{zsm1}. These enablers play a crucial role by abstracting physical resources into virtual ones, facilitating automated services, and enhancing network management.

While 5G is currently in its deployment phase, the research community has already started envisioning the sixth generation of wireless networks (6G). Network automation is a crucial facilitator for both 5G and 6G networks. It involves the application of advanced software to automate the management and orchestration of network services. Network automation enables efficient network management, reduces operational expenses, and enhances service delivery \cite{zsmml1}. The Zero-Touch Network and Service Management (ZSM) framework, proposed by the European Telecommunications Standards Institute (ETSI), is a state-of-the-art network automation framework \cite{zsm2}. The ZSM framework has been continuously updated and extended by ETSI. Artificial Intelligence (AI) and Machine Learning (ML) techniques serve as the backbone of state-of-the-art ZSM framework to enable Zero-Touch Networks (ZTNs) and security service automation \cite{ETSI_AI} \cite{ETSI_Sec}. A ZTN, integral to the ZSM framework, is conceptualized as a fully autonomous network system where all operational processes—from configuration, management, and optimization to the self-healing of network functions—are performed automatically, without manual intervention, leveraging advancements in AI/ML for decision-making processes\cite{zsm3}. 
This ZTN paradigm shifts towards fully automated network operations and services, which is critical in addressing the complexity and dynamic demands of next-generation networks, including 6G.

AI/ML and big data analytics techniques are also envisioned as key enablers of ZTNs and future networks \cite{ETSI_AI}. Through network data processing and analytics, AI/ML models can help enable enhanced automation capabilities, leading to substantial reductions in operational expenses, and minimizing human errors \cite{zsm1}. AI/ML techniques offer a wide range of network services and management functionalities, bolstering capabilities such as network behavior analysis, anomaly detection, traffic classification and prediction, mobility forecasting, and efficient resource allocation.

Despite the widespread utilization of AI/ML models in network applications, the development and deployment of traditional ML models often necessitate significant human intervention and extensive domain expertise, which can be a limiting factor in realizing fully autonomous networks or ZTNs \cite{myautoml}. Furthermore, traditional ML models often face limitations in handling data streams in dynamic and complex network environments. The performance of these models can degrade over time due to network environmental changes, device aging, and unpredictable events, leading to a phenomenon known as model drift \cite{dmdrift}. 

Automated ML (AutoML) has emerged as a promising solution for data-driven network services by addressing the challenges associated with traditional ML models and automating network data analytics \cite{myautoml}. AutoML aims to automate various data analytics and ML procedures, including Automated Data Pre-processing (AutoDP), Automated Feature Engineering (AutoFE), automated model selection, Hyper-Parameter Optimization (HPO), and automated model updating \cite{myautoml} \cite{automl1}. AutoDP and feature engineering are designed to improve network data quality for better data analytics results. Automated model selection and HPO aim to generate optimized ML models with improved performance. Furthermore, automated model updating or drift adaptation is a potential solution to address model drift issues by automatically updating the ML models to adapt to dynamic networking environments \cite{iotm}. AutoML techniques hold significant promise for realizing ZTNs by automating the complex and laborious tasks associated with ML model development and deployment.

The network revolution has introduced various cybersecurity threats that exploit vulnerabilities in modern networks. Common cyber-attacks can be categorized according to the 5-layer Internet Protocol Stack that provides guidelines for the creation of network protocols. The five layers, from bottom to top, are the physical, data link, network, transport, and application layers \cite{stack1}. Each layer has its own vulnerabilities, making it a potential target for cyber threats. The physical layer, which is responsible for the transmission and reception of raw bit streams over a physical medium, is vulnerable to attacks such as jamming, eavesdropping, and physical tampering. The data link layer, which provides node-to-node data transfer, is compromised by threats such as Media Access Control (MAC) address spoofing and flooding. The network layer, which handles data packet forwarding and routing, is targeted by Internet Protocol (IP) spoofing and Denial of Service (DoS) attacks. The transport layer, which provides end-to-end communication services, can be vulnerable to Transmission Control Protocol (TCP) Synchronization (SYN) flooding and User Datagram Protocol (UDP) flooding attacks. Lastly, the application layer, which enables services for end-user applications, is a common target for various forms of malware, phishing, and web attacks \cite{threat1} \cite{threat2}.

To protect networks from cyber-attacks, many cybersecurity solutions and mechanisms have been developed, particularly device authentication  and Intrusion Detection Systems (IDSs). Traditional cryptography-based authentication techniques implemented at the
upper layers have been commonly adopted for verifying users’
identities through key generation and detection processes.
However, in networks comprising a massive number of mobile
and heterogeneous devices, the distribution and management
of cryptographic keys is rather challenging. Additionally, the
computational complexity associated with key generation and
detection introduces latency issues that drain the battery life of
power-limited devices. These limitations motivated the development of innovative Physical Layer Authentication (PLA) schemes to verify the authenticity of network devices based on their physical layer characteristics, thereby preventing unauthorized devices from accessing the network \cite{pla1}. It is worth noting that, besides PLA, the physical layer encompasses other security techniques, including Physical Layer Security (PLS) and Physical Layer Key Generation (PLKG).  PLS techniques ensure confidential transmission between entities by leveraging the inherent features of the physical layer, without the need to share secret keys. PLKG extracts random keys based on the channel characteristics between legitimate entities. Nevertheless, PLA holds a higher significance as it serves as the first line of defense for authenticating the origin of received signals.

On the other hand, IDSs are designed to detect cyber-attacks in a network through network data analytics, and Cross-Layer IDSs (CLIDSs) are a special type of IDSs that integrates information from multiple layers of the protocol stack to detect and mitigate cyber threats. For instance, a CLIDS can utilize the physical layer's ability to detect anomalies in wireless signals, the network layer's capability to identify suspicious packet behaviors, and the application layer's capacity to recognize malicious software patterns \cite{clids1}. In modern networks, PLA mechanisms are usually deployed as the first layer of defense to authenticate devices, whereas CLIDSs can be utilized as the second layer of defense to detect malicious attacks that have breached the first layer of defense \cite{mythesis}. Although PLA and CLIDS operate at different layers of the network and are designed to address distinct cybersecurity challenges, both components are crucial for enhancing the security of complex and dynamic 5G/6G network environments. Although these solutions are promising, they are still under active research and development, as there is much room for improvement. They represent the ongoing efforts in the cybersecurity community to address the evolving challenges in securing 5G/6G networks.

Autonomous cybersecurity solutions, such as automated PLA and CLIDS frameworks, are crucial for enhancing network automation and security in ZTNs and future networks. By automating the process of authentication and threat detection, automated security mechanisms can significantly reduce the human labor and time to mitigate cyber attacks,  thereby minimizing potential damage caused by attacks. In this paper, an AuotoML-based autonomous cybersecurity framework is proposed to solve both PLA and CLIDS problems for protecting ZTNs and future networks. Specifically, the proposed framework consists of the following components: the Chebyshev over-sampling approach for automated data balancing and AutoDP, the Pearson Correlation Coefficient (PCC)-based Select-K-Best method for AutoFE, Adaptive Random Forest (ARF) and Streaming Random Patches (SRP) for base online model learning, the enhanced Successive Halving (SH) model for Combined Algorithm Selection and Hyper-parameter optimization (CASH), and the ADaptive WINdowing (ADWIN) and Early Drift Detection Method (EDDM) for model drift detection and automated model updating. The proposed AutoML framework can automatically adapt to any specific datasets related to cybersecurity problems and generate optimized ML models with optimal performance.

This manuscript presents a research paper introducing a novel automated cybersecurity framework and also provides a tutorial paradigm for achieving autonomous cybersecurity in similar applications in future works. To the best of our knowledge, this paper is the first work that proposes a comprehensive and automated cybersecurity framework to solve both PLA and CLIDS problems and achieve ZTNs for future networks. Specifically, this paper makes the following contributions:
\begin{enumerate}[label=\roman*)]
\item It reviews the existing cybersecurity threats and potential solutions at different Internet protocol stack layers for modern networks.
\item It proposes a novel automated cybersecurity framework, SH-CASH, that can address cyber threats among the physical layer and upper layers using AutoML techniques\footnote{
The code for this paper is publicly available at: \url{https://github.com/Western-OC2-Lab/Cross-Layer-Autonomous-Cybersecurity-Framework}}.
\item It evaluates the performance of the proposed automated cybersecurity framework using state-of-the-art PLA and IDS datasets, namely the Oracle Radio Frequency (RF) fingerprinting \cite{rfdata} and the Canadian Institute for Cybersecurity Intrusion Detection System 2017 (CICIDS2017) \cite{cic} datasets. 
\item It discusses the open challenges and research directions of developing cybersecurity solutions for ZTNs and future networks. 
\end{enumerate}

This paper is organized as follows. Section II provides an overview of ZTNs and their security requirements. Section III provides an introduction to 6G cybersecurity threats and mechanisms for different Internet protocol stack layers. Section IV presents the literature review of PLA, CLIDS, and automated cybersecurity methods. Section V introduces AutoML techniques and describes the proposed automated cybersecurity framework in detail. Section VI demonstrates and discusses the experimental results of evaluating the proposed framework for PLA and CLIDS tasks. Section VII presents the challenges and future directions of developing cybersecurity solutions for ZTNs and future networks. Section VIII provides a list of acronyms. Section IX concludes the paper.

\section{Zero-Touch Networks and Security Requirements}

\subsection{ZTN Overview}
ZTNs have emerged as a revolutionary paradigm in network management for next-generation networks, including 6G. They provide critical automation capabilities such as self-configuration, self-optimization, self-healing, and self-protection with minimal human involvement \cite{ETSI_AI} \cite{zsmsec1}.

ZTNs are integral to the ZSM framework (ETSI ZSM 002) that was developed by ETSI in October 2019 to revolutionize network and service management through end-to-end automation \cite{zsm2}. The ZSM framework has been significantly enhanced over the years. In June 2021, ETSI introduced specifications for closed-loop automation (ETSI ZSM 009-1), detailing key enablers for automated network management procedures \cite{ETSI_closed}. In December 2022, ETSI pushed the enablers for AI-based network and service automation (ETSI ZSM 012), focusing on AI integration within the ZSM framework \cite{ETSI_AI}. In March 2024, the release of ETSI ZSM 014 incorporated critical security assurance mechanisms, explicitly addressing security challenges, and providing comprehensive guidelines for enhancing ZTN robustness \cite{ETSI_Sec}. The evolution of the ZSM framework has resulted in a cutting-edge solution for end-to-end zero-touch automation for next-generation networks.

ZTNs offer various functionalities, including automated network configuration, performance optimization, fault detection, and security management \cite{mirna}. While ZTNs are a paradigm applicable to multiple generations of networks, including 5G, their principles align closely with the high automation demands of 6G. AI/ML techniques are key enablers for ZTN functionalities, allowing networks to learn from historical data and information, adapt to ever-changing environments, and make informed decisions.

\subsection{Existing Applications of ZTNs}
ZTNs have evolved from a conceptual framework to a practical solution, meeting the complex automation and dynamic demands of 5G/6G networks through advanced AI integration and real-world applications across diverse domains.

Radoglou-Grammatikis \textit{et al.} \cite{zsmapp1} position the ZSM framework as a practical, scalable solution for managing complex, multi-domain network environments, with implications extending to autonomous cybersecurity in 6G networks. This study introduces a use case analyzing the interactions between network attackers and defenders targeting AI-driven IDSs for ZSM. The cross-domain analytics demonstrate their value in orchestrating security strategies and identifying vulnerabilities in AI models, emphasizing the potential of the ZSM architecture in real-world applications.

Bello \textit{et al.} \cite{zsmapp2} delve into the application of ZSM principles to Beyond 5G (B5G) and 6G networks, with a detailed case study on Smart Traffic Management (STM) under the TrialsNet project. In the STM use case, ZSM principles are applied to enable intelligent traffic flow optimization and safety measures using edge-driven AI/ML techniques. The study showcases how ZSM facilitates automated management of network resources, enabling seamless orchestration of network functions and services for 6G networks.

Lacoboaiea \textit{et al.} \cite{zsmapp3} explore the design and deployment of zero-touch Wireless Local Area Networks (WLANs), a specific application of ZTNs, using deep reinforcement learning for autonomous resource management. This work highlights the potential of AI/ML-based automation for achieving scalable, adaptive, and zero-touch network management, which is essential for modern dense WLAN environments.

Mirna \textit{et al.} \cite{mirna} identify ZTNs as a next-generation solution and an advanced paradigm for automating the management and optimization of B5G networks within the framework of ZSMs. The study presents a detailed case study showcasing AutoML's application in predicting application throughput in B5G networks. The case study emphasizes that ZTNs, when augmented with AutoML, can deliver significant operational efficiencies by automating critical network functions. The AutoML-enabled ZTN system reduces manual intervention, minimizes latency, and enhances network adaptability, thereby supporting the diverse and dynamic requirements of B5G use cases.

In summary, the applications of ZTNs underscore their ability to automate, optimize, and secure next-generation networks across diverse and complex scenarios, such as resource management, traffic management, advanced cybersecurity, and adaptive network optimization. These use cases emphasize the significant role of ZTNs as a foundational paradigm for scalable, efficient, autonomous, and intelligent network management. By integrating advanced technologies such as AI/ML, AutoML, and other state-of-the-art methodologies, ZTNs provide robust and adaptable solutions to address the escalating complexity and dynamic requirements of B5G and 6G networks. 

\subsection{ZTN Security Requirements}

\begin{figure}[!t]
\centerline{
\includegraphics[width=8.1cm]{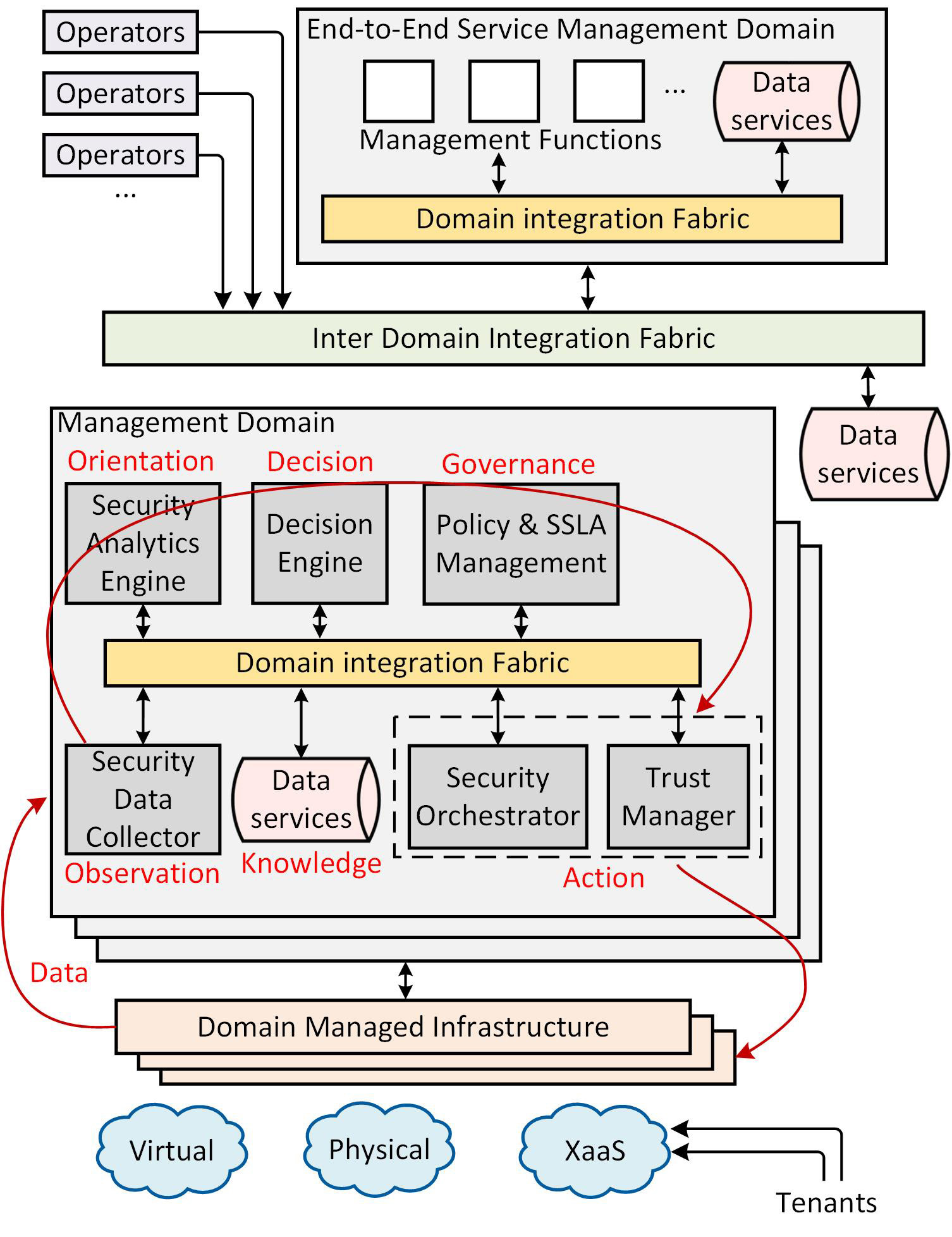}}
\caption{The closed-loop ZTN security framework \cite{zsmsec2}.}
\label{zsm}
\end{figure}

ZTNs aim to provide a comprehensive and autonomous security framework for next-generation networks. The domain-label closed-loop ZTN security framework is illustrated in Fig. \ref{zsm}, which consists of the following five stages for automation \cite{zsmsec2}:
\begin{enumerate}[label=\roman*)]
\item \textit{Observation}: The security data agents observe and collect data from various sources or devices.
\item \textit{Orientation}: The security analytics engine orients the network data to analyze its significance and meaning, which may involve data filtering, correlation analysis, and exploratory data analysis.
\item \textit{Decision}: The decision engine utilizes AI/ML algorithms to recognize data patterns and to make informed decisions/predictions regarding network service optimization.
\item \textit{Action}: The security orchestrator takes appropriate actions or plans for the current situation and environment after the root causes and recommendations are identified.
\item \textit{Learning}: The system gains knowledge from the detected data patterns and stores them in databases for improving future reactions or decisions.
\end{enumerate}

The proposed method aims to achieve autonomous cybersecurity solutions by considering the five stages of the ZTN framework. Additionally, for ZTN security framework development, there are several essential security requirements that must be met \cite{zsm1} \cite{zsm3} \cite{zsmsec1}: 
\begin{enumerate}[label=\roman*)]
\item \textit{Access Control and Authentication}:
The ZTN security framework aims to enhance the capabilities of authorized users and devices to access the network by employing robust authentication mechanisms. By doing so, it facilitates the automated application of suitable security policies that align with the unique security needs of various management services. Conversely, any unauthorized or illegitimate users or devices that do not pass the authentication process are promptly blocked, effectively mitigating potential cyber threats.
\item \textit{Intrusion Detection and Mitigation}: The ZTN security frameworks should be capable of automatically detecting cyber-attacks and determining appropriate mitigation mechanisms. This capability is critical for the rapid identification and mitigation of potential security threats with minimal human intervention, thus minimizing the impact of cyberattacks on networks services.
\item \textit{AI/ML Decision Supervision}: To safeguard against cyber-attacks, it is essential for ZTN frameworks to integrate mechanisms that supervise and potentially update security decisions made by AI/ML techniques. This enables AI/ML-driven models to consistently make precise decisions, effectively defending against emerging or previously unknown attacks. By incorporating this mechanism, organizations can enhance their security posture and stay resilient in the face of new or zero-day threats.
\end{enumerate}

Overall, effective ZTN security frameworks should be able to autonomously prevent, detect, and mitigate cyber-attacks in dynamic networking environments, safeguarding networks with minimal reliance on human intervention.

\section{6G Network Security}

This section provides a comprehensive overview of 6G network security, starting with an analysis of threats across the Internet protocol stack layers and followed by a discussion on security mechanisms. The focus is on understanding how layered architectures influence security threats and how cybersecurity mechanisms, such as PLA and CLIDSs, can be employed to mitigate these threats. This foundational discussion sets the stage for Section V, which delves into the proposed autonomous cybersecurity frameworks aligned with the ZTN paradigm to protect 6G networks.

\subsection{Network Security Threats of Internet Protocol Stack Layers}

\subsubsection{Internet Protocol Stack Layers}
\begin{figure*}[!t]
\centerline{
\includegraphics[width=13.3cm]{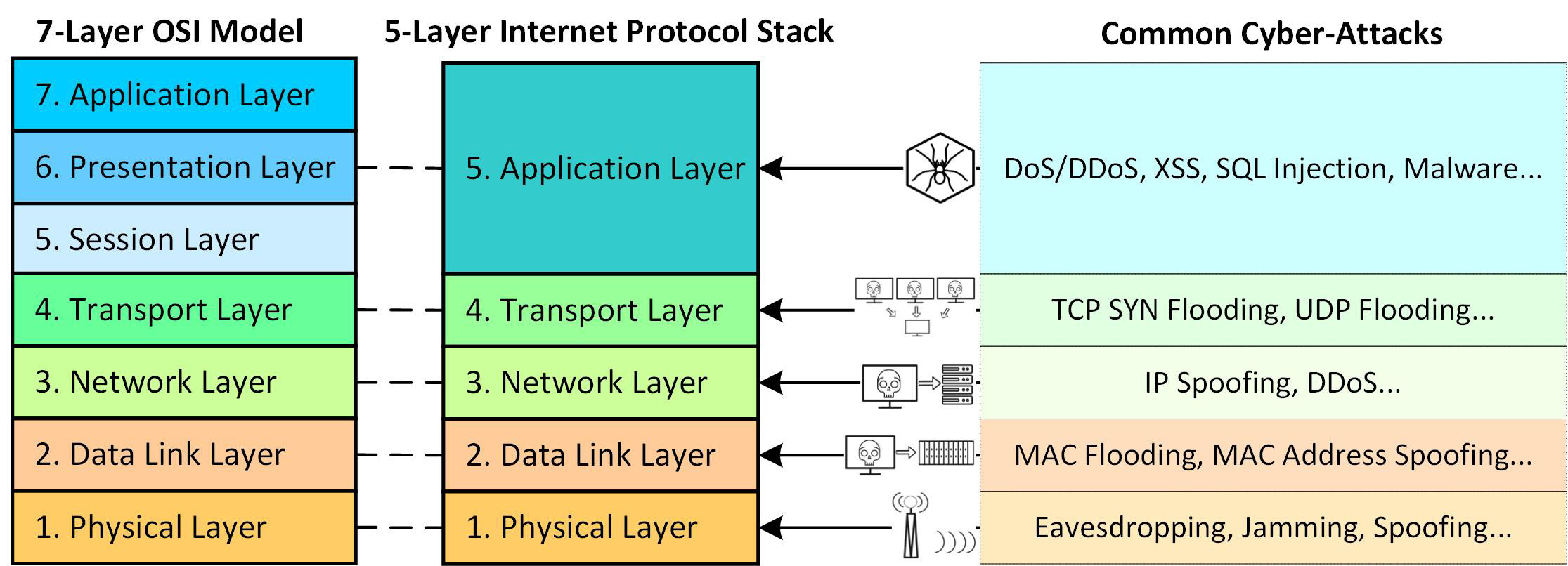}}
\caption{The OSI reference model and the Internet protocol stack with common cyber-attacks at each layer.}
\label{stack}
\end{figure*}

The layered protocol architectures are critical requirements for modern networks \cite{clids_g}. The Open Systems Interconnection (OSI) model, developed by the International Organization for Standardization (ISO), is a comprehensive seven-layer model that enables interoperability among diverse systems. The layers, from the lowest to the highest, include the physical, data link, network, transport, session, presentation, and application layers, as shown in Fig. \ref{stack} \cite{stack1}. The physical layer is responsible for transmitting and receiving raw bit streams over a physical medium. The data link layer facilitates data transfer between nodes, while the network layer handles packet routing and error checking. The transport layer offers end-to-end communication services, and the session layer manages connections between applications. The presentation layer ensures data is in a usable format, while the topmost layer, the application layer, provides network services to software applications.

The Internet protocol stack is a more practical model with five layers that forms the basis of the Internet \cite{stack1}. The layers, from the lowest to the highest, are the physical, data link, network, transport, and application layers, as presented in Fig. \ref{stack}. The physical, data link, network, and transport layers in the Internet protocol stack have similar functions to their counterparts in the OSI model. However, the application layer in the Internet protocol stack combines the functions of the session, presentation, and application layers in the OSI model, providing network services directly to the user's applications \cite{threat3}.

While the OSI model and the Internet protocol stack both serve the purpose of facilitating network communications, they differ in several aspects \cite{stack1}. The OSI model is known for its theoretical and comprehensive nature, comprising seven layers compared to the Internet protocol stack's five layers. It offers a detailed understanding of each layer's functionality, making it valuable for teaching and comprehending network communication concepts. However, due to its complexity, the OSI model is less commonly used in practical implementation.

On the other hand, the Internet protocol stack is renowned for its practicality and widespread adoption, especially in the implementation of the Internet. It exhibits simplicity and efficiency with its streamlined structure containing fewer layers than the OSI model. This is because the application layer in the Internet protocol stack combines the functions of the session, presentation, and application layers in the OSI model, which simplifies the application development process \cite{stack1}. Therefore, the 5-layer Internet protocol stacks are considered in this paper to explore security threats on different layers.

\subsubsection{Physical Layer Threats}
As the lowest layer of the Internet protocol stack, the physical layer deals with the physical characteristics of the communication medium. The inherent broadcast nature of the physical layer renders it susceptible to various forms of attacks aimed at exploiting the unique properties of wireless communications. Depending on the attackers' objectives, these encompass eavesdropping, jamming, spoofing, and contamination attacks \cite{phythreat1,phythreat2,phythreat3}. More specifically, in an eavesdropping or sniffing attack, an unauthorized user located in the coverage area of the source node intercepts the wireless data transmission \cite{phythreat1}. This type of intrusion poses a significant threat to data confidentiality, as the eavesdropper can gain unauthorized access to sensitive information without being detected or located by legitimate nodes. The absence of signal transmission from the eavesdropper allows them to covertly access data, further increasing the risk to confidentiality.  

On the other hand, in a jamming attack (also referred to as a physical layer DoS attack), a malicious node deliberately generates interference to disrupt the regular operation of wireless communications \cite{phythreat2}. This interference is intended to hinder data transmissions. It is important to note that jamming can take different forms: 1) Pilot jamming, launched during the channel state acquisition phase, where the jammer disrupts only the pilot signals, not the entire communication; 2) Proactive jamming, where the attacker consistently emits interfering signals regardless of legitimate communication; 3) Reactive jamming, triggered upon detecting a legitimate signal.  

In contrast, spoofing or impersonation attacks involve injecting fake signals or identities to deceive the network and gain unauthorized access \cite{phythreat1,phythreat2}. The attacker can either impersonate a legitimate user (identity spoofing) using a fake MAC/IP address, and then launch a more advanced attack or create multiple false identities (sybil attacks) to disrupt the network operations. Finally, contamination attacks occur at the channel state  acquisition phase, where the attacker seeks to contaminate the channel state information by sending the same pilot signals to the access point or injecting forged feedback to gain an unfair advantage in the subsequent communication phase \cite{phythreat2}. 

\subsubsection{Upper Layer Security Threats}
Apart from the physical layer, the Internet protocol stack involves four upper layers: the data link, network, transport, and application layers. The four upper layers are vulnerable to different cyber-attacks due to their different responsibilities in networks. The common cyber-attacks at each layer are demonstrated in Fig. \ref{stack}.

The data link layer ensures reliable transmission of data across physical network interfaces. At the data link layer, cyber-attacks occur by manipulating data packets in order to gain unauthorized access or disrupt the services of other machines in the network \cite{upperthreat1}. MAC-related attacks are primary attacks in the data link layer, including MAC flooding and MAC address spoofing attacks. MAC flooding, a type of DoS attack, refers to the flooding of packets into the MAC address table of a switch in order to grant access to unauthorized devices \cite{upperthreat2}. In MAC address spoofing, a device's MAC address is changed to impersonate a legitimate user on a network.

The network layer manages addressing and path selection as well as the routing of data packets across multiple interconnected networks. IP spoofing attacks are a common type of network layer attack that alters the source IP addresses in packets to impersonate other devices \cite{upperthreat3}. Distributed DoS (DDoS) is another type of network layer attack that occurs when multiple devices overwhelm a target device with a large amount of traffic, causing service congestion or unavailability \cite{ericsson}.

The transport layer is responsible for providing end-to-end communication services for applications. The transport layer is vulnerable to two DoS attacks, namely TCP SYN flooding and UDP flooding attacks \cite{upperthreat4}. In TCP SYN flooding attacks, a large volume of TCP SYN requests are sent to the target device to cause the device to become unresponsive. In UDP flooding attacks, a large amount of UDP packets are sent to the target device to hinder its operation. 

In the application layer, applications, software, and end-user processes are supported. The threat of DoS and DDoS attacks is common at the application level. Cross-Site Scripting (XSS) and Structured Query Language (SQL) injection are also employed as web-attacks at the application layer to disrupt or control web interfaces and computer systems \cite{myiotj}. Additionally, malicious software and infections can be launched at the application layer to harm a network or system.

\subsection{Network Security Mechanisms}

\subsubsection{Physical Layer Authentication}
\label{sec:pla}

Spoofing or impersonation attacks represent the first step for the intruder toward launching a variety of attacks, including DoS, session hijacking, Man-in-the-Middle (MITM), data modification, and sniffing. These attacks can severely degrade network performance. Moreover, spoofing or impersonation attacks are considered comparatively easier to initiate than other types of attacks, making them a preferred and the first choice for attackers \cite{Zeng2010}. In this regard, device authentication or identification serves as the first line of defense towards securing data and systems by verifying the legitimacy of users and devices before granting them access to the system or resources. As mentioned earlier, traditional cryptography-based techniques require complicated key distribution and management in massive device deployment and suffer from the computational complexity associated with key generation and detection, introducing latency issues that drain the battery life of power-limited devices. These limitations necessitate the development of innovative solutions to address the key distribution and management complexities and to enhance or complement traditional authentication protocols. As a consequence, PLA has emerged as an efficient and low-complexity security solution for devices' identification based on their unique physical characteristics \cite{pla2}. Specifically, PLA techniques use unique features extracted from the wireless signals transmitted by a device, such as channel impulse response or waveform characteristics, which can not be cloned, to authenticate the device. PLA techniques can be broadly categorized into two main categories: Radio Frequency (RF)/hardware-based PLA and location/channel-based PLA.

RF/hardware-based PLA techniques leverage the inherent imperfections or unique characteristics of hardware components during the process of circuit manufacturing, such as the turn-on transient features and steady-state features, to generate distinctive RF fingerprints. These fingerprints are utilized to identify and authenticate devices. By exploiting the subtle variations and irregularities within the RF signals, these techniques provide a means to establish device authenticity. On the other hand, location/channel-based PLA techniques exploit the distinct location-specific characteristics of wireless communication channels, particularly in rich multipath environments. Thus, when the distance between the adversary and legitimate user is greater than half of the wavelength, it is difficult to clone the same channel/location characteristics at the adversary.  These features include signal propagation delays, device location, signal strength fluctuations,  Channel Impulse Responses (CIR), and Channel Frequency Response (CFR) \cite{Xie2021}. 
\\
\subsubsection*{\textbf{\textit{Threshold-Based PLA}}} To provide the authentication decision, device-based and channel-based PLA schemes typically involve two phases, training and authentication, as shown in Fig. \ref{fig:ThresholdPLA}. In the system model under consideration, Alice serves as the legitimate transmitter who intends to communicate with Bob. Similarly, Bob represents the legitimate receiver who receives the transmitted signal and authenticates its source using a specific PLA scheme. Lastly, Eve is an adversary who either eavesdrops the wireless transmissions between Alice and Bob or attempts to deceive Bob by sending a forged signal while impersonating Alice. During the first phase, \textit{i.e.}, the training phase,  Bob receives a request message from Alice at time $t_0$. During this phase, Bob verifies the legitimacy of the transmitter by employing upper-layer authentication protocols. Once the transmitter's legitimacy is confirmed, Bob extracts a physical layer feature denoted as $X^{A\rightarrow B}_0$, which will be utilized in the subsequent message authentication stages. Bob then responds to Alice, acknowledging her request and permitting her to send the intended message. However, in the event that the transmitter is found to be illegitimate, Bob disregards the received request. In the second phase at time $t_1$, which is within the channel coherence time, Alice sends her message to Bob, who in turn extracts a physical layer feature from the received signal, \textit{i.e.}, $X^{\nu \rightarrow B}_1$,  and compares it with the features recorded in the previous stage. The authentication decision at Bob is modeled as a threshold-based hypothesis testing as follows:
\begin{figure}[t]
    \centering
    \includegraphics[width=7.5cm]{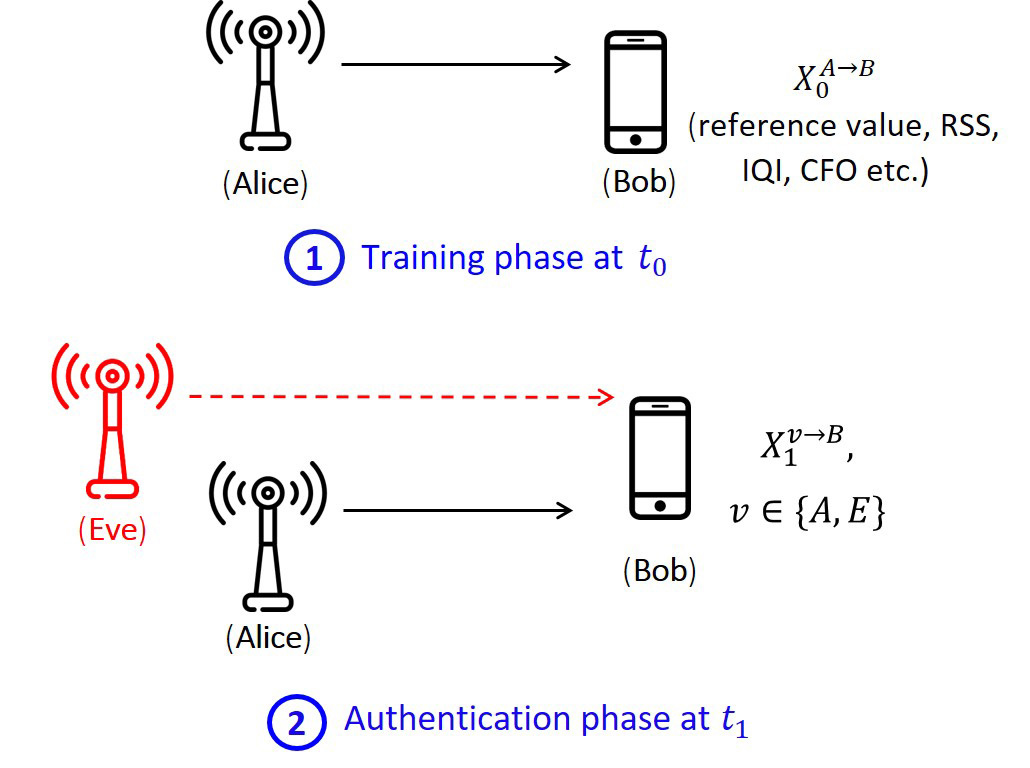}
    \caption{The phases involved in threshold-based PLA schemes.}
    \label{fig:ThresholdPLA}
\end{figure}


\begin{equation}
\label{eq:test}
\left\{\begin{array}{ll}
 \mathcal{H}_0,  & diff(X^{\nu \rightarrow B}_1-X^{A \rightarrow B}_0)\leq \gamma; \\ 
  \mathcal{H}_1,  & diff(X^{\nu \rightarrow B}_1-X^{A \rightarrow B}_0)\geq\gamma,
\end{array}\right.
\end{equation}
where $diff$ represents the difference between the feature values over two different time slots to decide on  $\mathcal{H}_0$ or $\mathcal{H}_1$ which are the hypotheses of deciding legitimate and forged signals, respectively.  The hypothesis decision problem is usually handled by employing the Likelihood Ratio Test
(LRT). Furthermore, $\gamma$ denotes the threshold value that effectively differentiates between legitimate and illegitimate transmitters,  while minimizing the risk of false positives or false negatives. Depending on the dynamics of the environment, the threshold might need to be adjusted to account for changes in the network or the presence of new transmitters. 

By employing hypothesis testing as shown in (\ref{eq:test}), the resilience of the PLA scheme is evaluated using various metrics, namely: 
\begin{itemize}
    \item Probability of Detection (PD), denoted as $P(\mathcal{H}_1|\mathcal{H}_1)$, which measures the probability that the PLA correctly detects the presence of an illegitimate transmitter when it is indeed present.
\item Probability of False Alarm (PFA), denoted as $P(\mathcal{H}_1|\mathcal{H}_0)$, which quantifies the probability that the PLA incorrectly identifies the presence of an illegitimate transmitter when it is not actually present.
\item Probability of Missed Detection (PMD), denoted as $P(\mathcal{H}_0|\mathcal{H}_1)$, which represents the probability that the PLA fails to detect the presence of an illegitimate transmitter when it is, in fact, present.
\end{itemize}

Another metric that serves to measure the robustness of the PLA scheme is the Receiver Operating Characteristic (ROC). Such a  curve can be obtained by plotting the PD against the PFA  for various threshold values, $\gamma$, to quantify the accuracy of the classier for different thresholds. 

It is essential to highlight that PLA schemes relying solely on a single feature typically exhibit low overhead, but may lack robustness and reliability. This limitation stems from the imperfect estimates of the particular feature and its limited-range distribution, which might not be sufficient to consistently differentiate between various devices. Thus, to enhance authentication accuracy, the development of PLA schemes based on multiple features becomes advantageous since it is difficult for the malicious node to imitate all the
features. Nevertheless, this improvement comes at the cost of increased communication overhead and higher computational complexity. Therefore, a trade-off must be considered between authentication accuracy and resource requirements when selecting an appropriate PLA scheme for a specific application. In order to determine the 
authenticity of a device based on multiple features,  hypothesis
testing is applied first based on the individual features separately. Then, 
a  device is authenticated only when
all the features-based tests claim the
same legitimate device. Alternatively, a combined device fingerprint can be formed by using a weighted combination of $M$ physical features. Subsequently, the LRT can be employed to authenticate the device. It is worth mentioning that the optimal weight for each selected feature must be optimized to achieve the highest detection probability. This process ensures an efficient and effective authentication mechanism based on the fusion of multiple physical characteristics.

\begin{figure*} [t]
\centering
\includegraphics[width=17.4cm]{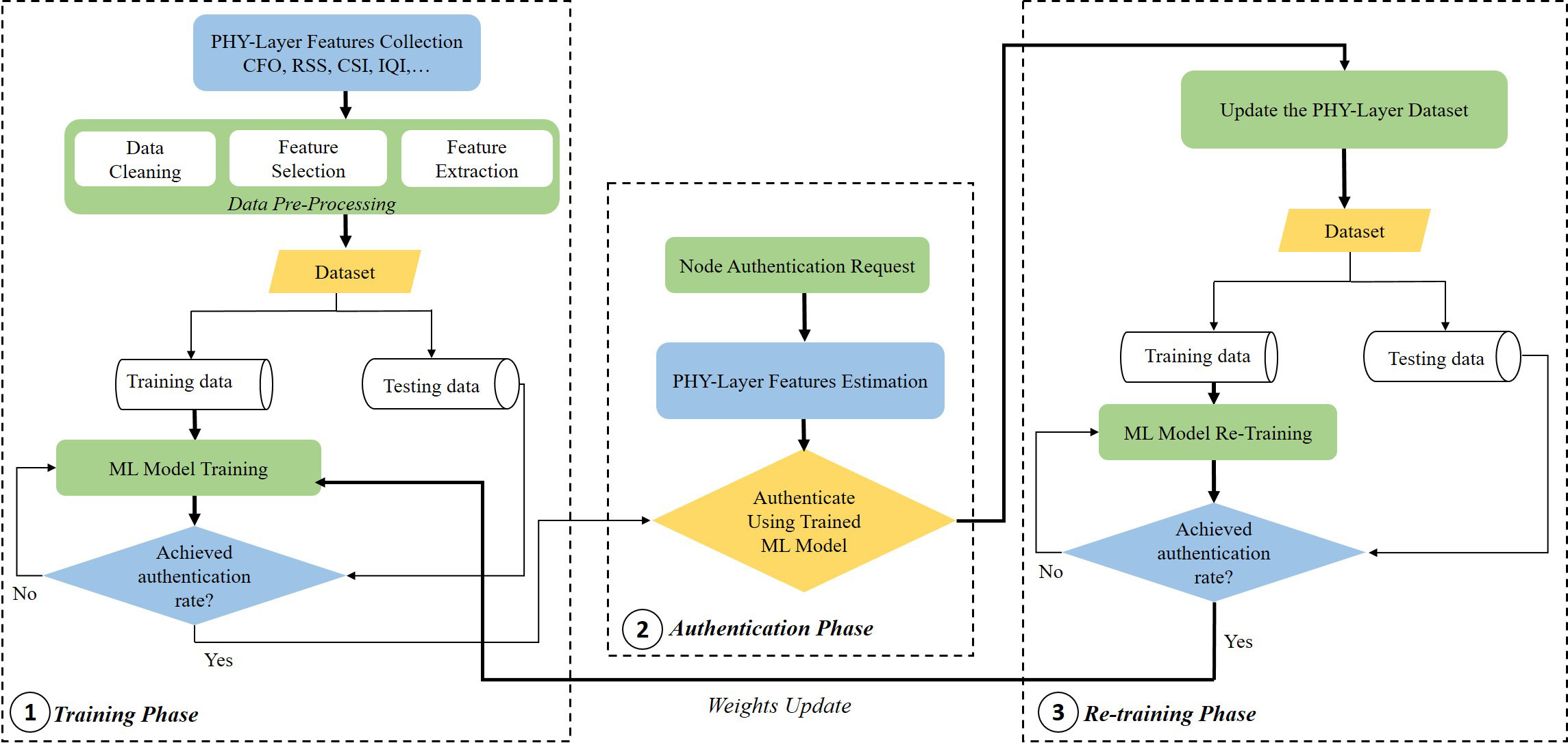} 
\caption{ML-aided physical layer authentication flow chart.}
\label{ML_Based_PLA}
\end{figure*}
\vspace{5pt}

\subsubsection*{\textbf{\textit{Machine Learning-Aided PLA}}} 
Despite the advantages of conventional threshold-based PLA techniques, such as their low computational requirements and network overhead, they encounter various challenges that impede their integration into future wireless networks that involve low-cost and power-constrained devices. These challenges include: 
\begin{enumerate}[label=\roman*)]
\item  \textit{Low Reliability of Single Feature}: Using a single feature for authentication encounters issues due to imperfect estimates, feature variations, and limited range distributions. This inadequacy can lead to inconsistent differentiation of transmitters; 
\item  \textit{Performance Deterioration in Dynamic Environments}: The authentication scheme's performance may degrade when deployed in complex and time-varying environments, impacting its accuracy and robustness; 
\item   \textit{Inability to Detect Varying Security Risks}: Conventional PLA schemes struggle to detect varying security risks effectively, mainly due to the random nature of communication between legitimate entities; 
\item  \textit{Optimum Threshold Determination}: In multi-attributes PLA schemes, finding the optimal threshold poses a challenging task due to the large search space involved. Additionally, in multi-node authentication scenarios, it becomes difficult to discriminate between multiple nodes simultaneously using the same threshold value. 
\item  \textit{Performance Evaluation}: Statistical properties and exact
models for the attributes are required to evaluate the authentication performance.
\end{enumerate}

In this regard, ML techniques have attracted significant interest in the PLA domain due to their inherent capability to learn the variations in the channel and device-based attributes. Thus, higher accuracy is attained when distinguishing between legitimate and illegitimate transmitters. Additionally, ML models exhibit adaptability to diverse environmental conditions, hardware variations, and interference, making ML-aided PLA more resilient in dynamic wireless environments and adaptable to new types of attacks. Finally, ML facilitates the development of PLA schemes that combine multiple physical layer features, leading to improved authentication performance.

The process of authenticating nodes based on ML typically involves three phases, as depicted in Fig. \ref{ML_Based_PLA}. In the first phase, \textit{Training Phase}, physical layer attributes are collected to develop the ML model. The selection of these attributes is crucial, prioritizing those that provide the most pertinent information for authentication. Attributes with a broad distribution range and higher estimation accuracy are particularly valuable as they can effectively distinguish between different transmitters, thereby enhancing authentication reliability. The collected raw data is then processed to maximize its utility, accomplished through data cleaning, feature selection, and feature extraction. Once the dataset is prepared, it is divided into training and testing data for the ML model to be trained effectively. Subsequently, In the second phase, \textit{Authentication Phase}, the well-trained ML model is utilized to authenticate the origin of received messages during the second time slot.  Finally, in the third phase, \textit{Re-training Phase}, the ML models are retrained to track attribute variations in time-varying environments and adapt accordingly. The model weights are updated based on these variations, ensuring continuous robustness in the face of dynamic environmental changes. 

It is crucial to emphasize that the choice of the ML algorithm should be made thoughtfully, considering the availability of the attacker's physical layer features. This is because certain ML algorithms are designed to work effectively with single-label data, while others are better suited for double-label data. For example, binary/multi-label ML classification techniques are ideal when the attacker's information is accessible. In contrast, single-label ML classification techniques and anomaly detection methods are more appropriate when the attacker's information is unavailable. Careful consideration of these factors ensures the selection of the most suitable ML approach to address specific authentication scenarios.

Those well-trained and properly selected ML models for PLA purposes can be then integrated into the ZTN provisioning system. Thus, when a new device attempts to connect to the network, the ZTN PLA framework can automatically initiate the authentication process. The device's physical layer data is collected and sent to the  ML model for classification.

\subsubsection{Cross-Layer Intrusion Detection}
Authentication mechanisms, such as PLA, serve as the first layer of network defense to prevent attacks by filtering unauthorized and suspicious network devices. However, there are certain malicious attacks (\textit{e.g.}, zero-day attacks) that can breach the first layer of defense and intrude on the compromised networks \cite{ericsson}. IDSs are critical components of modern cybersecurity frameworks as the second layer of defense to detect the cyber-attacks that have been launched on networks \cite{mythesis}. IDSs operate by monitoring network events and identifying data patterns or signatures that may indicate cyber-attacks. An IDS obtains network traffic data from the system being protected via network sniffers or Test Access Points (TAPs) \cite{myiotj}. The IDS analyzes incoming network traffic based on currently known types of attacks stored in the database, and then identifies malicious or suspicious network events or behaviors. Once a malicious attack is detected, the IDS will send alarms to the network to trigger countermeasures that can defend against the detected attack.

However, traditional IDSs are typically designed to monitor a single layer of the network protocol stack, primarily the application layer, which can limit their intrusion detection capabilities. For instance, an IDS monitoring the application layer might miss an attack targeting the network layer. Furthermore, traditional IDS often lack the ability to adapt to evolving network events or threats and can generate a large number of false alarms, as information in a single layer may be insufficient to comprehensively indicate these events or threats \cite{pwpae}. 

To address these limitations, the concept of cross-layer intrusion detection has emerged, attracting significant attention from researchers and practitioners alike. A Cross-Layer IDS (CLIDS) is an advanced security mechanism that integrates the information from multiple layers of the network stack to detect malicious activities, enabling the detection of complex, multi-layer attacks \cite{clids2}. For instance, a DDoS attack can target both the network and transport layers, and a CLIDS may intercommunicate between these two layers to detect all DDoS attacks effectively.

Compared to single-layer intrusion detection systems, CLIDSs offer several advantages \cite{cl1}:
\begin{enumerate}[label=\roman*)]
\item \textit{Improved Effectiveness And Comprehensive Protection}: Cross-layer IDSs enable attack detection across all protocol layers, providing comprehensive protection against cyber-attacks that can occur at different layers. By considering network behavior from multiple layers, these systems can make more accurate and reliable decisions, reducing false positives and enhancing the resiliency of the network to various types of cyber-attacks.
\item \textit{Efficient Resource Utilization}: Rather than implementing independent security solutions at each layer, a CLIDS allows for joint intrusion detection by leveraging the shared resources of different layers. This approach optimizes resource utilization, reducing redundancy and conserving power in resource-constrained wireless networks.
\item \textit{Collaborative Detection and Reduced False Alarms}: The collaborative nature of cross-layer intrusion detection enables effective fault diagnosis and decreases false alarms. By combining information from different layers, a holistic view of network behavior is obtained, enabling better identification of genuine attacks and reducing the likelihood of misclassifying legitimate network events.

\end{enumerate}


\section{Related Work}

This section provides a literature review of the current state-of-the-art cybersecurity solutions, focusing on PLA methods for physical layer security, and CLIDSs and automated IDSs for upper-layer security.

\subsection{Physical Layer Security Solutions: Physical Layer Authentication}

Motivated by the promising capabilities of PLA techniques, different research studies have been conducted in the open literature with the aim of enhancing and optimizing the implementation of PLA techniques to address diverse security challenges. Within this context, two distinct channel-based PLA schemes have been recognized in the open literature: statistical channel information-based PLA and instantaneous channel information-based PLA. The statistical channel information-based PLA scheme relies on metrics such as Received Signal Strength (RSS), which is influenced by path-loss and shadowing effects. Conversely, the instantaneous channel information-based PLA scheme incorporates metrics such as CFR and CIR, which consider the combined effects of path loss, shadowing, and small-scale fading.

In order to provide continuous authentication for network devices in wireless networks, several PLA schemes based on RSS have been developed by researchers as documented in \cite{Illi2022, xiao2018, Pei2014, yang2013}. However, different studies have demonstrated that the utilization of CFR and CIR in PLA schemes achieves superior security performance as compared to RSS-based approaches due to the additional uncertainty introduced by small-scale fading. For instance, in \cite{Xiao2007}, the authors proposed a CFR-based scheme for a single antenna system operating in a time-invariant channel and subsequently extended it to time-variant channels in \cite{Xiao2008}. Similarly, Tugnait \textit{et al.} presented a CIR-based scheme for a single carrier time-invariant channel in \cite{Tugnait2010}. Furthermore, this work was expanded to a multi-carrier transmission system by leveraging phase variation, as described in \cite{Wu2015}, and to a multipath time-variant channel, as detailed in \cite{Liu2011}.

Despite the uniqueness of the channel features, PLA schemes developed based on these features are sensitive to the variations of the channel and are susceptible to multipath fading, interference, and environmental changes. Thus, the performance of channel-based PLA schemes tends to deteriorate, and hence, affects the accuracy and reliability of the PLA process, leading to potential false positive or false negative alarms during the authentication process.  Consequently, researchers have resorted to the inherent device's turn-on transient and steady-state unique features, which are insensitive to channel variations.  It is worth noting that,  while the turn-on transient-based PLA schemes are highly secure, they have lower robustness due to the difficulty of extracting those features at the legitimate node. On the contrary,  steady-state features have relatively lower security but have higher robustness. For instance, the authors in \cite{Xie2022} have proposed a PLA scheme based on the multiple phase noise, where it achieved superior authentication performance compared to the prior schemes due to the elimination of the quantization algorithm. In \cite{Zhang2020}, the authors have incorporated both the channel gain and phase noise for the PLA,  and then they applied hypothesis testing and stochastic process to derive the closed-form expressions for PFA and PD with the consideration of quantization errors.

\begin{table*}
\caption{PLA related Work.}
\label{TablePLA}
\centering
\renewcommand{\arraystretch}{2}
\scalebox{0.82}{\begin{tabular}{m{40pt}m{180pt}m{160pt}m{80pt}m{80pt}}
\hline
\makecell{\textbf{Reference}} & \makecell{\textbf{Security Mechanism}} & \makecell{\textbf{Physical Layer Attribute}} & \makecell{\textbf{ Adaptation to Network \& } \\\textbf{Environment Dynamics}}&\makecell{\textbf{Automated
}\\\textbf{Solutions}}  \\
\hline 
\hline
\rowcolor{aliceblue}
\makecell{\cite{Illi2022}} & \makecell{Hypothesis testing } & \makecell{ RSS }&  \makecell{\tikzxmark } &\makecell{\tikzxmark}\\  \hline  
\rowcolor{aliceblue}

\makecell{\cite{Xiao2008,Tugnait2010}} & \makecell{ Hypothesis testing} & \makecell{ CFR} & \makecell{ \Checkmark}&\makecell{\tikzxmark} \\ \hline
\rowcolor{aliceblue}
\makecell{\cite{Liu2011}} & \makecell{ Hypothesis testing} & \makecell{ CIR} & \makecell{ \Checkmark}&\makecell{\tikzxmark} \\ \hline
\rowcolor{aliceblue}
\makecell{\cite{Xie2022}} & \makecell{Hypothesis testing} & \makecell{ Phase noise} & \makecell{ \tikzxmark}& \makecell{ \tikzxmark} \\ \hline
\rowcolor{aliceblue}
\makecell{\cite{Zhang2020}} & \makecell{Hypothesis testing} & \makecell{ Channel gains and phase noise} & \makecell{ \tikzxmark} & \makecell{ \tikzxmark} \\ \hline
\rowcolor{aliceblue}

\makecell{\cite{hou2012,hou2014}} & \makecell{Hypothesis testing} & \makecell{CFO} & \makecell{ \Checkmark}& \makecell{ \tikzxmark} \\ \hline
 
\rowcolor{aliceblue}
 
\makecell{\cite{Dolatshahi2010}} & \makecell{Hypothesis testing} & \makecell{Imperfections of RF power amplifier} & \makecell{ \tikzxmark} &\makecell{ \tikzxmark}\\ \hline
\rowcolor{aliceblue}
 
\makecell{\cite{Rahman2014}} & \makecell{Hypothesis testing} & \makecell{Time-varying clock offsets} & \makecell{ \tikzxmark} & \makecell{ \tikzxmark} \\ \hline
\rowcolor{aliceblue}

 \makecell{\cite{Hao2014,Hao_ICC_2014}} & \makecell{Hypothesis testing} & \makecell{IQI} & \makecell{ \tikzxmark} & \makecell{ \tikzxmark} \\ \hline

\rowcolor{paleaqua}
\makecell{\cite{xiao2018}} & \makecell{ Logistic regression} & \makecell{ RSS  with
multiple landmarks\\ each with multiple antennas} & \makecell{ \tikzxmark}& \makecell{ \tikzxmark}  \\ \hline

\rowcolor{paleaqua}
\makecell{\cite{Pei2014,yang2013}} & \makecell{Support vector machine (SVM) and \\ linear Fisher discriminant analysis} & \makecell{ Time-of-arrivals,  RSS,\\ and cyclic-features of the channel} & \makecell{ \tikzxmark} & \makecell{ \tikzxmark}\\ \hline

\rowcolor{paleaqua}
\makecell{\cite{pan2019}} & \makecell{k-nearest neighbour \\\& Decision and bagged trees} & \makecell{ CSI} & \makecell{ \tikzxmark} & \makecell{ \tikzxmark} \\ \hline

\rowcolor{paleaqua}
\makecell{\cite{Hoang2020}} & \makecell{ k-means, One-class SVM } & \makecell{ Mean \& varaince of the recieved signal} & \makecell{ \tikzxmark}& \makecell{ \tikzxmark} \\ \hline

\rowcolor{paleaqua}
\makecell{\cite{Qiu2018}} & \makecell{ Gaussian mixture model } & \makecell{ CSI} & \makecell{ \tikzxmark}& \makecell{ \tikzxmark} \\ \hline

\rowcolor{paleaqua}
\makecell{\cite{Liao2020}} & \makecell{ Neural network } & \makecell{ CFR} & \makecell{ \tikzxmark}& \makecell{ \tikzxmark} \\ \hline

\rowcolor{paleaqua}
\makecell{\cite{Qiu2020}} & \makecell{ Neural networks  } & \makecell{CFR} & \makecell{\Checkmark }& \makecell{\tikzxmark } \\ \hline

\rowcolor{paleaqua}
\makecell{\cite{Xiao2016}} & \makecell{ Reinforcement learning } & \makecell{RSS } & \makecell{ \Checkmark}& \makecell{ \tikzxmark}  \\ \hline

\rowcolor{paleaqua}
\makecell{\cite{Abdrabou2022}} & \makecell{ One-class SVM } & \makecell{Magnitude, real, and imaginary parts\\ of the received signal} & \makecell{ \Checkmark}& \makecell{ \tikzxmark} \\ \hline

\rowcolor{paleaqua}
\makecell{\cite{Chen2021}} & \makecell{ SVM with auto-labeling } & \makecell{CFR} & \makecell{ \tikzxmark}& \makecell{ \Checkmark} \\ \hline

\rowcolor{paleaqua}
\makecell{\cite{fang2019}} & \makecell{ Kernel-based  } & \makecell{CFO, CIR, RSS} & \makecell{\Checkmark }& \makecell{\tikzxmark } \\ \hline

\rowcolor{paleaqua}
\makecell{\cite{Wang2019}} & \makecell{Deep neural networks } & \makecell{CSI} & \makecell{\tikzxmark }& \makecell{\tikzxmark } \\ \hline

\rowcolor{palecyan}
\makecell{-} & \makecell{ Proposed AutoML-based PLA } & \makecell{RF fingerprint (IQI)} & \makecell{ \Checkmark} & \makecell{ \Checkmark} \\ \hline

\renewcommand{\arraystretch}{1}

\end{tabular}
}
\end{table*}

Furthermore, the work in \cite{hou2012} proposed a PLA scheme based on the carrier frequency offset (CFO) in a time-invariant orthogonal frequency division multiplexing (OFDM) system, which has been extended in \cite{hou2014} to a time-varying scenario, where the CFO was modeled as an Auto-Regressive random process. The imperfections of power amplifiers have been utilized by using the generalized likelihood ratio test (GLRT) and LRT in \cite{Dolatshahi2010}.  Clock offset and clock skew are unavoidable due to the device's imperfection; therefore, the authors in \cite{Rahman2014} have introduced PLA schemes that rely on time-varying clock offsets. To effectively track the clocks of nodes, the proposed approach incorporates two Kalman filters. Finally, the utilization of In-phase and Quadrature components Imbalance (IQI) samples as a feature for PLA purposes has been introduced in \cite{Hao2014,Hao_ICC_2014}. This novel approach leverages the inherent imbalances between the in-phase and quadrature components of the received signals to extract relevant information for the authentication process. A summary of the available contributions in the literature on threshold-based PLA is provided in Table \ref{TablePLA}.

As mentioned in Section \ref{sec:pla}, conventional threshold-based PLA schemes rely on hypothesis testing, which involves optimizing the test threshold to differentiate between devices. However, this optimization process becomes challenging when the features exhibit dynamic variations, particularly in complex and dynamic wireless environments. Consequently, there is a need for intelligent authentication approaches that can adapt automatically and learn implicitly from data without relying on specific attribute modes, thereby addressing these challenges for enhanced security and more efficient management in future wireless networks \cite{Fang2019}. Table \ref{TablePLA} illustrates a summary of the available contributions related to ML-aided PLA. It can be noticed that the majority of existing studies in the literature have primarily focused on utilizing the channel or device features as attributes for ML models without adaptation to the network dynamics or environmental changes \cite{Liao2020, pan2019, fang2019, Wong2018, jian2020, Merchant2018}. Additionally, none of the existing works in the literature have presented an automated PLA framework that is capable of addressing various threats in the physical layer using AutoML techniques through AutoDP, AutoFE, automated model selection, HPO, and automated model updating.


\subsection{Upper Layer Security Solutions: Cross-Layer and Automated Intrusion Detection Systems}

IDSs and anomaly detection systems are effective methods to identify cyber-attacks in modern networks. CLIDSs have emerged as a promising solution to address the limitations of traditional single-layer oriented IDSs. By integrating insights from multiple layers of the protocol stack, CLIDSs enable comprehensive detection of cyber attacks that span multiple layers. To effectively detect attacks across multiple layers, several methods have been dedicated to developing comprehensive CLIDS frameworks.

Zhu \textit{et al.} \cite{CLIDS_review1} proposed a cross-layer defense scheme for Communication-Based Train Control (CBTC) systems, a key Cyber-Physical System (CPS) in smart transportation. The framework employed a cross-layer approach, integrating the physical, cyber, and management layers to mitigate the impact of jamming attacks during train handoff processes. For example, a Model Predictive Control (MPC) algorithm at the physical layer optimized train profiles, while a stochastic game model at the cyber layer provided a randomized channel selection defense against attackers. Although this framework significantly mitigated communication disruptions in CBTC systems, its domain-specific nature and reliance on pre-defined defense models pose challenges for application in more dynamic networking environments.

Amouri \textit{et al.} \cite{CLIDS_review2} developed a two-stage CLIDS for Mobile Ad Hoc Networks (MANETs) and Wireless Sensor Networks (WSNs). The system used a heuristic-based detection mechanism called the Accumulated Measure of Fluctuation (AMoF) to identify blackhole and DDoS attacks. By analyzing fluctuations in Correctly Classified Instances (CCIs) collected from the MAC and network layers, the system achieved detection rates exceeding 98\% under high power and velocity conditions. This approach demonstrated robust performance in many simulated networking environments, including high power/node velocity and low power/node velocity scenarios.

Gandhimathi \textit{et al.} \cite{CLIDS_review3} proposed a hybrid two-stage IDS for WSNs. In the first stage, a flow-based anomaly detection model is used to detect abnormal network traffic. In the second stage, cross-layer features are extracted and correlated to improve detection performance. For example, the routing information extracted at the network layer and the medium access duration information collected at the MAC layer could be analyzed together to detect DoS attacks. The authors evaluated their proposed IDS in resource-constrained networks and reported that it achieved better performance than traditional single-layer IDSs in terms of detection accuracy and energy consumption.

While these examples demonstrate the effectiveness of CLIDSs in specific contexts, challenges such as scalability, adaptability to zero-day attacks, and real-time operation in dynamic networking environments remain unresolved. This work aims to address these gaps by proposing a comprehensive CLIDS capable of adapting to continuous network data streams and detecting novel attack patterns in real-time. The proposed SH-CASH framework integrates cross-layer features (e.g., physical-layer RF signatures, network-layer packet flows, and application-layer logs) to detect anomalies indicative of previously unseen attacks. The adaptability of the SH-CASH framework ensures that the CLIDS can learn and respond to novel threats, even in dynamic network environments.

\begin{table*}
\caption{CLIDS and Automated IDS Related Work.}
\label{IDS_liter}
\centering
\renewcommand{\arraystretch}{2}
\scalebox{0.82}{
\begin{tabular}{m{30pt}m{150pt}m{100pt}m{80pt}m{80pt}m{80pt}}
\hline
\makecell{\textbf{Reference}} & \makecell{\textbf{Security Mechanism}} & \makecell{\textbf{Dataset}} & \makecell{\textbf{Cross-Layer}  \\\textbf{Security}} & \makecell{\textbf{Automated}  \\\textbf{ Solutions}} & \makecell{\textbf{ Adaptation to Network} \\\textbf{Dynamics}} \\
\hline 
\hline
\rowcolor{aliceblue}
\makecell{\cite{CLIDS_review1}} & \makecell{Markov process} & \makecell{Simulated data} & \makecell{\Checkmark} & \makecell{\tikzxmark} & \makecell{\tikzxmark} \\ \hline
\rowcolor{aliceblue}
\makecell{\cite{CLIDS_review2}} & \makecell{ML models (RF)} & \makecell{Simulated data} & \makecell{\Checkmark} & \makecell{\tikzxmark} & \makecell{\tikzxmark} \\ \hline
\rowcolor{aliceblue}
\makecell{\cite{CLIDS_review3}} & \makecell{Statistical methods} & \makecell{Simulated data} & \makecell{\Checkmark} & \makecell{\tikzxmark} & \makecell{\tikzxmark} \\ \hline
\rowcolor{paleaqua}
\makecell{\cite{automl_review1}} & \makecell{AutoML-ID} & \makecell{Intrusion-Data-WSN} & \makecell{\tikzxmark} & \makecell{\Checkmark} & \makecell{\tikzxmark} \\ \hline
\rowcolor{paleaqua}
\makecell{\cite{automl_review2}} & \makecell{OE-IDS} & \makecell{UNSW-NB15, CICIDS2017} & \makecell{\tikzxmark} & \makecell{\Checkmark} & \makecell{\tikzxmark} \\ \hline
\rowcolor{paleaqua}
\makecell{\cite{automl_review3}} & \makecell{Double-PSO} & \makecell{CICIDS2017} & \makecell{\tikzxmark} & \makecell{\Checkmark} & \makecell{\tikzxmark} \\ \hline
\rowcolor{paleaqua}
\makecell{\cite{myautoml}} & \makecell{Existing ML models (RF, LightGBM, etc.)} & \makecell{CICIDS2017, IoTID20} & \makecell{\tikzxmark} & \makecell{\Checkmark} & \makecell{\Checkmark} \\ \hline
\rowcolor{palecyan}
\makecell{-} & \makecell{Proposed AutoML-based framework} & \makecell{CICIDS2017} & \makecell{\Checkmark} & \makecell{\Checkmark} & \makecell{\Checkmark} \\ \hline
\renewcommand{\arraystretch}{1}
\end{tabular}
}
\end{table*}

\begin{table*}[htbp]
\caption{A comprehensive comparison of ML models, their advantages and limitations, and their suitability for cybersecurity tasks.}
\setlength\extrarowheight{1pt}
\centering
\scriptsize
\begin{tabular}{p{2.5cm}|p{7cm}|p{7cm}}
\Xhline{1.2pt}
\textbf{ML Algorithm} & \textbf{Advantages and Limitations} & \textbf{Cybersecurity Task Suitability} \\ 
\hline
Logistic Regression & 
\begin{tabular}[t]{@{}p{7cm}} 
· Simple to implement and interpret. \\ 
· Works well with linearly separable data. \\ 
· Not suitable for complex non-linear relationships. 
\end{tabular} & 
\begin{tabular}[t]{@{}p{7cm}} 
Suitable for binary classification tasks, such as simple anomaly detection in PLA. 
\end{tabular} \\ 
\hline

SVM & 
\begin{tabular}[t]{@{}p{7cm}} 
· Effective in high-dimensional spaces. \\ 
· Works well with non-linear data using kernel functions. \\ 
· Requires careful parameter tuning and is not suitable for large datasets. 
\end{tabular} & 
\begin{tabular}[t]{@{}p{7cm}} 
Effective for both PLA and CLIDS tasks where data has complex patterns but with relatively smaller datasets. 
\end{tabular} \\ 
\hline

KNN & 
\begin{tabular}[t]{@{}p{7cm}} 
· Easy to implement. \\ 
· Effective for small datasets. \\ 
· Computationally expensive for large datasets and sensitive to noise. 
\end{tabular} & 
\begin{tabular}[t]{@{}p{7cm}} 
Suitable for small-scale PLA tasks where real-time classification is not critical. 
\end{tabular} \\ 
\hline

Decision Tree  & 
\begin{tabular}[t]{@{}p{7cm}} 
· Easy to interpret and visualize. \\ 
· Can handle non-linear data. \\ 
· Prone to overfitting, especially with deep trees. 
\end{tabular} & 
\begin{tabular}[t]{@{}p{7cm}} 
Useful for simple PLA tasks and initial stages of CLIDS, where interpretability is key. 
\end{tabular} \\ 
\hline

Random Forest  & 
\begin{tabular}[t]{@{}p{7cm}} 
· Reduces overfitting compared to single trees. \\ 
· Handles large datasets well. \\ 
· Can manage imbalanced datasets. 
\end{tabular} & 
\begin{tabular}[t]{@{}p{7cm}} 
Highly suitable for both PLA and CLIDS tasks, especially in environments with complex and high-dimensional data. 
\end{tabular} \\ 
\hline

LightGBM & 
\begin{tabular}[t]{@{}p{7cm}} 
· High efficiency and speed. \\ 
· Handles large datasets and high-dimensional features well. \\ 
· Sensitive to overfitting if not carefully tuned. 
\end{tabular} & 
\begin{tabular}[t]{@{}p{7cm}} 
Ideal for real-time PLA and CLIDS tasks, particularly where fast processing of large volumes of data is required. 
\end{tabular} \\ 
\hline

K-means & 
\begin{tabular}[t]{@{}p{7cm}} 
· Simple to implement. \\ 
· Low computational complexity. \\ 
· Assumes clusters are globular, which may not suit all data distributions. 
\end{tabular} & 
\begin{tabular}[t]{@{}p{7cm}} 
Suitable for unsupervised anomaly detection in PLA and CLIDS, particularly in simpler scenarios with clearly defined clusters. 
\end{tabular} \\ 
\hline

One-Class SVM & 
\begin{tabular}[t]{@{}p{7cm}} 
· Effective for anomaly detection in an unsupervised setting. \\ 
· Can handle high-dimensional data. \\ 
· Sensitive to parameter settings and requires careful tuning. 
\end{tabular} & 
\begin{tabular}[t]{@{}p{7cm}} 
Best suited for anomaly detection in both PLA and CLIDS when the task involves detecting outliers in high-dimensional data. 
\end{tabular} \\ 
\hline

Gaussian Mixture Model & 
\begin{tabular}[t]{@{}p{7cm}} 
· Provides probabilistic clustering. \\ 
· Can handle data that comes from a mixture of distributions. \\ 
· Requires the number of components to be specified, which can be complex. 
\end{tabular} & 
\begin{tabular}[t]{@{}p{7cm}} 
Suitable for anomaly detection and clustering in PLA and CLIDS, especially when data is expected to follow multiple distributions. 
\end{tabular} \\ 
\hline

Reinforcement Learning  & 
\begin{tabular}[t]{@{}p{7cm}} 
· Learns through interaction with the environment. \\ 
· Can be combined with NNs for complex tasks. \\ 
· High computational complexity and long training times. 
\end{tabular} & 
\begin{tabular}[t]{@{}p{7cm}} 
Best suited for adaptive cybersecurity strategies in CLIDS, particularly in dynamic environments where the model can continuously learn and adapt. 
\end{tabular} \\ 
\hline

Deep Neural Networks  & 
\begin{tabular}[t]{@{}p{7cm}} 
· Powerful for learning complex, high-dimensional patterns. \\ 
· Requires a large dataset and high computational power. \\ 
· Prone to overfitting, requires regularization techniques. 
\end{tabular} & 
\begin{tabular}[t]{@{}p{7cm}} 
Highly effective for advanced CLIDS tasks, especially in detecting sophisticated and evolving threats, provided the availability of large datasets and computing resources. 
\end{tabular} \\ 
\Xhline{1.2pt}
\end{tabular}
\label{ML}%
\end{table*}

On the other hand, in response to the growing complexity of cybersecurity challenges, there has been a surge of interest in the development of automated cybersecurity mechanisms. While this field is still in its early stage, a number of innovative studies have emerged that utilize AI/ML and AutoML techniques to develop IDSs.

Singh \textit{et al.} \cite{automl_review1} proposed an automated IDS, AutoML-ID, for WSNs, as traditional ML-aided IDSs often require extensive expertise and time for model selection and hyperparameter tuning. The AutoML-ID model can automatically select the best-performing ML model and the optimal hyperparameter values using Bayesian optimization, and achieve better performance than traditional ML models on the public Intrusion-Data-WSN dataset. However, this AutoML-ID model is only designed for known attack detection and supervised learning, and is incapable of zero-day attack detection in dynamic networking environments.  

Khan \textit{et al.} \cite{automl_review2} proposed an automated IDS, Optimized Ensemble (OE)-IDS, using a soft voting method for network intrusion detection. The AutoML framework is used in this study to select optimal supervised base models and develop optimal ensemble strategies. The OE-IDS model can achieve high intrusion detection rates on the University of New South Wales 2015 Network Benchmark (UNSW-NB15) and CICIDS2017 datasets. Similar to AutoML-ID, although this OE-IDS method can achieve good intrusion detection performance, it is designed for supervised models and lacks the capability for new attack detection in modern networks.

Elmasry \textit{et al.} \cite{automl_review3} proposed an advanced IDS based on a dual Particle Swarm Optimization (PSO) metaheuristic combined with deep learning techniques. The dual PSO method is employed for both feature selection and hyperparameter optimization, streamlining the IDS training process. The framework integrates three deep learning models to evaluate and enhance detection performance. Experimental results on benchmark datasets, including CICIDS2017, demonstrated significant improvements in detection accuracy and false alarm reduction. The study emphasizes the efficacy of leveraging PSO to automatically optimize deep learning architectures for robust network intrusion detection.

Yang \textit{et al.} \cite{myautoml}  comprehensively discussed how to apply AutoML techniques to network applications, \textit{i.e.}, the Internet of Things (IoT) systems. In this paper, a comprehensive review of AutoML methods in all procedures of the general ML or data analytics pipeline is conducted, including AutoDP, AutoFE, automated model selection, hyperparameter optimization, and automated model updating. This paper also conducted a case study on applying general AutoML methods (e.g., random forest and LightGBM) to CICIDS2017 and IoTID20 datasets for automated IoT anomaly detection. Although this paper has introduced AutoML to IoT systems, the specific and practical applications of AutoML in B5G/6G networks are still limited. 

ZTNs and network automation are key enablers in 6G networks, and AutoML techniques are promising solutions for achieving them. Thus, it is crucial to develop automated methods/frameworks for 6G network applications. Automated solutions can significantly enhance the security of these advanced networks by providing efficient and accurate detection of cyber threats. In this paper, we propose a novel automated cybersecurity framework to solve both PLA and CLIDS problems in dynamic networking environments. A summary of the available contributions in the literature on CLIDSs and automated IDSs is provided in Table \ref{IDS_liter}.

In summary, the advantages, limitations, and cybersecurity task suitability of the ML models in the literature are summarized in Table \ref{ML}. Appropriate ML models should be chosen according to specific problems and use cases.

\section{Proposed Automated Cybersecurity Framework} 
\subsection{AutoML Introduction and System Overview}

\begin{figure*}
     \centering
     \includegraphics[width=16.5cm]{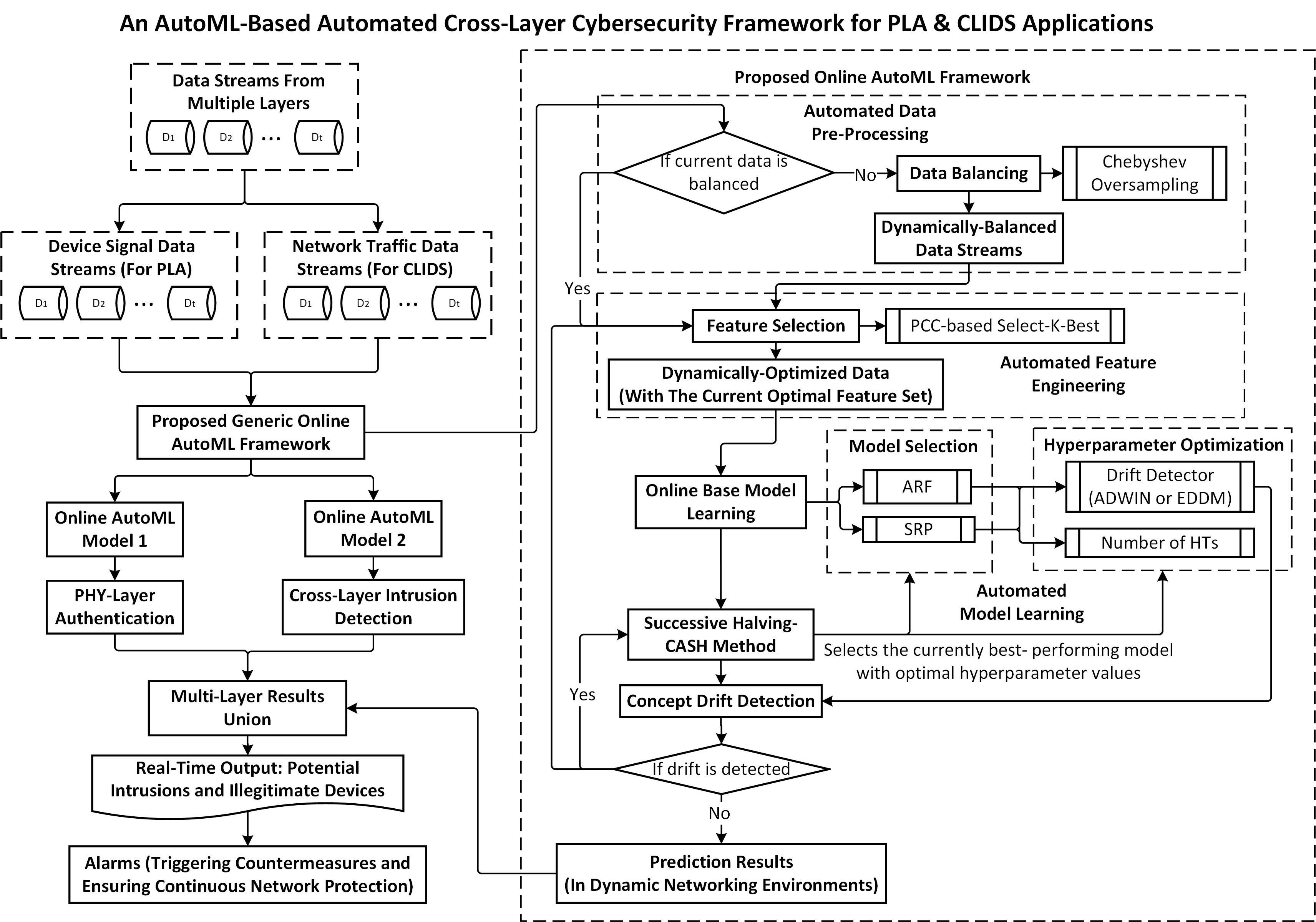}
     \caption{The framework of the proposed AutoML-based automated cybersecurity solutions. } \label{frame}
\end{figure*}

AI/ML techniques are pivotal in bolstering 6G security beyond the scope of traditional mechanisms such as authentication and IDSs. AI/ML techniques also facilitate efficient ZTN management in 6G networks, enabling automation and optimization across a variety of network functionalities \cite{zsmml1}. They have been proven as promising solutions to enhance cybersecurity across multiple layers of the Internet protocol stack, from the physical layer to the application layer \cite{pla2}. ML techniques can be classified as supervised learning, unsupervised learning, and reinforcement learning techniques, based on the type of learning tasks or available datasets \cite{ml1} \cite{lccde}.

Traditional AI/ML algorithms, despite their proven effectiveness in various fields, face several challenges when applied to cybersecurity tasks \cite{myautoml} \cite{automl2}:
\begin{enumerate}[label=\roman*)]
\item \textit{Time and Effort}: Traditional ML techniques necessitate extensive manual tasks, causing delays in the rapid development and deployment of ML models in 5G/6G networks.
\item \textit{Bias and Errors}: Manual development of ML models can introduce biases and errors, resulting in problems such as overfitting and underfitting, which negatively affect model performance.
\item \textit{Expertise Requirement}: Practical application of traditional ML algorithms often demands high levels of expertise, thereby creating barriers to interdisciplinary collaboration and increasing the need for skilled ML professionals.
\item \textit{Limited Adaptability to Dynamic Networks}: Traditional ML models exhibit limited adaptability and struggle to cope with dynamic networking environments, including real-time detection of unknown or zero-day attacks.
\end{enumerate}

AutoML has emerged as a comprehensive solution to address these challenges of traditional AI/ML models to fully leverage the potential of AI/ML techniques in enhancing security within the fast-evolving 6G network landscape \cite{automl1}. AutoML refers to a fully automated process of applying ML algorithms to real-world problems. The primary objective of AutoML is to empower domain specialists to develop ML models autonomously, without requiring comprehensive knowledge or expertise. Consequently, AutoML addresses the shortage of data scientists and enhances the performance and effectiveness of ML models by shortening work cycles and improving model performance \cite{automl1}. Additionally, online AutoML techniques can be used to address dynamic data stream analytics problems. As the demand for network automation rises, AutoML techniques have become necessary components in ZTNs and 6G networks.

The overview of the proposed automated security framework for ZTNs and future networks is demonstrated in Fig. \ref{frame}. The proposed framework is primarily designed to solve two critical cybersecurity problems, PLA and CLIDS, but can be extended to the development of various other data-driven cybersecurity methods. The proposed online AutoML framework is the core of the framework. This study constructs two online AutoML models, each designed to automatically adapt to the physical device signal and network traffic datasets (for PLA and CLIDS tasks, respectively) to achieve optimal performance. The proposed online AutoML framework consists of three primary stages: AutoDP, AutoFE, and automated model learning. All incoming dynamic network data streams are pre-processed in the AutoDP stage, which focuses on automated data balancing using the Chebyshev oversampling method to detect and address class-imbalance issues automatically. In the AutoFE stage, a PCC-based Select-K-Best method is employed to dynamically select the current important features from the original dataset and discard irrelevant features. In the automated model learning stage, ARF and SRP models with different hyperparameter values are used to construct the online learners for device authentication and intrusion detection. The SH-CASH model is then used to automatically select the real-time best-performing base models to achieve optimal model performance. The automated feature re-selection and model re-selection are triggered when model or concept drift is detected using two drift detectors, ADWIN or EDDM. Finally, the union results of PLA and CLIDS, including potential intrusions and illegitimate devices, are output to trigger further countermeasures to safeguard against the detected cyber-attacks. The details of each component in the proposed framework will be discussed in the following subsections.

The proposed framework aims to detect malicious activities and cyber-attacks at multiple layers of the Internet protocol stack to comprehensively and effectively safeguard networks against various security threats. In the proposed cross-layer cybersecurity framework, PLA serves as the first layer of defense, aimed at preventing unauthorized access at the physical layer by authenticating devices based on unique signal characteristics. This preventive mechanism is critical for ensuring that only legitimate devices can access network, significantly reducing the potential for cyber-attacks at the source. Following PLA, CLIDS serves as the second layer of defense, focused on the detection and mitigation of any cyber-attacks that might occur across various network layers. By implementing CLIDS, the proposed framework can identify and respond to threats that surpass PLA at the physical layer, ensuring a comprehensive view of cybersecurity. Additionally, the unified application of AutoML techniques across both PLA and CLIDS components ensures that the security measures are not only optimized for effectiveness but also dynamically adapt to evolving threats by analyzing both physical layer and upper layer abnormal patterns.

Moreover, the evaluation of the proposed AutoML framework on both PLA and CLIDS problems indicates its general usage and implementation, justifying its wide application in addressing other cybersecurity problems. This aligns with the primary objective of AutoML, which is to automatically design optimal solutions to various problems within the same domain.

\subsection{Automated Data Pre-Processing (AutoDP)}

Data pre-processing is a crucial step in the ML and data analytics pipeline, as it has a direct impact on input data quality and ML model performance. Automated data pre-processing, or AutoDP, is an initial and crucial stage of AutoML, which aims to automatically identify and mitigate the data quality issues in the datasets, such as outliers, missing values, and class imbalance \cite{automl3}. 

In the proposed framework, the AutoDP component focus on automated data balancing to address class imbalance issues. Class imbalance is a critical data quality issue for network data analytics, especially for cybersecurity and anomaly detection problems \cite{myiotj}. In real-world applications, data streams from network environments are often imbalanced, with a majority of samples representing normal traffic and a minority representing attacks. Imbalanced data often results in biased models that prioritize majority-class data (\textit{e.g.}, normal data) and perform poorly on minority-class data (\textit{e.g.}, attack data). Additionally, class distributions in dynamic network systems are dynamic variables that may change significantly over time \cite{tii}. Hence, it is usually challenging to maintain data balance in dynamic networking environments. The proposed AutoDP aims to address this imbalance by dynamically generating synthetic samples of the minority class. For example, in a dynamic wireless network, if data packets associated with physical-layer attacks (\textit{e.g.}, jamming) are underrepresented, AutoDP generates additional samples to ensure balanced data and unbiased models for the learning process. This enables the subsequent components to perform effectively, even under skewed class distributions.

Data resampling approaches, including over-sampling and under-sampling approaches, are commonly used for data balancing tasks \cite{sample1}. The Chebyshev Over-Sampling (COS) method \cite{cos} is an advanced over-sampling technique that generates synthetic instances of the minority class based on the Chebyshev inequality. The Chebyshev inequality defines a bound on the probability of a random variable deviating from its mean by an amount greater than a certain amount, denoted by \cite{cos}: 
\begin{equation}
P(|X - \mu| \geq k\sigma) \leq \frac{1}{k^2}
\end{equation}
where \(X\) is a random variable in the dataset, 
\(\mu\) and \(\sigma\) are the mean and the standard deviation of \(X\),  and \(k\) is a positive constant.

In the COS method, the Chebyshev inequality is used to calculate the distance between instances in the minority class. New synthetic samples are created for the minority class by interpolating between original samples and randomly chosen samples from the minority class until the data distributions are balanced. 

Unlike the Dynamic Random Over-Sampling (DROS) method, which randomly generates more samples for minority classes, the COS method generates synthetic instances that are more representative of the underlying data distribution. Additionally, using under-sampling methods, such as Dynamic Random Under-Sampling (DRUS), may lose important data patterns as they discard the majority class data samples to balance data \cite{sample1}. Thus, COS is selected for automated data balancing in the proposed framework to generate high-quality and representative minority samples without losing important information. For dynamic data stream analytics, the proposed data balancing module will continuously evaluate whether the current data is balanced and automatically trigger the COS method to generate synthetic data samples until the data is balanced.

\subsection{Automated Feature Engineering (AutoFE)}

After data pre-processing, the data quality is further improved by implementing Automated Feature Selection (AutoFS). AutoFS is a crucial procedure of AutoFE that aims to automatically identify and remove irrelevant or redundant features \cite{automl1}. As the original network features are usually not optimal for specific tasks, it is important to implement AutoFS to reduce data noise and improve learning efficiency. In the AutoFE component of the proposed framework, the Pearson Correlation Coefficient (PCC)-based Select-K-Best method is used for feature selection. For example, in a cross-layer intrusion detection scenario, AutoFE aims to automatically select the most important features, such as channel state information (CSI) for the physical layer and packet inter-arrival time for the network layer. By reducing irrelevant or redundant features, AutoFE ensures that the model focuses on high-impact attributes, leading to faster learning and improved accuracy. This step is critical in dynamic environments where network conditions and feature relevance may change over time.

The Select-K-Best method is a widely used feature selection method that ranks features according to their importance scores, which are determined by their correlation with the target variable \cite{tii}. PCC is a common metric used to calculate these correlations. The PCC between each feature $X$ and the target variable $y$ can be computed using the following formula \cite{fs1}:
\begin{equation}
PCC_{xy} = \frac{\sum_{i=1}^{n} (x_i - \bar{x})(y_i - \bar{y})}{\sqrt{\sum_{i=1}^{n} (x_i - \bar{x})^2 \sum_{i=1}^{n} (y_i - \bar{y})^2}}
\end{equation}
where \(x_i\) and \(y_i\) are individual instances of the input feature and the target variable, respectively, and \(\bar{x}\) and \(\bar{y}\) are their respective mean values.

PCC measures the correlation between two variables and comes in a range of -1.0 to 1.0, where -1.0 indicates a perfect negative correlation, 1.0 indicates a perfect positive correlation, and 0 indicates no correlation at all. This range allows for a quantitative comparison of the strength of the relationship between each feature and the target variable, thereby facilitating the assessment of feature importance. Moreover, the PCC-based Select-K-Best method is computationally efficient, making it suitable for large-scale network datasets. In contrast to wrapper methods that necessitate re-training the ML model, the PCC-based Select-K-Best method only requires straightforward calculations on the data to determine the PCC value for each feature \cite{tii}.

In the proposed framework, the PCC-based Select-K-Best method is automated by leveraging drift detection methods for dynamic network data stream analytics. The optimal feature set is generated using this method based on the originally available dataset, and then updated when a model drift is detected, which indicates a data distribution change and model performance degradation. Therefore, the PCC-based Select-K-Best method is continuously triggered by drift detectors in the proposed AutoFE process. The drift detection procedures are discussed in detail in the next subsection. 

The effectiveness of AutoFE relies heavily on the quality of data processed by AutoDP. For instance, in detecting zero-day attacks, AutoDP ensures that imbalanced attack samples are effectively balanced and represented in real-time, while AutoFE extracts the most discriminative features from these samples in real-time. Together, these components enable the SH-CASH framework to dynamically adapt to evolving network conditions and emerging attacks. This capability is particularly advantageous in real-time applications, such as PLA and CLIDS, where data evolves continuously.

\subsection{Automated Model Learning}
The challenge in real-world applications of ML algorithms is in selecting and configuring the appropriate ML model for optimal results. Hyperparameters, which define the architecture of ML models, play a critical role in this selection process. The problem of selecting an appropriate ML model and hyperparameter values is considered as a search problem, known as Combined Algorithm Selection and Hyper-parameter (CASH) optimization problem \cite{autoweka}. CASH involves two levels: automated model selection and HPO. Automated model selection aims to select suitable ML models for specific datasets/tasks, while hyper-parameter tuning or HPO aims to tune and optimize model-specific hyperparameters for achieving the best model performance \cite{automl3}. 

On the other hand, ML models need to be supervised and updated to maintain their prediction performance over time by adapting to data distribution changes. The automated model updating process helps ML models evolve with data distribution changes or model drift issues, thereby preventing model performance degradation \cite{automl3}. The automated model updating process includes two phases: drift detection and drift adaptation. The former monitors the model’s performance and triggers model updating when the performance degrades, while the latter updates the ML model when new concept data is detected \cite{tii}.

For automated model learning, the proposed framework involves two stages: base model learning with drift detection and CASH. Base model learning with drift detection is the process of continuously training online learning models on dynamic network data streams and updating these models when model drift is detected. Model drift indicates the learning model performance degradation due to data distribution changes \cite{dmdrift}. At the first stage of this base model learning process, the online learning models are trained on the originally available dataset as initial base models. The initial base models are then deployed in the network to learn and predict each new instance in the dynamic streams. Finally, the base models are updated on the new concept dataset if model drift occurs. CASH is the process of automatically updating the base model list (\textit{i.e.}, re-selecting the base models) and tuning these models’ hyper-parameters. 

\subsubsection{Drift Detection}
Model drift can be divided into two broad categories: sudden drifts and gradual drifts \cite{tii}. The sudden drift occurs when the data distribution data changes rapidly within a short period of time, while the gradual drift takes place when a new distribution of data gradually replaces an existing distribution. To detect model drift for automated model updating, two drift detectors, ADaptive WINdowing (ADWIN) \cite{adwin} and Early Drift Detection Method (EDDM) \cite{eddm}, are used in the proposed framework. 

ADWIN is a distribution-based drift detection method that employs a dynamic sliding window to detect concept drift \cite{adwin}. This method identifies variations in data distribution by calculating and comparing the characteristic values of the old and new distributions, such as the mean and variance values. ADWIN has a significant advantage when dealing with gradual drifts and long-term changes, as the adaptable sliding window size allows it to accommodate such distribution changes effectively. However, the statistical properties of the data windows may not reflect actual drifts, causing false alarms \cite{mythesis}.

EDDM is a performance-based drift detection method that monitors changes in model performance metrics \cite{eddm}. It tracks performance changes based on the change rate of a learning model’s error rate and standard deviation, using a drift threshold and a warning threshold. Specifically, the EDDM algorithm operates as follows \cite{eddm}:
\begin{enumerate}[label=\roman*)]
\item It monitors the running average and maximum values of the error rate changes and the standard deviations at each time $t$, denoted as $p_{t}^{\prime}$, $p_{\max}^{\prime}$, $s_{t}^{\prime}$, and $s_{\max}^{\prime}$, respectively.
\item If the ratio $(p_{t}^{\prime}+2 * s_{t}^{\prime}) / (p_{\max}^{\prime}+2 * s_{\max}^{\prime})$ falls below a warning threshold $\alpha$, it triggers a warning.
\item If the same ratio falls below a drift threshold $\beta$, it signifies a drift.
\end{enumerate}

EDDM is especially effective in detecting sudden drifts. If the error rate of a model increases dramatically, it indicates a degradation in model performance, signaling the occurrence of model drift. Thus, EDDM can detect all real drifts that have degraded model performance. However, its performance in detecting gradual drifts is inferior to distribution-based methods like ADWIN \cite{tii}. Overall, as EDDM and ADWIN have their own advantages and limitations for the detection of different types of drifts (\textit{i.e.}, sudden and gradual drifts), they are jointly used in the proposed framework to effectively detect both sudden and gradual model drift.

\subsubsection{Base Model Learning}
In this subsection, we discuss the drift-adaptive ML algorithms used to build the base online learners in the proposed framework. Hoeffding Trees (HT), also known as Very Fast Decision Trees (VFDT), are a type of decision tree designed for data stream mining \cite{ht}. HTs are constructed according to Hoeffding's inequality, which provides a bound on the probability that the mean of a random sample will deviate from the expected value. The Hoeffding bound is defined as \cite{myautoml}:
\begin{equation}\epsilon=\sqrt{\frac{R^2 \ln (1 / \delta)}{2 n}},\end{equation}
where $R$ is the range of the variable (\textit{i.e.}, the difference between the maximum and minimum), $\delta$ is a confidence level, and $n$ is the number of instances processed.

The Adaptive Random Forest (ARF) algorithm \cite{arf} is an ensemble learning method specifically designed for streaming data analytics. It leverages the power of the HT algorithm, which forms the base learners of the ensemble, and incorporates concept drift detection mechanisms at two levels: the tree level and the forest level. At the tree level, each HT in the ensemble is equipped with a drift detection method, typically ADWIN or EDDM, to monitor the performance of the tree. If a drift is detected, the affected subtree is replaced with a new one. At the forest level, a background learner is maintained for each tree in the ensemble. When a drift is detected at the forest level, the background learner replaces the original tree.

The Streaming Random Patches (SRP) algorithm \cite{srp} is another HT-based ensemble method for evolving data streams. It extends the idea of random patches and random subspace methods to the streaming setting. In SRP, each base learner (\textit{i.e.}, HT) is trained on a subset of features and instances, which are selected randomly. This diversity, combined with the use of base online learners, makes SRP robust to model drifts.

Both ARF and SRP are well-suited for online network data stream analytics. They can process high-speed data streams effectively due to their incremental nature, making them suitable for real-time network intrusion detection, traffic classification, and anomaly detection \cite{pwpae}. They can also handle sudden and gradual drifts by using different drift detectors (\textit{e.g.}, ADWIN and EDDM). 

The primary advantage of ARF and SRP is their efficiency and robustness to model drift. The time complexity of both algorithms is $O(nt)$, where $n$ is the number of instances, and $t$ is the number of base learners or HTs. This is because each instance needs to be processed only once, and the decision to expand a node is based on a statistical test, rather than an exhaustive search. The SRP model, while slightly more complex due to the maintenance of random patches, still maintains a time complexity of $O(nt)$, making it a viable option for large-scale, real-time applications. 

Since ARF and SRP may achieve varying performances for specific tasks and data streams or distributions, and their drift detectors may also behave differently for various types of model drifts (\textit{i.e.}, sudden or gradual drifts), we propose an automated model selection and hyperparameter optimization component, or a CASH module, to obtain the optimal model for network data stream analytics. The number of base learners or HTs '\textit{n\_models}', and the type of drift detectors '\textit{drift\_detector}', selecting from ADWIN or EDDM, are two hyperparameters to be tuned for base model optimization \cite{tii}.
 In the proposed framework, multiple ARF and SRP models with different hyperparameter configurations are constructed as the search space. The CASH model, which will be discussed in the next subsection, is utilized to select the best-performing model at any given time.

\subsubsection{Proposed SH-CASH Model for Automated Model Selection and Hyper-Parameter Tuning}
CASH is the process of implementing automated model selection and hyperparameter optimization in AutoML. It can be denoted as follows: \cite{myautoml}:
 \begin{equation}
A^{\star}, \boldsymbol{\lambda}_{\star} \in \underset{A(j) \in \mathcal{A}, \boldsymbol{\lambda} \in \Lambda^{(j)}}{\operatorname{argmin}} \frac{1}{K} \sum_{i=1}^{K} \mathcal{L}\left(A_{\boldsymbol{\lambda}}^{(j)}, D_{\text {train}}^{(i)}, D_{\text {valid}}^{(i)}\right)
\end{equation}
where $A^{\star} \in A$ is the ML algorithm to be selected, $\lambda$ is the hyperparameters to be tuned, $D_{train}$ and $D_{valid}$ are the training and validation sets, $K$ indicates k-fold cross-validation. In the CASH process, a specific ML algorithm with a certain hyperparameter combination is referred to as a configuration. 

The Successive Halving (SH) method \cite{sh1} is chosen for implementing CASH in the proposed framework. SH is a bandit-based algorithm that operates by iteratively eliminating the half-worst-performing configurations based on a certain budget, such as computation time or the number of iterations, and gradually identifying the optimal configuration \cite{sh1}. The principle behind SH involves allocating more resources to currently-promising configurations and discarding poor-performing configurations in each iteration, allowing a more efficient exploration of the model and hyperparameter configuration space. 

The SH method begins by uniformly sampling $n$ configurations from the configuration space at random. Then, each configuration is assessed using a small budget. The configurations are evaluated according to a suitable performance metric (\textit{e.g.}, accuracy), and only half-better-performing configurations are retained for the next iteration. Each remaining configuration's budget is doubled in the next iteration, and this procedure is repeated until the best configuration is identified \cite{sh2}.

The proposed AutoML framework extends the traditional SH method to an enhanced SH-CASH method. The details of the SH-CASH method are described in Algorithm \ref{algo:sh}. SH-CASH evaluates multiple candidate ML models in parallel, dynamically allocating computational resources to high-performing models and discarding underperforming ones in successive iterations. This ensures that only the most promising models are retained for further exploration, reducing computational overhead while maintaining performance. In addition to model selection, SH-CASH integrates hyperparameter tuning by leveraging successive halving, which iteratively refines the hyperparameter configurations of retained models. This combination enables the algorithm to identify both optimal models and their hyperparameters simultaneously, streamlining the AutoML pipeline.
Unlike traditional methods such as Grid Search (GS) or Random Search (RS), which evaluate all configurations equally, SH-CASH employs an adaptive approach to focus resources on promising configurations early in the process.

A primary novel contribution of the proposed SH-CASH model is to enable dynamic data stream analytics by incorporating drift detection and online learning techniques. The proposed AutoML framework will continuously monitor model performance and automatically re-implement the CASH processes when a new drift is detected by ADWIN and EDDM detectors. The primary objective of the SH-CASH method is to identify the best-performing base learner along with its optimal hyperparameter configuration for each new concept or new data distributions. Consequently, the optimal online learner can be obtained at any time to adapt to each model drift in dynamic networking environments. 

Existing optimization methods, such as GS, RS, Bayesian optimization, and Particle Swarm Optimization (PSO), are widely used for HPO and model selection \cite{hpome}. However, these methods are primarily designed for batch learning in static datasets and lack adaptability to dynamic, real-time data streams. In contrast, SH-CASH leverages successive halving to dynamically allocate resources based on real-time performance metrics and continuously produces the current optimal model by updating the search process in response to drift. This capability establishes SH-CASH as the first method specifically tailored for dynamic networking environments.

\begin{algorithm}[t!]
    {\footnotesize
    \caption{SH-CASH: Enhanced Successive Halving for Combined Algorithm Selection and Hyper-parameter Optimization for Dynamic Data Stream Analytics}
    \label{algo:sh}
    \LinesNumbered
    \KwIn{ 
    $M$: the set of candidate online ML models;\\
    $H$: the set of possible hyperparameters for each model;\\
    $B$: initial budget;\\
    $n$: number of configurations to be evaluated;\\
    $d$: model drift indicator;\\
    $S$: data stream;
    }
    \KwOut{
    $C$: the set of best configurations for each concept;\\
    $P$: prediction results for all data samples in $S$;
    }
    $\Omega \leftarrow M \times H$, the set of all combinations of models and their possible hyperparameters;\\
    Initialize a sliding window $W$ of size $w$ on the data stream $S$;\\
    Initialize concept drift detectors (\textit{i.e.}, ADWIN and EDDM);\\
    Initialize $C \leftarrow \emptyset$, $P \leftarrow \emptyset$, $d \leftarrow 1$;\\
    \While{data samples remain in $S$}{
        \If{the concept drift detector signals a drift (d==1)}{
        Sample $n$ configurations uniformly at random from $\Omega$. Each configuration is a pair $(m, h)$, where $m$ is a model in $M$, and $h$ is a set of hyperparameters from $H$ for $m$;\\
        $b \leftarrow B/n$;\\

        \While{$n > 1$}{
            \For{each configuration $c$ in the current set of configurations}{
                Train the model in $c$ on the data in the sliding window $W$ using the hyperparameters in $c$;\\
                Evaluate the performance of $c$ on the next $w$ data points in $S$ with budget $b$;\\
            }
            Rank the configurations based on their performance;\\
            Discard the worst $n/2$ configurations;\\
            $n \leftarrow n/2$;\\
            $b \leftarrow 2b$;\\
            Double the size of the sliding window $W$;\\
        }
        Let $L$ be the remaining configuration (best learner);\\
        Add $L$ to $C$;\\
        Update $d \leftarrow 0$ to finish the current drift adaptation and start monitoring the next drift; \\
        }
        Predict the next $w$ data points in $S$ using $L$ and add the results to $P$;\\
        Feed the prediction errors of $L$ to the concept drift detector;\\
    }
    \KwRet $C$, $P$;
    }

\end{algorithm}

The SH method has a time complexity of $O(B \log n) $, where $B$ is the budget, and $n$ is the number of configurations. This is because the SH method performs $\log n$ iterations to identify the optimal configuration, and in each iteration, it allocates a budget $B$ \cite{hpome}.

The proposed SH-CASH method is designed for automated model selection in dynamic data stream analytics due to the following reasons:
\begin{enumerate}[label=\roman*)]
\item \textit{Efficient Resource Allocation}: Model learning efficiency is crucial for network data stream analytics to achieve fast detection and response to cyber-attacks. The SH method allocates more computational resources to promising configurations, enabling it to rapidly identify well-performing models with hyperparameter values and discard underperforming ones.
\item \textit{Scalability}: The SH method has a low time complexity of $O(B \log n) $, making it scalable to large configuration spaces. This is crucial when dealing with complex learning models that have many hyperparameters.
\item \textit{Parallelizability}: The evaluations of different configurations in the SH method can be performed in parallel, enabling further speed-ups when multiple processing units are available.
\item \textit{Performance Enhancement}: The proposed SH-CASH method can continuously identify and select the best-performing online learner for dynamic data streams with model drifts to achieve high performance at any time. 
\end{enumerate}

\subsection{Proposed Framework Deployment}
The proposed automated framework is a generic cybersecurity solution that can be deployed for various data-driven cybersecurity tasks. It automates the process of designing and selecting appropriate data balancing methods, input features, and online learning models with optimal hyperparameter configurations. By learning the data patterns of specific datasets, the framework ensures a tailored approach to effectively address the unique challenges of each dataset and task.
\subsubsection{Physical Layer Authentication}
For PLA tasks, the physical layer features that reflect the unique characteristics and properties of physical devices, such as the IQI, can be used as the input data for the proposed AutoML model. These attributes are collected over time to form a time-series signal dataset. The proposed PLA model can then be deployed in real network environments by automatically integrating the RF signals in a certain period to provide a comprehensive view of each device, allowing for effective authentication.  

The authentication process can be adapted for cellular network architectures, such as Open Radio Access Network (O-RAN). Specifically, monitoring probes for capturing RF features can be strategically placed within the RAN components, including the Distributed Unit (DU) and Central Unit (CU) \cite{ranfl1}. Additionally, the Near-Real-Time RAN Intelligent Controller (Near-RT RIC) can host AI-driven authentication models (e.g., xApps) to process real-time RF signal attributes and identify anomalous device behavior \cite{ranfl2} \cite{rancl}. By leveraging such modular placements, the framework ensures efficient and localized anomaly detection.

In the authentication process, the proposed AutoML model is utilized to analyze the incoming RF signals as device property indicators. If the proposed model detects abnormal signals in real-time that may indicate illegitimate or compromised devices under potential cyber-attacks, it sends an alarm to alert the network administrators to take corresponding countermeasures, such as blocking traffic from unauthorized devices, in order to prevent potential attacks. This process is continuously carried out, and new illegitimate devices can be identified using the suggested drift detection approaches (\textit{i.e.}, ADWIN and EDDM), and subsequently included in the PLA database. These automated model updating and drift detection capabilities enhance the generalization ability of the model, allowing it to adapt to new and evolving threats, even when initially trained on limited datasets.

\subsubsection{Cross-Layer Intrusion Detection Systems}
CLIDSs typically extract and integrate information from multiple layers of the Internet protocol stack. In general, CLIDSs can be classified into two types of architectures: the direct per-layer interaction-based architecture and the shared database-based architecture \cite{clids_g}. In the direct per-layer interaction-based architecture, network information is exchanged directly between adjacent or non-adjacent layers, which can result in fast anomaly detection and high detection accuracy. However, this architecture introduces internal overhead due to the frequent data exchanges across layers and may lead to system instability if not properly managed. Additionally, the direct interaction among layers can compromise the modularity of the network protocol stack, potentially affecting overall network optimization.

The shared database-based architecture is another generalized approach for implementing a cross-layer IDS. This architecture utilizes a shared database, where every layer of the protocol stack interacts with an IDS server by collecting and sharing audit data in the database. The shared database-based architecture for cross-layer IDS offers simplicity and ease of management due to the significant differences between protocol layers. This approach also enhances system stability and maintains the modularity of the protocol stack by limiting direct interactions between layers, thus avoiding potential feedback loops and unnecessary complexities. In the proposed framework, the shared database-based architecture is selected for CLIDS, as it has higher system stability and lower protocol \& implementation complexities than the direct per-layer interaction-based architecture. Additionally, in the shared database-based architecture, only the necessary information is shared across layers, preserving the privacy and parallel evolution of each layer \cite{clids_g}.

To deploy the proposed AutoML framework to CLIDS tasks, the network traffic from multiple layers is combined in a shared database and then processed by the AutoML framework to provide a comprehensive view of network conditions. The collected network traffic is collected and sent to the model as data streams, and the proposed model will continuously analyze the incoming data instances and send alarms once abnormal traffic and potential cyber-attacks are detected. This architecture is particularly suited to dynamic real-world networking environments, where scalability and robustness are critical.

To make the proposed framework applicable to cellular network environments, monitoring probes can be strategically deployed at various components of the RAN, such as the DU, CU, and Core Network. Furthermore, RICs in O-RAN play a pivotal role in cross-layer intrusion detection \cite{rancl}. Specifically, Near-RT RICs, hosting xApps for real-time analytics, and Non-RT RICs, leveraging rApps for long-term policy optimization, can enable efficient data collection, feature extraction, and anomaly detection using cross-layer inputs \cite{ranad}. For example, Det-RAN’s approach of utilizing cross-layer features from PHY and RRC layers can enhance the IDS functionality, ensuring dynamic adaptability to new network threats \cite{rancl}.

The generalization ability of the proposed model is enhanced through its automated model updating and concept drift adaptation capabilities. These mechanisms allow the model to continuously adapt to new data and changes in the network environment, maintaining high detection accuracy even when initially trained on limited datasets. While this improves generalization, the model's performance in a real-world deployment may still be affected by the variability and quality of the network data.

\subsubsection {Cost and Feasibility of Deployment}
Integrating AI/ML models into 6G networks introduces significant computational demands, particularly for real-time data processing and continuous model updates. There is a particular relevance to these challenges in dynamic real-world environments where data streams must be adapted nearly instantaneously without disrupting network operations. To address these demands, the proposed online AutoML framework is designed to be lightweight and efficient, processing small-sized online data streams continuously without necessitating large-scale data storage or batch processing. The SH-CASH model within this framework achieves a low computational complexity of $O(B \log n)$, where $B$ represents the computational budget and $n$ is the number of configurations evaluated. This ensures that the framework operates within practical resource limits for real-world deployment scenarios \cite{hpome}. Moreover, leveraging AutoML technologies can facilitate the development of robust security solutions with minimal human intervention, thus enhancing resource efficiency \cite{mirna}. It is in line with the vision of the 6G network, which requires network automation and optimization.

The financial costs associated with deploying AI/ML models in 6G environments are considerable, but they can be managed through the use of modular strategies. This framework, for instance, provides the capability to deploy incrementally, allowing organizations to prioritize critical areas such as industrial IoT clusters or high-risk zones before expanding across the entire network. As a result of this strategy, cost-efficiency is assured by concentrating resources where they can have the most significant impact. Additionally, the lightweight design of the framework facilitates integration with existing infrastructures, minimizing the need for extensive hardware upgrades and reducing overall deployment costs \cite{tii}.

Due to the dynamic nature of 6G networks, AI/ML models are required that can adapt to changing conditions and detect emerging threats \cite{aml1}. The proposed online AutoML framework incorporates online learning techniques, enabling models to update continuously as new data becomes available. It ensures continued performance despite the emergence of new network behaviors, minimizing the need for manual intervention and the need for hardware modifications. Moreover, the framework's adaptability contributes to enhanced network resilience, as well as proactive management and rapid response to anomalies.

With its real-time adaptability, the proposed framework is capable of responding to changing network conditions as well as emerging threats without requiring manual intervention. With this flexibility, potential operational disruptions can be minimized, thus ensuring seamless integration with existing network functions. For latency-sensitive applications such as augmented reality or telemedicine, its real-time processing capabilities preserve Quality of Service (QoS) standards\cite{cost1}. This is achieved through efficient resource allocation mechanisms that prioritize low-latency operations.

Automated mechanisms, including drift detection and model retraining, ensure the framework maintains high performance over time \cite{dmdrift}. By automating updates, we address concept drift and incorporate new data without requiring significant human intervention, reducing operational costs and allowing our experts to concentrate on strategic decision-making rather than routine maintenance. Its modular design ensures minimal disruption during deployment phases, facilitating the transition to an automated network management system.

In summary, the proposed online AutoML framework addresses key deployment challenges, including computational demands, cost-effectiveness, adaptability, and operational continuity. Its design makes it a practical and scalable solution for integrating AI/ML into next-generation networks while ensuring network performance and security.

\subsection{Countermeasures to Detected Attacks}
After implementing the proposed AutoML framework to address PLA and CLIDS tasks, the experimental results from these two tasks are combined as multi-layer union results to reflect the network security conditions. The combined results are then used to trigger appropriate countermeasures to mitigate the detected threats and maintain robust network security. Based on the cross-layer intrusion detection results, the following security measures can be implemented at different layers.

At the physical layer, if compromised or affected physical devices are detected by the PLA model, isolating these devices is an effective method to prevent the propagation of malicious traffic and attacks \cite{ericsson}. This can be achieved by disabling specific physical ports exposed to attacks or by implementing network segmentation strategies to protect critical sub-networks. These strategies can minimize the impact of malicious devices and attacks on the overall network/system integrity.

At the data link layer, if attacks such as MAC address spoofing or MAC flooding are identified, malicious or compromised MAC addresses can be mitigated by implementing dynamic MAC address filtering or blacklisting measures \cite{measure1}. Dynamic filtering restricts network access to trusted devices by validating their MAC addresses against predefined lists, while blacklisting known malicious addresses prevents the re-establishment of unauthorized connections.

At the network layer, after identifying the attacks and potential attacker IP addresses, maintaining and updating a blacklist of such IP addresses can prevent unauthorized access and reduce the risk of attacks like DoS/DDoS \cite{ericsson}.  A critical component of mitigating network layer attacks is blocking traffic from known malicious IP addresses, and the implementation of automatic blacklist management can enhance the efficiency of this countermeasure.

At the transport layer, when attacks such as SYN flooding or other DoS attacks are detected, implementing countermeasures like terminating suspicious connections and rate limiting can effectively mitigate their impact \cite{measure2}. For the detected attacks targeting Transport Layer Security (TLS) protocols, such as MITM attacks, incorporating robust cryptographic algorithms can protect communications integrity and confidentiality.

When application-layer attacks like XSS or SQL injection are identified, enforcing strict input validation ensures that only properly formatted data is processed by the application layer \cite{measure3}. Additionally, after detecting the compromised web applications that are vulnerable to certain cyber-attacks, uninstalling or updating these applications can prevent attackers from exploiting application weaknesses.

Overall, the proposed PLA and CLIDS frameworks can detect anomalies and attacks across different layers, and corresponding countermeasures can be implemented at each layer to safeguard the networks against such attacks.

\section{Performance Evaluation} 
This section presents the experimental results and discussion for evaluating the proposed AutoML framework on two representative use cases in ZTNs and future network security applications: PLA and CLIDS. As this paper is the first work to propose such a comprehensive cybersecurity framework for PLA and CLIDS, there is no existing dataset that provides data for both PLA and CLIDS within the same environment. Therefore, the experiments focus on individual tests of these two functionalities. The first subsection provides the experimental setup and results of the PLA case study, which focuses on solving physical layer security problems. In this case, the proposed AutoML framework builds a PLA model that serves as the first layer of network defense to authenticate devices. The second subsection discusses the experimental results of intrusion detection problems, which involve addressing cybersecurity issues on upper layers. In this case, the proposed AutoML framework constructs a CLIDS model that operates as the second layer of defense to detect cyber-attacks that breach the first layer of defense.

\subsection{Use Case 1: Automated Physical Layer Authentication}
\subsubsection{Experimental Setup}
The proposed AutoML framework was developed in Python 3.10 by extending the River \cite{river} library on a machine equipped with an Intel i7-8700 Central Processing Unit (CPU) and 16 GigaByte (GB) of memory, representing a cloud server machine in 5G/6G networks for large-scale network data stream analytics.

The Oracle RF Fingerprinting \cite{rfdata}, a set of public datasets, is utilized to assess the proposed AutoML framework for PLA tasks. The Oracle RF Fingerprinting dataset \cite{rfdata} was created by the Genesys Lab at Northeastern University to support the development and validation of device fingerprinting techniques for PLA problems. This dataset includes IQI samples collected from 100 USRP X310 radios, with each radio functioning both as a transmitter and receiver pair. The radios were configured to use two different daughterboards, namely UBX-160 and CBX-120, and they transmitted at a power level of 10 dBm. Each radio generated signals based on identical Binary Phase-Shift Keying (BPSK) modulation scheme, and the signals were then captured by another USRP X310 radio acting as a receiver. It is important to mention that we opted for the IQI feature due to its distinct variations among devices and its resistance to complete predictability and control during the manufacturing of electronic components. Additionally, IQI features are readily available in most wireless receivers, making them a practical choice for our application.

The combined dataset \#2 in the Oracle RF Fingerprinting datasets, “KRI-16IQImbalances-DemodulatedData” is employed to evaluate the proposed AutoML framework. The dataset contains RF signals from 16 different physical devices, used for authentication. To evaluate the proposed AutoML framework for cybersecurity problems (\textit{i.e.}, anomaly detection), devices \#2, \#6, \#10, and \#14 are labeled as potential illegitimate/unauthorized devices, and the other 12 devices are labeled as authorized devices. At the initial stage of the online learning experiments, we assumed that only the RF signals of the first two devices (\#1 and \#2) were available for building the initial ML models. The RF signals from other devices (\#3 - \#16) are utilized as test data for online data stream analytics. Hence, there are three model drifts in the datasets that necessitate model updates due to the presence of previously-unseen abnormal devices (\#6, \#10, and \#14). To offer a comprehensive perspective on device IQI characteristics, every 100 consecutive samples are combined as a new data sample in the generated data streams, symbolizing a segment of RF signals. For this study, a representative Oracle RF Fingerprinting subset consisting of 37,888 instances is selected for model evaluation. 

The proposed SH-CASH framework is specifically designed to dynamically select the optimal base model between ARF and SRP models at any given time, considering their various hyper-parameter configurations. The framework incorporates two key hyperparameters: '\textit{n\_models}' and '\textit{drift\_detector}' for ARF and SRP models. The '\textit{n\_models}' hyperparameter is fine-tuned within the range of 3 to 5, striking a balance between model complexity and performance. On the other hand, the '\textit{drift\_detector}' hyperparameter is selected from two methods, namely ADWIN and EDDM, to effectively handle both gradual and sudden drifts in the data. Additionally, the framework's performance in handling PLA problems is evaluated by comparing it with state-of-the-art online ML models, including HT \cite{ht}, Leveraging Bagging (LB) \cite{lb}, Aggregated Mondrian Forest (AMF) \cite{amf}, Extremely Fast Decision Tree (EFDT) \cite{efdt}, Hoeffding Adaptive Tree (HAT) \cite{hat}, ARF \cite{arf}, and SRP \cite{srp}. The learning process and data used for evaluation are the same across all methods. 

The experimental setup for evaluating SH-CASH and baseline models follows the prequential evaluation method, ensuring consistency and fairness. 
Prequential validation \cite{prequential}, also known as test-and-train validation, is used to assess the proposed model for online data stream analytics. At the start of prequential evaluation, it is assumed that the model is initialized in a dynamic networking environment with no pre-existing data, but the model begins to collect and process newly generated data as it becomes available. This setup is aligned with the complexity and dynamic characteristics of 5G and next-generation networks. In prequential validation, each incoming data sample or RF signal is first used to test the online models, monitoring their real-time performance. Subsequently, the models are updated and adapted based on the learned information from the new data, enabling automated model updating and drift adaptation.

Lastly, to provide a comprehensive assessment of the model performance and mitigate the impact of class imbalance on performance comparison, the experiments utilize four standard classification performance metrics: accuracy, precision, recall, and F1-score. Additionally, the average learning time per data sample, which is calculated by dividing the total model testing and training/updating time by the number of samples, is employed to evaluate the overall efficiency of the proposed framework and to compare it with other methods.

Overall, the experiments simulate real-time dynamic networking environments where physical-layer data streams continuously evolve. Unlike traditional static datasets, these environments involve changes in network topology, traffic patterns, and the previously-unseen or zero-day attack scenarios that the physical devices are continuously compromised by attacks. The proposed SH-CASH framework operates under these dynamic conditions, leveraging its ability to adaptively update models and optimize hyperparameters in real time. The dynamic datasets used in this study emulate scenarios where normal RF signals is interspersed with diverse attack patterns, mimicking real-world network behavior for the PLA process. Many existing state-of-the-art methods, such as \cite{Illi2022} - \cite{yang2013} \cite{Xie2022} \cite{Dolatshahi2010} \cite{Rahman2014} \cite{Hao2014} \cite{Hao_ICC_2014} \cite{Liao2020} \cite{pan2019} \cite{Hoang2020} \cite{Qiu2018} \cite{Chen2021} \cite{Wang2019} described in Section IV-A, primarily focus on static learning environments and PLA of known devices. In contrast, the proposed SH-CASH framework is explicitly designed for and evaluated in real-time dynamic networking environments.

\subsubsection{Results Discussion}

\begin{figure}[!t]
\centerline{
\includegraphics[width=8.9cm]{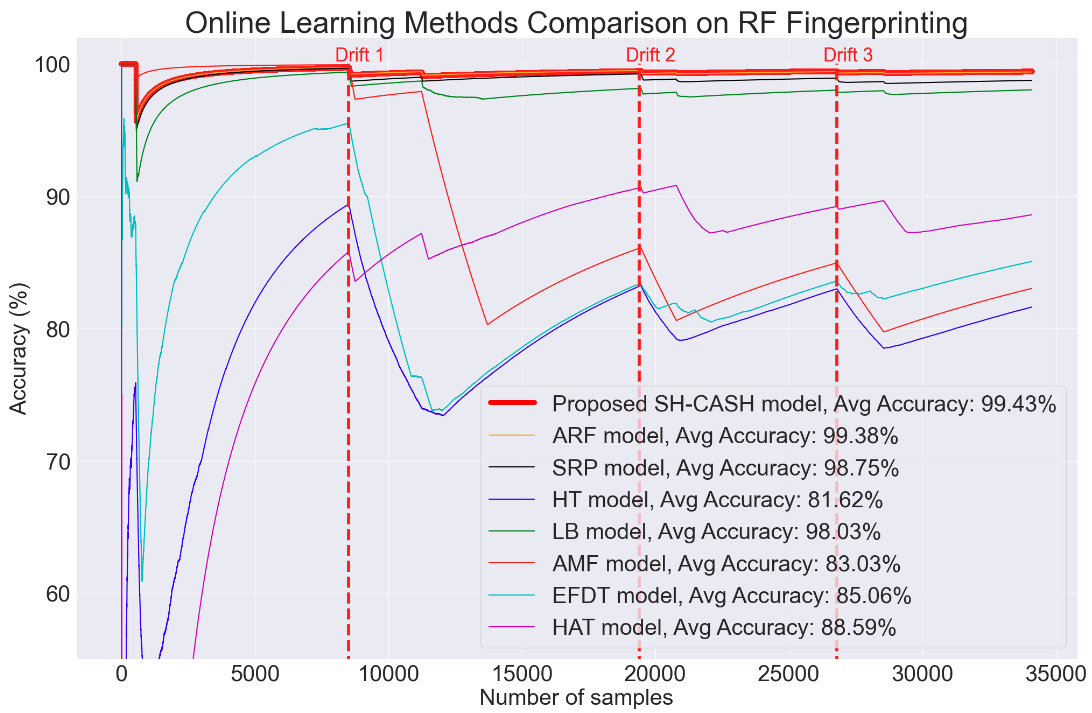}}
\caption{Accuracy comparison of state-of-the-art drift adaptation methods on the RF Fingerprinting dataset for PLA tasks.}
\label{rf_figure}
\end{figure}

\begin{table}[!t]
\caption{Performance Comparison of Online Learning Methods on The Oracle RF Fingerprinting Dataset for PLA Tasks.}
\centering
\setlength\extrarowheight{1pt}
\scalebox{0.82}{
\begin{tabular}{|>{\centering\arraybackslash}p{5em}|>{\centering\arraybackslash}p{4.5em}|>{\centering\arraybackslash}p{4.0em}|>{\centering\arraybackslash}p{3.5em}|>{\centering\arraybackslash}p{3.5em}|>{\centering\arraybackslash}p{6.5em}|}
\hline
\textbf{Method}         & \textbf{Accuracy (\%)} & \textbf{Precision (\%)} & \textbf{Recall (\%)} & \textbf{F1 (\%)} & \textbf{Avg Execution Time Per Sample (ms)}  \\ 
\hline
HT \cite{ht} & 81.616 & 51.141 & 24.377 & 33.016 & 1.3 \\
\hline
LB \cite{lb} & 98.032 & 93.187 & 96.466 & 94.798 & 14.0 \\
\hline
AMF \cite{amf} & 83.026 & 55.002 & 47.712 & 57.098 & 6.0 \\
\hline
EFDT \cite{efdt} & 85.062 & 64.393 & 43.941 & 52.237 & 4.9 \\
\hline
HAT \cite{hat} & 88.592 & 64.134 & 87.630 & 74.063 & 1.9 \\
\hline
ARF \cite{arf} & 99.384 & 98.527 & 98.154 & 98.340 & 0.4 \\
\hline
SRP \cite{srp} & 98.748 & 97.524 & 95.693 & 96.600 & 11.2 \\
\hline
\textbf{Proposed SH-CASH} & \textbf{99.431} & \textbf{98.470} & \textbf{98.470} & \textbf{98.470} & 0.9 \\
\hline
\end{tabular}
}
\label{rf_table}%
\end{table}

Figure \ref{rf_figure} and Table \ref{rf_table} present an extensive performance evaluation of the proposed SH-CASH model in comparison to various state-of-the-art online learning methods on the Oracle RF Fingerprinting dataset for PLA tasks. It is observed that, due to the inherent characteristics of the prequential evaluation process, the testing accuracy fluctuates at the beginning of the experiment. However, as the model encounters diverse samples, including anomalies, the accuracy gradually stabilizes. As shown in Table \ref{rf_table}, The proposed SH-CASH model not only leads in accuracy, attaining 99.431\%, but also excels in the F1-score with 98.470\%. This supremacy in both metrics suggests a balanced performance in both maximizing the detection rate and minimizing the false alarm rate, which is essential for real-world PLA applications. Additionally, the ARF method closely follows with an accuracy of 99.384\% and a F1-score of 98.340\%, and the SRP method achieves an accuracy of 98.748\% and a F1-score of 96.600\%. This highlights the benefits of selecting ARF and SRP as base learners within the SH-CASH framework. In contrast, the HT, LB, AMF, EFDT, and HAT methods exhibit relatively lower F1-scores, ranging from 33.016\% to 94.798\%. This difference underlines the proposed SH-CASH model's superior capability in selecting the best-performing base model with the optimal hyperparameter configuration in real time.

In the Oracle RF Fingerprinting dataset, three model drifts occurred in the PLA experiments, namely Drift 1, Drift 2, and Drift 3, indicating the occurrence of new unauthorized devices \#6, \#10, and \#14, respectively, as shown Fig. \ref{rf_figure}. As the signals of these three new unauthorized devices were previously unseen by the evaluated models, the accuracy of these models dropped to varying degrees in the early stages of each drift. However, after the drift detection and adaptation procedures, most of the online learners resumed their performance, with the proposed SH-CASH model showing the best adaptation to the three drifts and maintaining high accuracy. This demonstrates the proposed SH-CASH model's ability to handle model drift issues in dynamic data streams, making it suitable for dynamic networking environments. Furthermore, the ARF and SRP models were the second and third best-performing models in terms of adapting to drift and resuming performance, which supports the reasons for selecting them as base learners in the proposed framework. 

In terms of model efficiency, ARF demonstrates the fastest execution time at 0.4 ms per sample, followed closely by the proposed SH-CASH model with an average execution time of 0.9 ms per sample. Despite ARF's slightly faster execution, the SH-CASH model outperforms in terms of accuracy, precision, recall, and F1-score, while still maintaining a relatively efficient execution time. This is achieved by the feature selection of the proposed AutoFE component and the budget allocation strategy employed by the proposed SH-CASH model. By evaluating these models on smaller subsets, the SH-CASH model intelligently assigns larger budgets only to the best-performing base models. This approach reduces computational complexity, optimizes resource utilization, and enhances overall model efficiency. 

In summary, the proposed SH-CASH method outperforms the other compared methods in terms of accuracy, precision, recall, and F1-score on the Oracle RF Fingerprinting dataset for PLA tasks, while maintaining competitive execution times. This demonstrates the effectiveness and efficiency of the SH-CASH method in processing this PLA dataset, and suggests its potential utility in PLA and real-time cybersecurity applications. 
Additionally, experimental results demonstrate that the proposed framework can be extended to scenarios with highly dynamic channel conditions or hardware impairments, where device behavior evolves over time, aligning with real-world networking environments.

\begin{table}[!t]
\caption{Attack Type, Layer, Drift Number, and Size of the Sampled CICIDS2017 Dataset.}
\centering
\setlength\extrarowheight{1pt}
\scalebox{0.82}{
\begin{tabular}{|>{\centering\arraybackslash}p{6em}|>{\centering\arraybackslash}p{5em}|>{\centering\arraybackslash}p{15em}|>{\centering\arraybackslash}p{4em}|}
\hline
\textbf{Class Label} & \textbf{Number of Samples} & \textbf{Layers Belong To} & \textbf{Drift No.} \\
\hline
BENIGN & 107,226 & All Layers & - \\
\hline
Brute-Force & 1,384 & Application Layer & - \\
\hline
DoS & 16,942 & Physical, Network, Transport, and Application Layers & 1 \\
\hline
Web-Attack & 2,180 & Application Layer & 2 \\
\hline
Infiltration & 1,800 & Network and Application Layers & 3 \\
\hline
Botnet & 1,966 & Network and Application Layers & 4 \\
\hline
Sniffing & 7,947 & Physical and Network Layers & 5 \\
\hline
DDoS & 2,092 & Physical, Network, Transport, and Application Layers & 6 \\
\hline
\end{tabular}
}
\label{dataset_table}
\end{table}

\subsection{Use Case 2: Automated Cross-Layer Intrusion Detection}
\subsubsection{Experimental Setup}
A public network traffic dataset, CICIDS2017 \cite{cic}, is used to evaluate the proposed framework for CLIDS problems. CICIDS2017 is a state-of-the-art IDS dataset developed by the Canadian Institute for Cybersecurity (CIC) and widely used in various cybersecurity applications. The dataset is recognized for its variety and realism, containing benign and up-to-date common attack traffic for developing and evaluating IDSs. It includes cyber-attacks from multiple layers, such as DoS \& DDoS attacks, botnets, brute-force attacks, sniffing/eavesdropping attacks, port-scan attacks, web-attacks, and infiltration, making it suitable for CLIDS problems \cite{cicdata}. For this work, we use a representative subset of 141,537 records from the CICIDS2017 dataset. The size and corresponding layers of each attack/class in the CICIDS2017 dataset are provided in Table \ref{dataset_table}.

For the purpose of zero-day attack detection, at the beginning of the experiments, we assume that there are two classes of data available: benign/normal and brute-force attacks. All other types of attack samples in the CICIDS2017 dataset are used for evaluating the proposed model for online data stream analytics. Therefore, there are six model drifts in chronological order, namely DoS, web-attacks, infiltration, botnets, sniffing, and DDoS attacks, as illustrated in Table \ref{dataset_table}.

Similar to the PLA use case, prequential validation is employed to evaluate the online data stream analytics and model drift adaptation capabilities of the proposed framework in CLIDS problems. The same five metrics - accuracy, precision, recall, F1-score, and average learning time per sample - are used to assess the effectiveness and efficiency of the proposed framework in intrusion detection tasks. Furthermore, the performance of the proposed framework is compared with that of the same state-of-the-art online learning methods used in the PLA experiments, including HT \cite{ht}, LB \cite{lb}, AMF \cite{amf}, EFDT \cite{efdt}, HAT \cite{hat}, ARF \cite{arf}, and SRP \cite{srp}. To ensure fairness, the proposed SH-CASH framework was subjected to identical testing conditions as the baseline methods. The prequential evaluation ensures that no method has access to future data, and all algorithms process samples sequentially in a real-time dynamic networking environment.

Similar to the PLA evaluation environments, the experiments on CLIDS evaluation simulate real-time dynamic networking environments with evolving network traffic and activities. Additionally, the evaluation includes scenarios that simulate zero-day attacks at different layers, where the proposed framework encounters malicious activities and attacks not present during training. Existing state-of-the-art methods \cite{CLIDS_review1} - \cite{automl_review3} are typically designed for static environments and focus only on detecting known attack patterns. However, real-world networks are usually dynamic environments with evolving attacks, so zero-day attack detection is critical for next-generation network security. Therefore, the proposed AutoML framework is designed for and evaluated in real-time dynamic networking environments with continuous zero-day attacks. This highlights the novel contributions and real-world applicability of the proposed framework.

\subsubsection{Results Discussion}

\begin{figure}[!t]
\centerline{
\includegraphics[width=8.9cm]{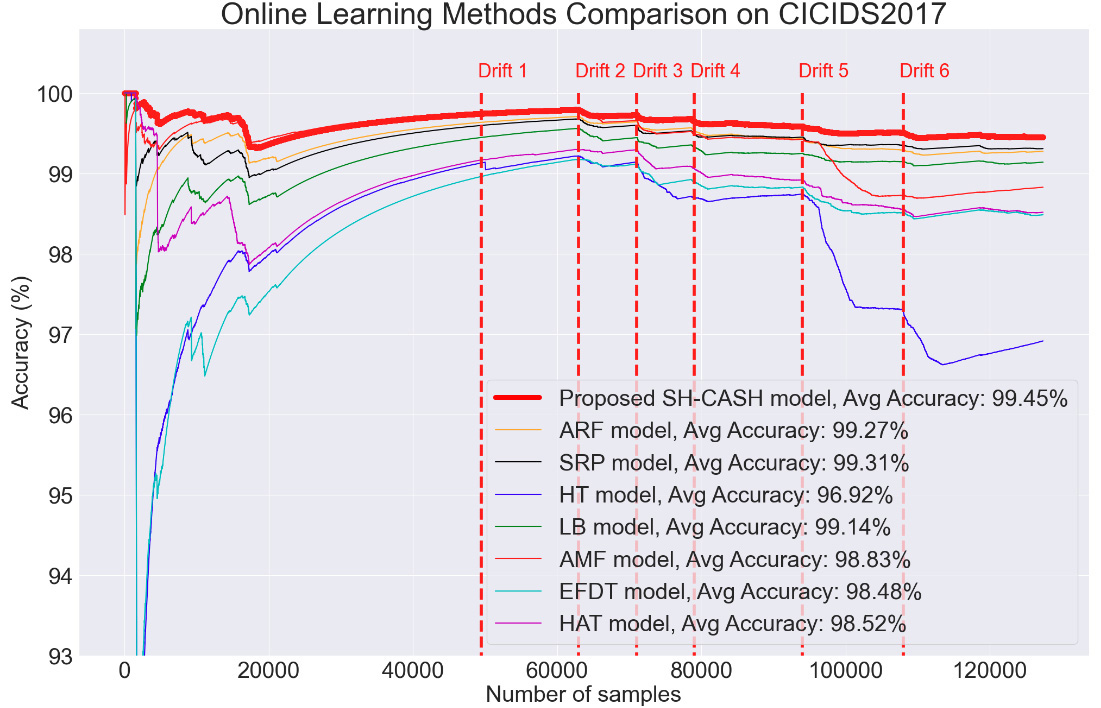}}
\caption{Accuracy comparison of state-of-the-art drift adaptation methods on the CICIDS2017 dataset for CLIDS tasks.}
\label{cic_figure}
\end{figure}

For CLIDS tasks, the performance of the proposed SH-CASH framework is compared to other state-of-the-art drift-adaptive methods using the CICIDS2017 dataset. The results are illustrated in Fig. \ref{cic_figure} and summarized in Table \ref{cic_table}. As shown in Table \ref{cic_table}, the SH-CASH algorithm consistently outperforms the other models across all metrics, including accuracy, precision, recall, and F1-score, demonstrating its superior performance in intrusion detection tasks.

The SH-CASH model demonstrates exceptional overall performance across all key performance metrics, including accuracy (99.450\%), precision (99.244\%), recall (97.470\%), and F1-score (98.349\%). These values underline the model's precision in cyber-attack detection, balanced handling of false positives and negatives, and overall effectiveness in accurately detecting instances.

A comprehensive comparison between the SH-CASH model and state-of-the-art online learning methods in the literature, including HT, LB, AMF, EFDT, HAT, ARF, and SRP, clearly demonstrates the superior performance of the proposed model. While the HT method exhibits a swift execution time of 0.1 ms per sample, it lags behind in terms of accuracy (96.916\%) and F1-score (90.486\%). On the other hand, the LB, AMF, EFDT, and HAT methods show higher F1-scores than HT, ranging from 95.475\% to 97.410\%. However, they still do not match the results achieved by the SH-CASH model. The ARF and SRP models achieve high F-scores of 97.818\% and 97.922\% on the CICIDS2017 dataset, respectively, ranking second and third in terms of performance. This further reinforces the selection of ARF and SRP as base models in the proposed SH-CASH framework, given their strong performance.

An essential advantage of the proposed SH-CASH model is its ability to automatically recognize and adapt to model drift, thus enhancing detection accuracy. As shown in Fig. \ref{cic_figure}, six model drifts occurred in the CLIDS experiments, including sudden drifts (Drifts 1, 2, 4, and 6) and gradual drifts (Drifts 2 and 5), corresponding to the six new attacks illustrated in Table \ref{dataset_table}. It is noticeable in Fig. \ref{cic_figure} that the proposed SH-CASH model can effectively adapt to all the six drifts and maintain high accuracy and F1-scores in the detection of zero-day attacks. However, certain online learning methods, like HT and AMF, are significantly impacted by model drifts. The robust drift-handling capability of the proposed SH-CASH model ensures consistently high-quality results, reflecting the model's resilience against changing patterns and trends within dynamic networking environments.

\begin{table}[!t]
\caption{Performance Comparison of Online Learning Methods on The CICIDS2017 Dataset for CLIDS Tasks.}
\centering
\setlength\extrarowheight{1pt}
\scalebox{0.82}{
\begin{tabular}{|>{\centering\arraybackslash}p{5em}|>{\centering\arraybackslash}p{4.5em}|>{\centering\arraybackslash}p{4.0em}|>{\centering\arraybackslash}p{3.5em}|>{\centering\arraybackslash}p{3.5em}|>{\centering\arraybackslash}p{6.5em}|}
\hline
\textbf{Method}         & \textbf{Accuracy (\%)} & \textbf{Precision (\%)} & \textbf{Recall (\%)} & \textbf{F1 (\%)} & \textbf{Avg Execution Time Per Sample (ms)}  \\ 
\hline
HT \cite{ht} & 96.916 & 93.993 & 87.231 & 90.486 & 0.1 \\
\hline
LB \cite{lb} & 99.139 & 98.534 & 96.312 & 97.410 & 2.0 \\
\hline
AMF \cite{amf} & 98.827 & 99.376 & 93.613 & 96.408 & 1.4 \\
\hline
EFDT \cite{efdt} & 98.485 & 95.889 & 95.065 & 95.475 & 0.3 \\
\hline
HAT \cite{hat} & 98.516 & 97.083 & 94.001 & 95.517 & 0.4 \\
\hline
ARF \cite{arf} & 99.274 & 98.837 & 96.821 & 97.818 & 0.4 \\
\hline
SRP \cite{srp} & 99.308 & 98.826 & 97.035 & 97.922 & 1.6 \\
\hline
\textbf{Proposed SH-CASH} & \textbf{99.450} & \textbf{99.244} & \textbf{97.470} & \textbf{98.349} & 0.5 \\
\hline
\end{tabular}
}
\label{cic_table}%
\end{table}

Regarding the execution times, the SH-CASH model illustrates a favorable trade-off between computational efficiency and classification performance, with an average execution time of 0.5 ms per sample, which is significantly lower than the execution times of LB, AMF, and SRP. This is primarily due to the AutoFE component and the resource allocation functionality of the proposed SH-CASH model. The low execution speed, when complemented by its robustness against model drift and remarkable intrusion detection capabilities, establishes the SH-CASH model as a viable and promising solution for real-time network data analytics in cybersecurity applications.

In conclusion, the proposed SH-CASH method not only achieves the highest accuracy and F1-scores among the competing online learning methods but also maintains a balance between accuracy and efficiency. The superior results of the SH-CASH model, alongside its drift adaptability and streamlined execution time, manifest it as a promising and innovative cybersecurity solution for both PLA and CLIDS tasks in real-world networking environments.

\section{Open Challenges and Future Research Directions} 

The application of AI/ML models in automated cybersecurity solutions for ZTNs and future networks represents a promising area for future research. However, to fully realize the potential of these technologies, several open challenges need to be addressed. In this section, we will discuss the open challenges and future research directions in the development of AI/ML models for enhancing cybersecurity.

\subsection{Model Performance and Robustness}
The performance and robustness of AI/ML models are critical factors in their application to automated cybersecurity solutions. The AI/ML models must be capable of consistently and accurately identifying and mitigating cyber-attacks in dynamic and evolving network environments. However, traditional AI/ML models encounter several challenges:
\begin{enumerate}[label=\roman*)]
\item \textit{Model Generalization}: The performance of AI/ML models can be affected by their ability to generalize from training data to unseen data. This is particularly challenging in cybersecurity applications, due to the dynamic and evolving nature of cyber threats. ML models trained on current network data may not perform well when confronted with new or unknown threats. Therefore, To enhance model generalization, researchers should investigate advanced methods such as transfer learning and continual learning techniques \cite{continual}.
\item \textit{Model Efficiency and Real-Time Threat Detection}: As ZTNs and future networks will involve a large number of devices and high data rates, the deployed AI/ML models must be efficient and scalable. Given the high data rates and low latency requirements of ZTNs and future networks, the ML models should be capable of detecting and mitigating cyber-attacks in real-time.  Complex ML models, such as deep learning models, are computationally intensive, limiting their scalability and efficiency. Thus, future research should investigate methods for improving ML model efficiency, such as Tiny Machine Learning (TinyML) \cite{tinyml} and edge computing techniques \cite{sam1}.
\item \textit{Model Interpretability}: The black-box nature of many AI/ML models can hinder their application in cybersecurity. To build trust in AI/ML models, model interpretability or explainability is essential for cybersecurity applications. Network administrators must comprehend the decision-making process of AI/ML models if they rely on the models' predictions to take actions. Techniques such as Explainable AI (XAI) can be employed to make the models more transparent and understandable \cite{xai}.
\end{enumerate}

\subsection{Availability and Richness of Datasets}
The availability and richness of datasets are crucial for the effective training and validation of AI/ML models in the development of automated cybersecurity solutions for ZTNs and future networks. However, there are several challenges regarding this issue: 
\begin{enumerate}[label=\roman*)]
\item \textit{Data Collection Challenges}: Due to the diverse and dynamic nature of networking environments, the data collected at different locations or nodes in ZTNs and future networks are non-Independent and Identically Distributed (non-IID) data, making it challenging to collect a comprehensive and representative dataset for training effective AI/ML models. Data sampling and concept drift adaptation methods are potential solutions to generating representative datasets and training reliable ML models that can process dynamic network data streams \cite{tii}. 
\item \textit{Data Quality and Richness}: The quality and richness of the collected network data are crucial for the development of effective AI/ML models. However, ensuring data quality and richness in ZTNs and future networks is challenging due to the high dimensionality and complexity of network data. There are various types of network data, such as network traffic data, user behavior data, device behavior data, and sensor data. Various types of network data may necessitate distinct data pre-processing and feature engineering procedures to ensure data quality, owing to their unique characteristics.
\item \textit{Need for Publicly Available Datasets}: The limited availability of publicly available datasets for 5G/6G networks poses a challenge to the development and evaluation of AI/ML models, particularly for physical layer security datasets. Publicly available datasets serve as a common foundation for researchers to develop and compare their models. Thus, there is a need for increased efforts to create and share such datasets for ZTNs and future networks.
\end{enumerate}

\subsection{Adversarial Attacks: AI/ML Model Security Threats}
While AI/ML techniques enhance network automation capabilities, such as self-optimization and self-protection, they also expose ZTNs and future networks to novel threats that target AI/ML models directly. These threats are known as Adversarial Machine Learning (AML) attacks \cite{aml1}. Although AML attacks are not specifically designed for breaching ZTNs and future networks, their prominence has increased due to the growing reliance on AI/ML techniques for network automation. This growing reliance on AI/ML models poses a significant challenge in ensuring the security of ZTNs and future networks, especially considering the lack of human supervision in such environments.

Based on their occurrence in the ML process, AML attacks can be categorized into data poisoning, evasion, and inference attacks \cite{aml2}. Data poisoning attacks, which occur during the training phase, involve the introduction of malicious data into the training set. Evasion attacks occur post-training, where attackers manipulate the model to degrade its performance. Inference attacks take place during model deployment, where tampered prediction data can induce specific outcomes or reveal sensitive information. These threats underscore the need for robust AML defense measures.

The presence of AML attacks can directly impact the progression of cybersecurity mechanisms. Traditional ML techniques have shown vulnerability to malicious AML attacks, leading to a degradation in the performance of AI/ML-driven cybersecurity models. For instance, in the context of physical layer security, AML attacks introduce uncertainty into the underlying data, making it difficult for PLA models to effectively learn and accurately characterize RF signals \cite{physec1}. Similarly, in cross-layer security, AML attacks can degrade the performance of ML-based CLIDS models in identifying malicious cyber-attacks, thereby enabling breaches and leading to severe security concerns.

To defend against AML attacks and protect AI/ML-based cybersecurity models, several defense mechanisms require further research, such as adversarial sample detection and removal, adversarial training, defense Generative Adversarial Networks (GANs), and concept drift adaptation \cite{zsmsec1}.

\subsection{Privacy Issues}
The integration of AI/ML models in ZTNs and future networks presents a new set of privacy challenges. As modern networks become more complex, the exponential growth in generated data raises significant privacy concerns. Traditional AI/ML models, which often rely on centralized data collection and processing, are now facing bottlenecks due to these privacy concerns. The privacy issues can be categorized into three main areas: network privacy, data privacy, and ML model privacy.
\begin{enumerate}[label=\roman*)]
\item \textit{Network Privacy Issues}: ZTNs and future networks pose significant privacy threats due to their inherent openness and accessibility. Common network privacy threats include unauthorized access, eavesdropping, and MITM attacks. These threats can expose confidential network information such as network topology, network traffic, and network configuration, which can be used by adversaries to launch further attacks \cite{pri1}.
\item \textit{Data Privacy Issues}: The massive volume of data generated and processed by ZTNs and future networks poses critical data privacy concerns. Network data, which usually contains sensitive network or user information, can be intercepted during transmission or extracted from databases with inadequate security. Many important applications of 6G networks, such as smart healthcare, wearable devices, and smart homes, are expected to collect more sensitive information about users and organizations to provide reliable services \cite{pri2}. 
\item \textit{ML Model Privacy Issues}: ML model privacy issues refer to the potential exposure of sensitive information embedded in the models themselves. The use of AI/ML models in networks for tasks like user profiling and behavior prediction can pose privacy risks if the models are trained on sensitive data. Many cyber-attacks, such as model inversion and membership inference attacks, can be launched to extract sensitive information from ML model updates \cite{pri3}.
\end{enumerate}

To mitigate these privacy threats, several privacy preservation methods have been proposed, such as Federated Learning (FL) and blockchain. FL is a distributed ML approach that allows ML models to be trained on decentralized data from multiple parties without sharing their original data \cite{dmdrift}. Blockchain technology can play a significant role in enhancing the privacy of AI/ML models in ZTNs. By leveraging the decentralized and tamper-resistant nature of blockchain, a secure and transparent framework for data and model management can be created \cite{pri4}. For instance, blockchain can be used to track and verify the origin of data and model updates, thereby preventing unauthorized access and manipulation. In conclusion, further research is needed to comprehensively address these security and privacy challenges and optimize potential solutions for use in ZTNs and future networks.

\subsection{AI Ethical and Legal Challenges}

By integrating AI/ML technologies into ZTNs and next-generation networks, significant advancements in automation and cybersecurity can be achieved. However, this integration presents a number of complex ethical and legal challenges that must be addressed in order to ensure a safe and effective deployment. The following are typical ethical and legal challenges.
\begin{enumerate}[label=\roman*)]
\item \textit{Bias and Fairness}: In many AI/ML models, biases occurring in their training data can affect the decision-making process in ways that are unfair or discriminatory. In network automation and security applications, such biases may result in unequal treatment of users or devices, affecting the principles of fairness and equity \cite{fair1}. As an example, a ML or AutoML model handling network security may be unable to adequately protect certain locations or types of devices/users if it is trained on data that underrepresents those locations or types of devices/users. The fairness assessment and bias mitigation strategies required to address these issues must be implemented to ensure that decisions that are made using AI are equitable and impartial. Using techniques such as transparent auditing of quantitative bias in AI models, it is possible to identify and address biases in the automated decision-making process, thereby promoting fairness \cite{fair2}.
\item \textit{Regulatory Compliance}: Adoption of AI/ML technologies in network automation must be in compliance with existing laws and regulations, including data protection and privacy laws, as well as emerging legislation relating to AI \cite{fair1}. Failure to comply may result in legal penalties, reputational damage, and the loss of user trust. A number of authorities, including the General Data Protection Regulation (GDPR) of the European Union and the Financial Industry Regulatory Authority of the United States (FINRA), have imposed strict requirements on the use of AI and data protection \cite{law1}. Organizations must be aware of these regulations as well as implement compliance measures to stay on top of the complexities of the legal environment. This includes conducting risk assessments, ensuring transparency in AI decision-making, and maintaining robust data governance practices.
\item \textit{Liability and Accountability}: The determination of responsibility when AI/ML models cause harm or fail to perform as intended presents significant legal challenges \cite{account1}. As part of network automation, it is important to establish the accountability of the system (e.g., the developers, the deployers, or the operators) in the event that an ML or AutoML model makes an incorrect decision that results in a security breach or service disruption. It is important to establish clear guidelines and frameworks so that liability can be assigned appropriately. To facilitate identification of responsible parties, it is crucial to develop systems that facilitate the auditing and tracing of actions and decisions. AI accountability can be achieved by implementing explainable AI/ML models, maintaining detailed records of AI decision-making processes, and implementing organizational policies that define accountability mechanisms \cite{account1}.
\end{enumerate}

The integration of AI/ML technologies into next-generation networks requires a comprehensive approach to address these ethical and legal challenges. AI-driven automation can be enhanced while remaining ethically and legally compliant by proactively implementing bias mitigation strategies, ensuring regulatory compliance, and establishing clear accountability frameworks.

\section{List of Acronyms} 
\begin{acronym}[CICIDS2017]
\acro{1G/2G/.../6G}{First to Sixth Generation Mobile Networks}
\acro{ADWIN}{ADaptive WINdowing}
\acro{AI}{Artificial Intelligence}
\acro{AMF}{Aggregated Mondrian Forest}
\acro{AML}{Adversarial Machine Learning}
\acro{AMoF}{Accumulated Measure of Fluctuation}
\acro{ARF}{Adaptive Random Forest}
\acro{AutoDP}{Automated Data Pre-processing}
\acro{AutoFE}{Automated Feature Engineering}
\acro{AutoFS}{Automated Feature Selection}
\acro{AutoML}{Automated Machine Learning}
\acro{B5G}{Beyond Fifth Generation Mobile Networks}
\acro{BPSK}{Binary Phase-Shift Keying}
\acro{CASH}{Combined Algorithm Selection and Hyper-parameter}
\acro{CBTC}{Communication-based Train Control}
\acro{CCI}{Correctly Classified Instance}
\acro{CFO}{Carrier Frequency Offset}
\acro{CFR}{Channel Frequency Response}
\acro{CIC}{Canadian Institute for Cybersecurity}
\acro{CICIDS2017}{Canadian Institute for Cybersecurity Intrusion Detection System 2017}
\acro{CIR}{Channel Impulse Responses}
\acro{CLIDS}{Cross-Layer Intrusion Detection System}
\acro{COS}{Chebyshev Over-Sampling}
\acro{CPS}{Cyber-Physical Systems}
\acro{CU}{Central Unit}
\acro{DDoS}{Distributed Denial of Service}
\acro{DoS}{Denial of Service}
\acro{DROS}{Dynamic Random Over-Sampling}
\acro{DRUS}{Dynamic Random Under-Sampling}
\acro{DU}{Distributed Unit}
\acro{EDDM}{Early Drift Detection Method}
\acro{EFDT}{Extremely Fast Decision Tree}
\acro{ETSI}{European Telecommunications Standards Institute}
\acro{FINRA}{Financial Industry Regulatory Authority}
\acro{FL}{Federated Learning}
\acro{GAN}{Generative Adversarial Network}
\acro{GDPR}{General Data Protection Regulation}
\acro{GLRT}{Generalized Likelihood Ratio Test}
\acro{GS}{Grid Search}
\acro{HAT}{Hoeffding Adaptive Tree}
\acro{HPO}{Hyper-Parameter Optimization}
\acro{HT}{Hoeffding Trees}
\acro{IDS}{Intrusion Detection System}
\acro{IID}{Independent and Identically Distributed}
\acro{IP}{Internet Protocol}
\acro{IQI}{In-phase and Quadrature components Imbalance}
\acro{ISO}{International Organization for Standardization}
\acro{LB}{Leveraging Bagging}
\acro{LRT}{Likelihood Ratio Test}
\acro{MAC}{Media Access Control}
\acro{MANET}{Mobile Ad Hoc Network}
\acro{MITM}{Man-in-the-Middle}
\acro{ML}{Machine Learning}
\acro{MPC}{Model Predictive Control}
\acro{Near-RT RI}{Near-Real-Time RAN Intelligent Controller}
\acro{NFV}{Network Function Virtualization}
\acro{O-RAN}{Open Radio Access Network}
\acro{OE}{Optimized Ensemble}
\acro{OSI}{Open Systems Interconnection}
\acro{PCC}{Pearson Correlation Coefficient}
\acro{PD}{Probability of Detection}
\acro{PFA}{Probability of False Alarm}
\acro{PLA}{Physical Layer Authentication}
\acro{PLKG}{Physical Layer Key Generation}
\acro{PLS}{Physical Layer Security}
\acro{PMD}{Probability of Missed Detection}
\acro{PSO}{Particle Swarm Optimization}
\acro{QoS}{Quality of Service}
\acro{RF}{Radio Frequency}
\acro{ROC}{Receiver Operating Characteristic}
\acro{RS}{Random Search}
\acro{RSS}{Received Signal Strength}
\acro{SDN}{Software-Defined Networking}
\acro{SH}{Successive Halving}
\acro{SH-CASH}{Successive Halving for Combined Algorithm Selection and Hyper-parameter Optimization}
\acro{SRP}{Streaming Random Patches}
\acro{SQL}{Structured Query Language}
\acro{STM}{Smart Traffic Management}
\acro{SYN}{Synchronization}
\acro{TAP}{Test Access Point}
\acro{TCP}{Transmission Control Protocol}
\acro{TinyML}{Tiny Machine Learning}
\acro{TLS}{Transport Layer Security}
\acro{UDP}{User Datagram Protocol}
\acro{UNSW-NB15}{University of New South Wales 2015 Network Benchmark}
\acro{VFDT}{Very Fast Decision Tree}
\acro{WLAN}{Wireless Local Area Network}
\acro{WSN}{Wireless Sensor Network}
\acro{XAI}{Explainable AI}
\acro{XSS}{Cross-Site Scripting}
\acro{ZSM}{Zero-touch Network and Service Management}
\acro{ZTN}{Zero-Touch Network}
\end{acronym}

\section{Conclusion} 
The evolution towards 6G networks demands network automation and ZTNs to meet the escalating demands for high data rates, ultra-low latency, and seamless integration with advanced technologies, including AI/ML and AutoML techniques. This paper proposes a comprehensive and automated cybersecurity framework to address the cybersecurity and network automation challenges associated with ZTNs and future networks. The proposed framework leverages AutoML, online learning, and the proposed enhanced SH-CASH techniques to enable the automated construction of PLA methods and CLIDS models in dynamic networking environments, effectively mitigating cyber threats across multiple layers of the Internet protocol stack. Experimental results demonstrate that the proposed AutoML framework achieves high accuracy of 99.431\% and 99.450\% on the public Oracle RF fingerprinting and CICIDS2017 datasets, respectively. The proposed SH-CASH framework represents a significant step forward in autonomous cybersecurity for next-generation networks. By addressing both PLA and CLIDS tasks effectively, it demonstrates exceptional robustness in complex, real-world scenarios, including dynamic networking environments and zero-day attacks. Future work will explore its applicability to additional network conditions and further enhance its capabilities for multi-layer defense.


\begin{thebibliography}{1}
\bibliographystyle{IEEEtran}
\bibitem{6gsec1} P. Porambage, G. Gür, D. P. M. Osorio, M. Liyanage, A. Gurtov, and M. Ylianttila, "The Roadmap to 6G Security and Privacy," \textit{IEEE Open J. Commun. Soc.}, vol. 2, pp. 1094–1122, 2021.
\bibitem{zsm1} M. Liyanage \textit{et al.}, "A survey on Zero touch network and Service Management (ZSM) for 5G and beyond networks," \textit{J. Netw. Comput. Appl.}, vol. 203, p. 103362, Jul. 2022.
\bibitem{zsmml1}	J. Gallego-Madrid, R. Sanchez-Iborra, P. M. Ruiz, and A. F. Skarmeta, “Machine learning-based zero-touch network and service management: a survey,” \textit{Digit. Commun. Networks}, vol. 8, no. 2, pp. 105–123, Apr. 2022.
\bibitem{zsm2} ETSI, “GS ZSM 002 - V1.1.1 - Zero-touch network and Service Management (ZSM); Reference Architecture,” \textit{Gr. Specif. ETSI GS ZSM}, vol. 2, 2019.

\bibitem{ETSI_AI} ETSI, “GS ZSM 012 - V1.1.1 - Zero-touch network and Service Management (ZSM); Enablers for Artificial Intelligence-based Network and Service Automation,” \textit{Gr. Specif. ETSI GS ZSM}, 2022.
\bibitem{ETSI_Sec} ETSI, “GS ZSM 014 - V1.1.1 - Zero-touch network and Service Management (ZSM); ZSM security aspects,”  \textit{Gr. Specif. ETSI GS ZSM}, 2024.


\bibitem{zsm3} C. Benzaid and T. Taleb, “AI-Driven Zero Touch Network and Service Management in 5G and Beyond: Challenges and Research Directions,” \textit{IEEE Netw.}, vol. 34, no. 2, pp. 186–194, 2020.
\bibitem{myautoml} L. Yang and A. Shami, "IoT Data Analytics in Dynamic Environments: From An Automated Machine Learning Perspective," \textit{Eng. Appl. Artif. Intell.}, vol. 116, pp. 1–33, 2022.
\bibitem{dmdrift} D. M. Manias, I. Shaer, L. Yang, and A. Shami, “Concept Drift Detection in Federated Networked Systems,” in \textit{Proc. IEEE Global Commun. Conf (GLOBECOM)}, 2021, pp. 1–6.
\bibitem{automl1} F. Hutter, L. Kotthoff, and J. Vanschoren, "Automated machine learning: methods, systems, challenges," \textit{Springer}, 2019.
\bibitem{iotm} L. Yang and A. Shami, “A Lightweight Concept Drift Detection and Adaptation Framework for IoT Data Streams,” \textit{IEEE Internet Things Mag.}, vol. 4, no. 2, pp. 96–101, 2021.
\bibitem{stack1}	J. F. Kuros and K. W. Ross, Computer Networking: A Top-Down Approach, Boston, MA, USA:Pearson, 2007.
\bibitem{threat1} S. Deep, X. Zheng, A. Jolfaei, D. Yu, P. Ostovari, and A. Kashif Bashir, “A survey of security and privacy issues in the Internet of Things from the layered context,” \textit{Trans. Emerg. Telecommun. Technol.}, vol. 33, no. 6, p. e3935, Jun. 2022.
\bibitem{threat2}	C. Hennebert and J. Dos Santos, “Security protocols and privacy issues into 6LoWPAN stack: A synthesis,” \textit{IEEE Internet Things J.}, vol. 1, no. 5, pp. 384–398, Oct. 2014.
\bibitem{pla1} L. Bai, L. Zhu, J. Liu, J. Choi, and W. Zhang, “Physical Layer Authentication in Wireless Communication Networks: A Survey,” \textit{J. Commun. Inf. Networks}, vol. 5, no. 3, pp. 237–264, 2020.
\bibitem{clids1}	N. Kumar Trivedi, A. Kumar, A. Anand, and S. Maheshwari, “Cross-Layer Intrusion Detection in Mobile Ad Hoc Networks-A Survey,” \textit{Ann. Rom. Soc. Cell Biol.}, vol. 25, pp. 09–20, Jan. 2021.
\bibitem{mythesis} L. Yang, "Optimized and Automated Machine Learning Techniques Towards IoT Data Analytics and Cybersecurity," \textit{Electron. Thesis Diss. Repos.}, no. 8734, 2022, [Online]. Available: https://ir.lib.uwo.ca/etd/8734
\bibitem{rfdata} K. Sankhe, M. Belgiovine, F. Zhou, S. Riyaz, S. Ioannidis, and K. Chowdhury, “ORACLE: Optimized Radio clAssification through Convolutional neuraL nEtworks,” \textit{Proc. - IEEE INFOCOM}, vol. 2019-April, pp. 370–378, Apr. 2019.
\bibitem{cic} I. Sharafaldin, A. H. Lashkari, and A. A. Ghorbani, “Toward Generating a New Intrusion Detection Dataset and Intrusion Traffic Characterization,” in \textit{Proc. Int. Conf. Inf. Syst. Secur. Privacy}, 2018, pp. 108–116.


\bibitem{zsmsec1} C. Benzaid and T. Taleb, "ZSM Security: Threat Surface and Best Practices," \textit{IEEE Netw.}, vol. 34, no. 3, pp. 124–133, May 2020. 

\bibitem{ETSI_closed} ETSI, “GS ZSM 009-1 - V1.1.1 - Zero-touch network and Service Management (ZSM); Closed-Loop Automation; Part 1: Enablers,”  \textit{Gr. Specif. ETSI GS ZSM}, 2021. 
\bibitem{mirna} M. El Rajab, L. Yang, and A. Shami, “Zero-touch networks: Towards next-generation network automation,” \textit{Comput. Networks}, vol. 243, p. 110294, Apr. 2024.
\bibitem{zsmapp1} P. Radoglou-Grammatikis et al., “Trustworthy Analytics in ETSI ZSM: A 5G Security Case Study,” IEEE Open J. Commun. Soc., 2024.
\bibitem{zsmapp2} R. C. Bello, N. Slamnik-Krijestorac, and J. M. Marquez-Barja, “Zero-touch Service Management for 6G verticals: Smart Traffic Management Case Study,” Proc. - IEEE Consum. Commun. Netw. Conf. CCNC, pp. 582–585, 2024.
\bibitem{zsmapp3} O. Iacoboaiea, J. Krolikowski, Z. Ben Houidi, and D. Rossi, “From Design to Deployment of Zero Touch Deep Reinforcement Learning WLANs,” IEEE Commun. Mag., vol. 61, no. 2, pp. 104–109, Feb. 2023.



\bibitem{zsmsec2} S. Jayasinghe, Y. Siriwardhana, P. Porambage, M. Liyanage, and M. Ylianttila, "Federated Learning based Anomaly Detection as an Enabler for Securing Network and Service Management Automation in Beyond 5G Networks," \textit{2022 Jt. Eur. Conf. Networks Commun. 6G Summit, EuCNC/6G Summit 2022}, pp. 345–350, 2022.
\bibitem{clids_g} S. Halder and A. Ghosal, “Cross Layer–Based Intrusion Detection Techniques in Wireless Networks: A Survey,” \textit{State Art Intrusion Prev. Detect.}, pp. 379–408, Dec. 2013.
\bibitem{threat3}	A. S. Sastry, Shazia Sulthana;, and Dr. S Vagdevi;, “Security Threats in Wireless Sensor Networks in Each Layer,” \textit{Int. J. Adv. Netw. Appl.}, vol. 4, no. 4, pp. 1657–1661, 2013.
\bibitem{phythreat1} Y. Liu, H. H. Chen, and L. Wang, “Physical Layer Security for Next Generation Wireless Networks: Theories, Technologies, and Challenges,” \textit{IEEE Commun. Surv. Tutorials}, vol. 19, no. 1, pp. 347–376, Jan. 2017.
\bibitem{phythreat2}	N. Wang, P. Wang, A. Alipour-Fanid, L. Jiao, and K. Zeng, “Physical-Layer Security of 5G Wireless Networks for IoT: Challenges and Opportunities,” \textit{IEEE Internet Things J.}, vol. 6, no. 5, pp. 8169–8181, Oct. 2019.
\bibitem{phythreat3}	L. Alhoraibi \textit{et al.}, “Physical Layer Authentication in Wireless Networks-Based Machine Learning Approaches,” \textit{Sensors 2023, Vol. 23, Page 1814}, vol. 23, no. 4, p. 1814, Feb. 2023.
\bibitem{upperthreat1}	S. N. Uke, A R Mahajan, and R C Thool, “UML Modeling of Physical and Data Link Layer Security Attacks in WSN,” \textit{Int. J. Comput. Appl.}, vol. 70, no. 11, pp. 975–8887, 2013.
\bibitem{upperthreat2}	M. Dasari, “Real time detection of MAC layer DoS attacks in IEEE 802.11 wireless networks,” \textit{2017 14th IEEE Annu. Consum. Commun. Netw. Conf. CCNC 2017}, pp. 939–944, Jul. 2017.
\bibitem{upperthreat3}	G. Dayanandam, T. V. Rao, D. Bujji Babu, and S. Nalini Durga, “DDoS attacks—Analysis and prevention,” \textit{Lect. Notes Networks Syst.}, vol. 32, pp. 1–10, 2019.
\bibitem{ericsson}	L. Yang \textit{et al.}, “Multi-Perspective Content Delivery Networks Security Framework Using Optimized Unsupervised Anomaly Detection,” \textit{IEEE Trans. Netw. Serv. Manag.}, vol. 19, no. 1, pp. 686–705, 2022.
\bibitem{upperthreat4}	N. M. Yungaicela-Naula, C. Vargas-Rosales, and J. A. Perez-Diaz, “SDN-based architecture for transport and application layer DDoS attack detection by using machine and deep learning,” \textit{IEEE Access}, vol. 9, pp. 108495–108512, 2021.
\bibitem{myiotj}	L. Yang, A. Moubayed, and A. Shami, “MTH-IDS: A Multitiered Hybrid Intrusion Detection System for Internet of Vehicles,” \textit{IEEE Internet Things J.}, vol. 9, no. 1, pp. 616–632, 2022.


\bibitem{Zeng2010} K. Zeng, K. Govindan, and P. Mohapatra, "Non-cryptographic authentication and identification in wireless networks [Security and Privacy in Emerging Wireless Networks]," in \textit{IEEE Wireless Commun.}, vol. 17, no. 5, pp. 56-62, Oct. 2010.
\bibitem{pla2} V. L. Nguyen, P. C. Lin, B. C. Cheng, R. H. Hwang, and Y. D. Lin, “Security and Privacy for 6G: A Survey on Prospective Technologies and Challenges,” \textit{IEEE Commun. Surv. Tutorials}, vol. 23, no. 4, pp. 2384–2428, 2021.
\bibitem{Xie2021} N. Xie, Z. Li and H. Tan, "A Survey of Physical-Layer Authentication in Wireless Communications," in \textit{IEEE Commun. Surv. Tutorials}, vol. 23, no. 1, pp. 282-310, Firstquarter 2021.


\bibitem{pwpae}	L. Yang, D. M. Manias, and A. Shami, “PWPAE: An Ensemble Framework for Concept Drift Adaptation in IoT Data Streams,” in \textit{Proc. IEEE Global Commun. Conf (GLOBECOM)}, 2021, pp. 1–6.
\bibitem{clids2}	Z. A. Khan and P. Herrmann, “Recent advancements in intrusion detection systems for the Internet of Things,” \textit{Secur. Commun. Networks}, vol. 2019, 2019.
\bibitem{cl1}	V. Srivastava and M. Motani, “Cross-layer design: A survey and the road ahead,” \textit{IEEE Commun. Mag.}, vol. 43, no. 12, pp. 112–119, Dec. 2005.


\bibitem{Illi2022} E. Illi, A. Pandey, L. Bariah, G. Singh, J. -P. Giacalone and S. Muhaidat, "Physical Layer Continuous Authentication for Wireless Mesh Networks: An Experimental Study," \textit{2022 IEEE Int. Mediterr. Conf. Commun. Netw. (MeditCom)}, Athens, Greece, 2022, pp. 136-141.
\bibitem{xiao2018} L. Xiao, X. Wan, and Z. Han, “PHY-layer authentication with multiple landmarks with reduced overhead,” \textit{IEEE Trans. Wireless Commun.}, vol. 17, no. 3, pp. 1676–1687, Mar. 2018.
\bibitem{Pei2014} C. Pei, N. Zhang, X. S. Shen, and J. W. Mark, “Channel-based physical layer authentication,” in \textit{Proc. IEEE Global Commun. Conf (GLOBECOM)}, Austin, TX, USA, Dec. 2014, pp. 4114–4119.
\bibitem{yang2013} J. Yang, Y. Chen, W. Trappe, and J. Cheng, “Detection and localization of multiple spoofing attackers in wireless networks,” \textit{IEEE Trans. Parallel Distrib. Syst.}, vol. 24, no. 1, pp. 44–58, Jan. 2013.
\bibitem{Xiao2007} L. Xiao, L. Greenstein, N. Mandayam, and W. Trappe, “Fingerprints in the ether: Using the physical layer for wireless authentication,” in \textit{Proc. IEEE Int. Conf. Commun. (ICC)}, Glasgow, U.K., Jun. 2007, pp. 4646–4651.
\bibitem{Xiao2008} L. Xiao, L. Greenstein, N. Mandayam, and W. Trappe, “Using the physical layer for wireless authentication in time-variant channels,” \textit{IEEE Trans. Wireless Commun.}, vol. 7, no. 7, pp. 2571–2579, Jul. 2008.
\bibitem{Tugnait2010} J. K. Tugnait and H. Kim, “A channel-based hypothesis testing approach to enhance user authentication in wireless networks,” in \textit{Proc. Int. Conf. Commun. Syst. Netw. (COMSNETS)}, Bengaluru, India, 2010, pp. 1–9.
\bibitem{Wu2015} X. Wu and Z. Yang, “Physical-layer authentication for multi-carrier transmission,” \textit{IEEE Commun. Lett.}, vol. 19, no. 1, pp. 74–77, Jan. 2015.
\bibitem{Liu2011} F. J. Liu, X. Wang, and H. Tang, “Robust physical layer authentication using inherent properties of channel impulse response,” in \textit{Proc. IEEE Mil. Commun. Conf. (MILCOM)}, Baltimore, MD, USA, Nov. 2011, pp. 538–542.
\bibitem{Xie2022} N. Xie, W. Xiong, J. Chen, P. Zhang, L. Huang and J. Su, "Multiple Phase Noises Physical-Layer Authentication," in \textit{IEEE Trans. Commun.}, vol. 70, no. 9, pp. 6196-6211, Sept. 2022.

\bibitem{hou2012} W. Hou, X. Wang, and J. Chouinard, “Physical layer authentication in OFDM systems based on hypothesis testing of CFO estimates,” in \textit{Proc. Int. Conf. Commun. (ICC)}, Ottawa, ON, Canada, 2012, pp. 3559–3563.
\bibitem{hou2014} W. Hou, X. Wang, J. Y. Chouinard, and A. Refaey, “Physical layer authentication for mobile systems with time-varying carrier frequency offsets,” \textit{IEEE Trans. Commun.}, vol. 62, no. 5, pp. 1658–1667, May 2014.
\bibitem{Dolatshahi2010} S. Dolatshahi, A. C. Polak, and D. L. Goeckel, “Identification of wireless users via power amplifier imperfections,” in \textit{Proc. Asilomar Conf. Signals Syst. Comput. (ASILOMAR)}, Pacific Grove, CA, USA, 2010, pp. 1553–1557.
\bibitem{Rahman2014} M. M. U. Rahman, A. Yasmeen, and J. Gross, “PHY layer authentication via drifting oscillators,” in \textit{Proc. Global Commun. Conf. (GLOBCOM)}, Austin, TX, USA, 2014, pp. 716–721.
\bibitem{Kohno2005} T. Kohno, A. Broido, and K. C. Claffy, “Remote physical device fingerprinting,” \textit{IEEE Trans. Depend. Secure Comput.}, vol. 2, no. 2, pp. 93–108, Apr.–Jun. 2005.

\bibitem{Hao2014} P. Hao, X. Wang and A. Behnad, "Relay authentication by exploiting I/Q imbalance in amplify-and-forward system," \textit{Proc. IEEE Glob. Commun. Conf.}, Austin, TX, USA, 2014, pp. 613-618.
\bibitem{Hao_ICC_2014} P. Hao, X. Wang and A. Behnad, "Performance enhancement of I/Q imbalance based wireless device authentication through collaboration of multiple receivers," \textit{IEEE Int. Conf. Commun. (ICC)}, Sydney, NSW, Australia, 2014, pp. 939-944.


\bibitem{Fang2019} H. Fang, X. Wang and S. Tomasin, "Machine Learning for Intelligent Authentication in 5G and Beyond Wireless Networks," \textit{IEEE Wireless Commun.}, vol. 26, no. 5, pp. 55-61, October 2019.
\bibitem{Liao2020} R.-F. Liao \textit{et al.}, "Multiuser Physical Layer Authentication in Internet of Things With Data Augmentation," \textit{IEEE Internet Things J.}, vol. 7, no. 3, pp. 2077-2088, March 2020.
\bibitem{pan2019} F. Pan \textit{et al.}, "Threshold-Free Physical Layer Authentication Based on Machine Learning for Industrial Wireless CPS," \textit{IEEE Trans. Ind. Informatics}, vol. 15, no. 12, pp. 6481-6491, Dec. 2019.
\bibitem{fang2019} H. Fang, X. Wang and L. Hanzo, "Learning-Aided Physical Layer Authentication as an Intelligent Process," \textit{IEEE Trans. Commun.}, vol. 67, no. 3, pp. 2260-2273, March 2019.
\bibitem{Wong2018} L. J. Wong, W. C. Headley, and A. J. Michaels, “Emitter identification using CNN IQ imbalance estimators,” \textit{arXiv preprint arXiv:1808.02369}, 2018.
\bibitem{jian2020} T. Jian \textit{et al.}, "Deep Learning for RF Fingerprinting: A Massive Experimental Study," \textit{IEEE Internet Things Mag.}, vol. 3, no. 1, pp. 50-57, March 2020.
\bibitem{Merchant2018} K. Merchant, S. Revay, G. Stantchev and B. Nousain, "Deep Learning for RF Device Fingerprinting in Cognitive Communication Networks," \textit{IEEE J. Sel. Top. Signal Process.}, vol. 12, no. 1, pp. 160-167, Feb. 2018.


\bibitem{Zhang2020} P. Zhang, Y. Shen, X. Jiang and B. Wu, "Physical Layer Authentication Jointly Utilizing Channel and Phase Noise in MIMO Systems," in \textit{IEEE Trans. Commun.}, vol. 68, no. 4, pp. 2446-2458, April 2020.
\bibitem{Hoang2020} T. M. Hoang, N. M. Nguyen and T. Q. Duong, "Detection of Eavesdropping Attack in UAV-Aided Wireless Systems: Unsupervised Learning With One-Class SVM and K-Means Clustering," in \textit{IEEE Wirel. Commun. Lett.}, vol. 9, no. 2, pp. 139-142, Feb. 2020.
\bibitem{Senigagliesi2019} L. Senigagliesi, L. Cintioni, M. Baldi and E. Gambi, "Blind Physical Layer Authentication over Fading Wireless Channels through Machine Learning," \textit{2019 IEEE Int. Workshop Inf. Forensics Secur. (WIFS)}, Delft, Netherlands, 2019, pp. 1-6.
\bibitem{Qiu2018} X. Qiu, T. Jiang, S. Wu and M. Hayes, "Physical Layer Authentication Enhancement Using a Gaussian Mixture Model," in \textit{IEEE Access}, vol. 6, pp. 53583-53592, 2018.
\bibitem{Xiao2016} L. Xiao, Y. Li, G. Han, G. Liu and W. Zhuang, "PHY-Layer Spoofing Detection With Reinforcement Learning in Wireless Networks," in \textit{IEEE Trans. Veh. Technol.}, vol. 65, no. 12, pp. 10037-10047, Dec. 2016.
\bibitem{Abdrabou2022} M. Abdrabou and T. A. Gulliver, "Adaptive Physical Layer Authentication Using Machine Learning With Antenna Diversity," in \textit{IEEE Trans. Commun.}, vol. 70, no. 10, pp. 6604-6614, Oct. 2022.
\bibitem{Chen2021} S. Chen, Z. Pang, H. Wen, K. Yu, T. Zhang and Y. Lu, "Automated Labeling and Learning for Physical Layer Authentication Against Clone Node and Sybil Attacks in Industrial Wireless Edge Networks," in \textit{IEEE Trans. Ind. Informatics}, vol. 17, no. 3, pp. 2041-2051, March 2021.
\bibitem{Qiu2020} X. Qiu, J. Dai and M. Hayes, "A Learning Approach for Physical Layer Authentication Using Adaptive Neural Network," in \textit{IEEE Access}, vol. 8, pp. 26139-26149, 2020.
\bibitem{Fang2020} H. Fang, X. Wang and L. Xu, "Fuzzy Learning for Multi-Dimensional Adaptive Physical Layer Authentication: A Compact and Robust Approach," in \textit{IEEE Trans. Wireless Commun.}, vol. 19, no. 8, pp. 5420-5432, Aug. 2020.
\bibitem{Wang2019} Q. Wang, H. Li, D. Zhao, Z. Chen, S. Ye and J. Cai, "Deep Neural Networks for CSI-Based Authentication," in \textit{IEEE Access}, vol. 7, pp. 123026-123034, 2019.


\bibitem{CLIDS_review1} L. Zhu, Y. Li, F. R. Yu, B. Ning, T. Tang and X. Wang, “Cross-Layer Defense Methods for Jamming-Resistant CBTC Systems,” \textit{IEEE Trans. Intell. Transp. Syst.}, vol. 22, no. 11, pp. 7266-7278, Nov. 2021.
\bibitem{CLIDS_review2} A. Amouri, V. T. Alaparthy, and S. D. Morgera, “A Machine Learning Based Intrusion Detection System for Mobile Internet of Things,” \textit{Sensors 2020, Vol. 20, Page 461}, vol. 20, no. 2, p. 461, Jan. 2020.
\bibitem{CLIDS_review3} L. Gandhimathi and G. Murugaboopathi, “A Novel Hybrid Intrusion Detection Using Flow-Based Anomaly Detection and Cross-Layer Features in Wireless Sensor Network,” \textit{Autom. Control Comput. Sci.}, vol. 54, no. 1, pp. 62–69, Jan. 2020.

\bibitem{automl_review1} A. Singh, J. Amutha, J. Nagar, S. Sharma, and C. C. Lee, “AutoML-ID: automated machine learning model for intrusion detection using wireless sensor network,” \textit{Sci. Reports 2022 121}, vol. 12, no. 1, pp. 1–14, May 2022.
\bibitem{automl_review2} M. A. Khan, N. Iqbal, Imran, H. Jamil, and D. H. Kim, “An optimized ensemble prediction model using AutoML based on soft voting classifier for network intrusion detection,” \textit{J. Netw. Comput. Appl.}, vol. 212, p. 103560, Mar. 2023.
\bibitem{automl_review3} W. Elmasry, A. Akbulut, and A. H. Zaim, “Evolving deep learning architectures for network intrusion detection using a double PSO metaheuristic,” \textit{Comput. Networks}, vol. 168, p. 107042, Feb. 2020.


\bibitem{ml1} N. Afshan and R. K. Rout, “Machine Learning Techniques for IoT Data Analytics,” \textit{Big Data Anal. Internet Things}, pp. 89–113, 2021.
\bibitem{lccde} L. Yang and A. Shami, “LCCDE: A Decision-Based Ensemble Framework for Intrusion Detection in The Internet of Vehicles,” in \textit{Proc. IEEE Global Commun. Conf (GLOBECOM)}, Rio de Janeiro, Brazil, 2022, pp. 1–6.
\bibitem{hpome} L. Yang and A. Shami, “On hyperparameter optimization of machine learning algorithms: Theory and practice,” \textit{Neurocomputing}, vol. 415, pp. 295–316, 2020.
\bibitem{automl2} X. He, K. Zhao, and X. Chu, “AutoML: A survey of the state-of-the-art,” \textit{Knowledge-Based Syst.}, vol. 212, no. January, p. 106622, 2021.
\bibitem{automl3} K. Chauhan \textit{et al.}, “Automated Machine Learning: The New Wave of Machine Learning,” \textit{2nd Int. Conf. Innov. Mech. Ind. Appl. ICIMIA 2020 - Conf. Proc.}, no. Icimia, pp. 205–212, 2020.
\bibitem{tii} L. Yang and A. Shami, “A Multi-Stage Automated Online Network Data Stream Analytics Framework for IIoT Systems,” \textit{IEEE Trans. Ind. Informatics}, vol. 19, no. 2, pp. 2107–2116, 2023.
\bibitem{sample1} P. Kaur and A. Gosain, “Comparing the Behavior of Oversampling and Undersampling Approach of Class Imbalance Learning by Combining Class Imbalance Problem with Noise,” in \textit{ICT Based Innovations}, Springer, Singapore, 2018, pp. 23–30.
\bibitem{cos} E. Aminian, R. P. Ribeiro, and J. Gama, “Chebyshev approaches for imbalanced data streams regression models,” \textit{Data Min. Knowl. Discov.}, vol. 35, no. 6, pp. 2389–2466, Nov. 2021.
\bibitem{fs1} S. Pande, A. Khamparia, and D. Gupta, “Feature selection and comparison of classification algorithms for wireless sensor networks,” \textit{J. Ambient Intell. Humaniz. Comput.}, 2021.

\bibitem{autoweka} C. Thornton, F. Hutter, H. H. Hoos, and K. Leyton-Brown, “Auto-WEKA: Combined selection and hyperparameter optimization of classification algorithms,” \textit{Proc. ACM SIGKDD Int. Conf. Knowl. Discov. Data Min.}, vol. Part F1288, pp. 847–855, 2013.

\bibitem{adwin} A. Bifet and R. Gavaldà, “Learning from time-changing data with adaptive windowing,” \textit{Proc. 7th SIAM Int. Conf. Data Min.}, pp. 443–448, 2007.
\bibitem{eddm} M. Baena-García, J. del Campo-Ávila, R. Fidalgo, A. Bifet, R. Gavaldà, and R. Morales-Bueno, “Early Drift Detection Method,” \textit{4th ECML PKDD Int. Work. Knowl. Discov. from Data Streams}, vol. 6, pp. 77–86, 2006.
\bibitem{ht} J. Lu, A. Liu, F. Dong, F. Gu, J. Gama, and G. Zhang, “Learning under Concept Drift: A Review,” \textit{IEEE Trans. Knowl. Data Eng.}, vol. 31, no. 12, pp. 2346–2363, 2019.
\bibitem{arf} H. M. Gomes \textit{et al.}, “Adaptive random forests for evolving data stream classification,” \textit{Mach. Learn.}, vol. 106, no. 9–10, pp. 1469–1495, 2017.
\bibitem{srp} H. M. Gomes, J. Read, and A. Bifet, “Streaming random patches for evolving data stream classification,” \textit{Proc. - IEEE Int. Conf. Data Mining, ICDM}, vol. 2019-Novem, no. Icdm, pp. 240–249, 2019.
\bibitem{sh1} K. Jamieson and A. Talwalkar, “Non-stochastic Best Arm Identification and Hyperparameter Optimization,” in \textit{Proceedings of the 19th International Conference on Artificial Intelligence and Statistics}, 2016, pp. 240–248.
\bibitem{sh2} L. Li, K. Jamieson, G. DeSalvo, A. Rostamizadeh, and A. Talwalkar, “Hyperband: A novel bandit-based approach to hyperparameter optimization,” \textit{J. Mach. Learn. Res.}, vol. 18, pp. 1–52, 2018.

\bibitem{ranfl1} 	C. Benzaïd, F. M. Hossain, T. Taleb, P. M. Gómez, and M. Dieudonne, “A Federated Continual Learning Framework for Sustainable Network Anomaly Detection in O-RAN,” IEEE Wirel. Commun. Netw. Conf. WCNC, 2024.
\bibitem{ranfl2} 	D. Attanayaka, P. Porambage, M. Liyanage, and M. Ylianttila, “Peer-to-Peer Federated Learning Based Anomaly Detection for Open Radio Access Networks,” IEEE Int. Conf. Commun., vol. 2023-May, pp. 5464–5470, 2023.
\bibitem{rancl} 	A. Scalingi, S. D’Oro, F. Restuccia, T. Melodia, and D. Giustiniano, “Det-RAN: Data-Driven Cross-Layer Real-Time Attack Detection in 5G Open RANs,” Proc. - IEEE INFOCOM, pp. 41–50, 2024.
\bibitem{ranad} 	T. Sundqvist, M. Bhuyan, and E. Elmroth, “Robust Procedural Learning for Anomaly Detection and Observability in 5G RAN,” IEEE Trans. Netw. Serv. Manag., vol. 21, no. 2, pp. 1432–1445, Apr. 2024.

\bibitem{aml1} L. Yang, M. El Rajab, A. Shami, and S. Muhaidat, “Enabling AutoML for Zero-Touch Network Security: Use-Case Driven Analysis,” \textit{IEEE Trans. Netw. Serv. Manag.}, vol. 21, no. 3, pp. 3555–3582, 2024.

\bibitem{cost1} Y. Li, N. Raison, S. Ourselin, T. Mahmoodi, P. Dasgupta, and A. Granados, “AI solutions for overcoming delays in telesurgery and telementoring to enhance surgical practice and education,” J. Robot. Surg., vol. 18, no. 1, p. 403, Dec. 2024.

\bibitem{measure1} P. Sinha, V. K. Jha, A. K. Rai, and B. Bhushan, “Security vulnerabilities, attacks and countermeasures in wireless sensor networks at various layers of OSI reference model: A survey,” \textit{Proc. IEEE Int. Conf. Signal Process. Commun. ICSPC 2017}, vol. 2018-January, pp. 288–293, Jul. 2017.
\bibitem{measure2} A. de L. Leandro and S. P. Gustavo, “Provisioning and Recovery in Flexible Optical Networks using Ant Colony Optimization,” \textit{2021 IFIP/IEEE Int. Symp. Integr. Netw. Manag.}, pp. 677–681, 2021.
\bibitem{measure3} B. S. Lakshmi, D. Kovvuri, H. N. V. Bolisetti, D. S. Chikkala, S. Karri, and G. Yadlapalli, “A Proactive Approach for Detecting SQL and XSS Injection Attacks,” \textit{Proc. 3rd Int. Conf. Appl. Artif. Intell. Comput. ICAAIC 2024}, pp. 1415–1420, 2024.



\bibitem{river} J. Montiel \textit{et al.}, “River: Machine learning for streaming data in python,” \textit{J. Mach. Learn. Res.}, vol. 22, pp. 1–8, 2021.

\bibitem{prequential} J. I. G. Hidalgo, B. I. F. Maciel, and R. S. M. Barros, “Experimenting with prequential variations for data stream learning evaluation,” \textit{Comput. Intell.}, vol. 35, no. 4, pp. 670–692, 2019.
\bibitem{lb} A. Bifet, G. Holmes, and B. Pfahringer, “Leveraging bagging for evolving data streams,” \textit{Lect. Notes Comput. Sci. (including Subser. Lect. Notes Artif. Intell. Lect. Notes Bioinformatics)}, vol. 6321 LNAI, no. PART 1, pp. 135–150, 2010.
\bibitem{amf} J. Mourtada, S. Gaïffas, and E. Scornet, “AMF: Aggregated Mondrian Forests for Online Learning,” \textit{J. R. Stat. Soc. Ser. B Stat. Methodol.}, vol. 83, no. 3, pp. 505–533, Jul. 2021.
\bibitem{efdt} C. Manapragada, G. I. Webb, and M. Salehi, “Extremely fast decision tree,” \textit{Proc. ACM SIGKDD Int. Conf. Knowl. Discov. Data Min.}, pp. 1953–1962, 2018.
\bibitem{hat} A. Bifet and R. Gavaldà, “Adaptive learning from evolving data streams,” \textit{Lect. Notes Comput. Sci. (including Subser. Lect. Notes Artif. Intell. Lect. Notes Bioinformatics)}, vol. 5772 LCNS, pp. 249–260, 2009.
\bibitem{cicdata} R. Panigrahi and S. Borah, “A detailed analysis of CICIDS2017 dataset for designing Intrusion Detection Systems,” \textit{Int. J. Eng. Technol.}, vol. 7, no. January, pp. 479–482, 2018.



\bibitem{continual} G. I. Parisi, R. Kemker, J. L. Part, C. Kanan, and S. Wermter, “Continual lifelong learning with neural networks: A review,” \textit{Neural Networks}, vol. 113, pp. 54–71, May 2019.
\bibitem{tinyml} F. Dehrouyeh, L. Yang, F. B. Ajaei, and A. Shami, “On TinyML and Cybersecurity: Electric Vehicle Charging Infrastructure Use Case,” \textit{IEEE Access}, vol. 12, pp. 108703–108730, Apr. 2024.
\bibitem{sam1} S. Aleyadeh, A. Moubayed, and A. Shami, “Mobility Aware Edge Computing Segmentation Towards Localized Orchestration,” \textit{2021 Int. Symp. Networks, Comput. Commun. ISNCC 2021}, 2021.
\bibitem{xai} A. Barredo Arrieta \textit{et al.}, “Explainable Artificial Intelligence (XAI): Concepts, taxonomies, opportunities and challenges toward responsible AI,” \textit{Inf. Fusion}, vol. 58, pp. 82–115, Jun. 2020.

\bibitem{aml2} Y. E. Sagduyu, Y. Shi, and T. Erpek, “IoT Network Security from the Perspective of Adversarial Deep Learning,” in \textit{2019 16th Annual IEEE International Conference on Sensing, Communication, and Networking (SECON)}, 2019, pp. 1–9.
\bibitem{physec1} Y. Zou, J. Zhu, X. Wang, and L. Hanzo, “A Survey on Wireless Security: Technical Challenges, Recent Advances, and Future Trends,” \textit{Proc. IEEE}, vol. 104, no. 9, pp. 1727–1765, Sep. 2016.

\bibitem{pri1} U. Gorrepati, P. Zavarsky, and R. Ruhl, “Privacy Protection in LTE and 5G Networks,” \textit{ICSCCC 2021 - Int. Conf. Secur. Cyber Comput. Commun.}, pp. 382–387, May 2021.
\bibitem{pri2} Y. Sun, J. Liu, J. Wang, Y. Cao, and N. Kato, “When Machine Learning Meets Privacy in 6G: A Survey,” \textit{IEEE Commun. Surv. Tutorials}, vol. 22, no. 4, pp. 2694–2724, 2020.
\bibitem{pri3} C. Benzaïd and T. Taleb, “AI for beyond 5G Networks: A Cyber-Security Defense or Offense Enabler?,” \textit{IEEE Netw.}, vol. 34, no. 6, pp. 140–147, Nov. 2020.
\bibitem{pri4} I. T. Javed, F. Alharbi, T. Margaria, N. Crespi, and K. N. Qureshi, “PETchain: A Blockchain-Based Privacy Enhancing Technology,” \textit{IEEE Access}, vol. 9, pp. 41129–41143, 2021.

\bibitem{fair1} G. Giannopoulos et al., “Fairness in AI: challenges in bridging the gap between algorithms and law,” Proc. - 2024 IEEE 40th Int. Conf. Data Eng. Work. ICDEW 2024, pp. 217–225, 2024.
\bibitem{fair2} C. C. R. Yuan and B. Y. Wang, “Ensuring Fairness with Transparent Auditing of Quantitative Bias in AI Systems,” Proc. 2024 Pacific Neighborhood Consort. Annu. Conf. Jt. Meet. Green, Flow, Signal Reconnecting Fragm. Communities Age Digit. Transform. PNC 2024, pp. 25–32, 2024.
\bibitem{law1} A. Deshpande, “Regulatory Compliance and AI: Navigating the Legal and Regulatory Challenges of AI in Finance,” 2024 Int. Conf. Knowl. Eng. Commun. Syst. ICKECS 2024, 2024.
\bibitem{account1} A. K. Raja and J. Zhou, “AI Accountability: Approaches, Affecting Factors, and Challenges,” Computer (Long. Beach. Calif)., vol. 56, no. 4, pp. 61–70, Apr. 2023.































\end{thebibliography}


\begin{IEEEbiography}[{\includegraphics[width=1in,height=1.25in,clip,keepaspectratio]{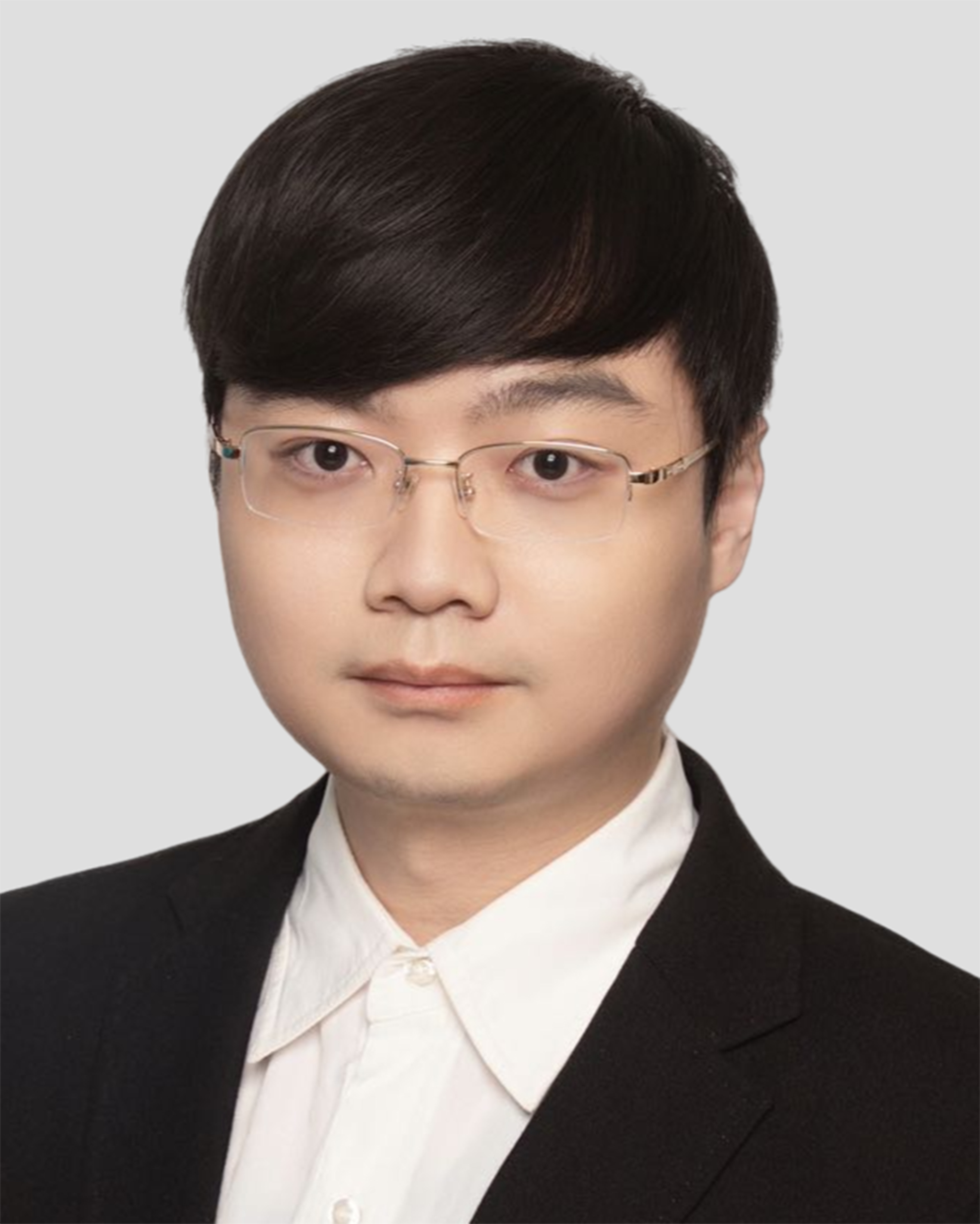}}]{Li Yang} (Member, IEEE) is currently an Assistant Professor in the Faculty of Business and Information Technology at Ontario Tech University, and an Adjunct Research Professor in the Department of Electrical and Computer Engineering at Western University. He received his Ph.D. in Electrical and Computer Engineering from Western University in 2022. He was the vice chair of the IEEE Computer Society, London Section, Canada, from 2022 to 2023. He was also on the technical program committee for IEEE GlobeCom 2023 and 2024, the workshop chair for SMC-IoT 2023, and the technical session chair for IEEE CCECE 2020. His paper and code publications have received thousands of Google Scholar citations and GitHub stars. His research interests include cybersecurity, machine learning, deep learning, AutoML, model optimization, network data analytics, Internet of Things (IoT), intrusion detection, anomaly detection, concept drift, continual learning, and adversarial machine learning. Li Yang is also included in Stanford University/Elsevier's List of the World's Top 2\% Scientists. He was ranked among the world's Top 0.5\% of researchers in 'Networking \& Telecommunications' in 2024, and 52nd in Canada.
\end{IEEEbiography}

\begin{IEEEbiography}[{\includegraphics[width=1in,height=1.25in,clip,keepaspectratio]{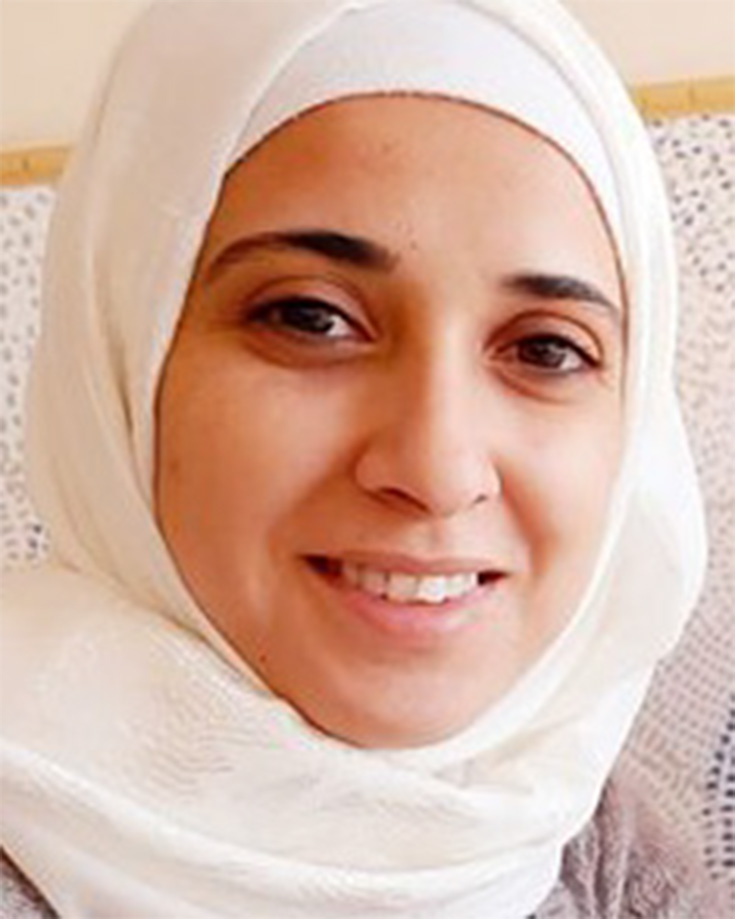}}]{Shimaa Naser} (Member, IEEE) received the M.Sc. degree in Electrical Engineering from Jordan University of Science and Technology, Jordan, 2015. Also, she received her Ph.D. degree in Electrical and Computer Engineering from Khalifa University, Abu Dhabi, UAE, 2022. She is currently a Postdoctoral fellow with the 6G Research Center, Department of Computer and Information Engineering, Khalifa University. Dr. Naser was a session chair for local conferences and a member of the technical program Committee for multiple IEEE conferences such as IEEE VTC 2022, IEEE ICC 2023, and IEEE 6GNet 2023. Also, she participated in the peer-review process in multiple top IEEE journals such as Transactions on Communications, Transactions on Wireless Communications Communication Letters, and Photonics Journals. Dr. Naser has authored/co-authored 35+ journal and conference publications and is involved in local and international research collaborations with world-class universities in Canada and UK. Her research interests include advanced digital signal processing, convex optimization, mobile communication networks, optical wireless communications, ultra-low power networks, MIMO-based communication, and orthogonal/non-orthogonal multiple  access.
\end{IEEEbiography}

\begin{IEEEbiography}[{\includegraphics[width=1in,height=1.25in,clip,keepaspectratio]{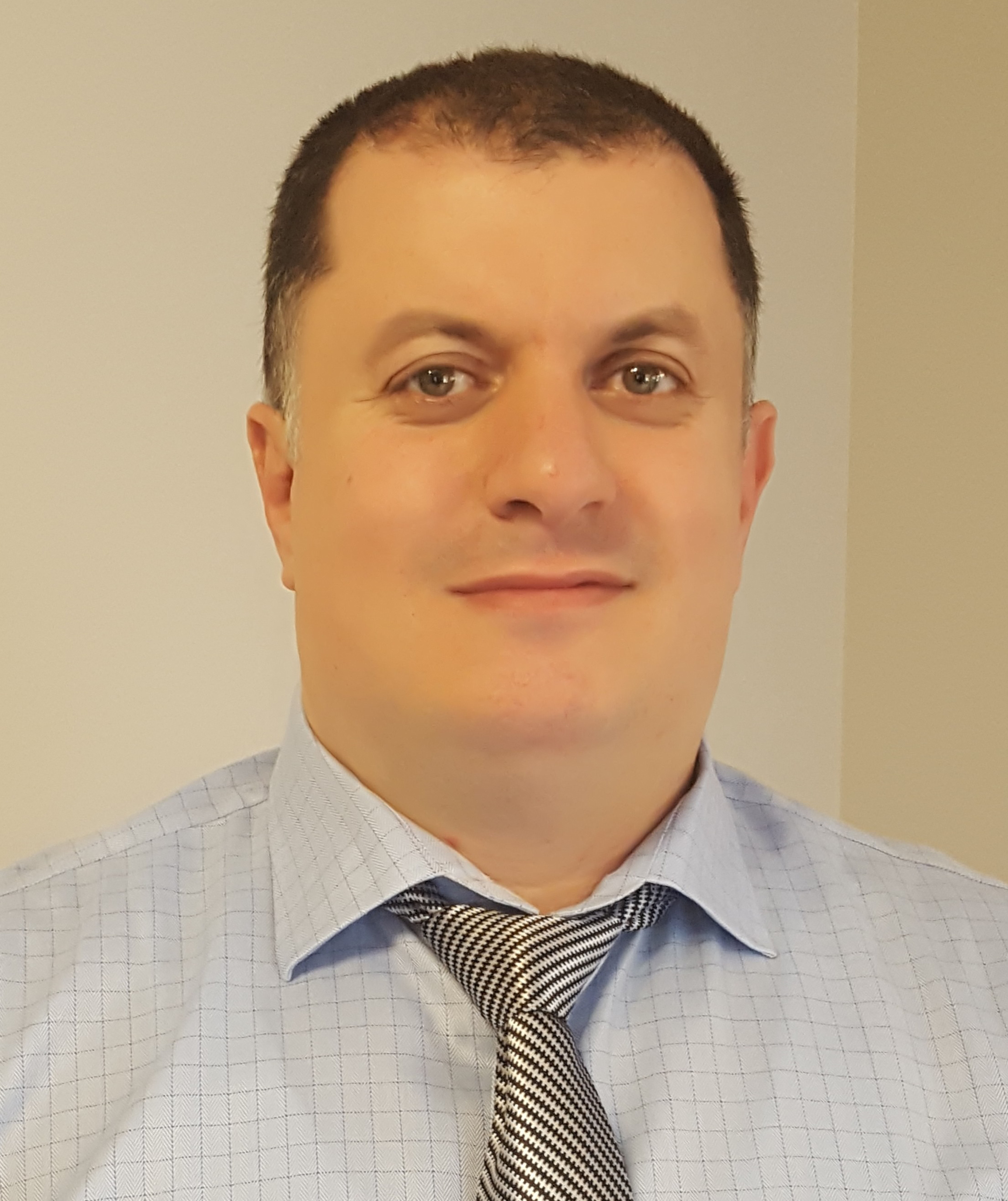}}] {Abdallah Shami} (Fellow, IEEE) is currently a Professor and Chair of the Department of Electrical and Computer Engineering, Western University, London, ON, Canada, where he is also the Director of the Optimized Computing and Communications Laboratory. Dr. Shami has chaired key symposia for the IEEE GLOBECOM, IEEE International Conference on Communications, and IEEE International Conference on Computing, Networking and Communications. He was the elected Chair for the IEEE Communications Society Technical Committee on Communications Software and the IEEE London Ontario Section Chair. He is currently an Associate Editor of the IEEE Transactions on Information Forensics and Security, IEEE Transactions on Network and Service Management, and IEEE Communications Surveys and Tutorials journals. Dr. Shami is a Fellow of IEEE, a Fellow of the Canadian Academy of Engineering (CAE), and a Fellow of the Engineering Institute of Canada (EIC).
\end{IEEEbiography}
\begin{IEEEbiography}[{\includegraphics[width=1in,height=1.25in,clip,keepaspectratio]{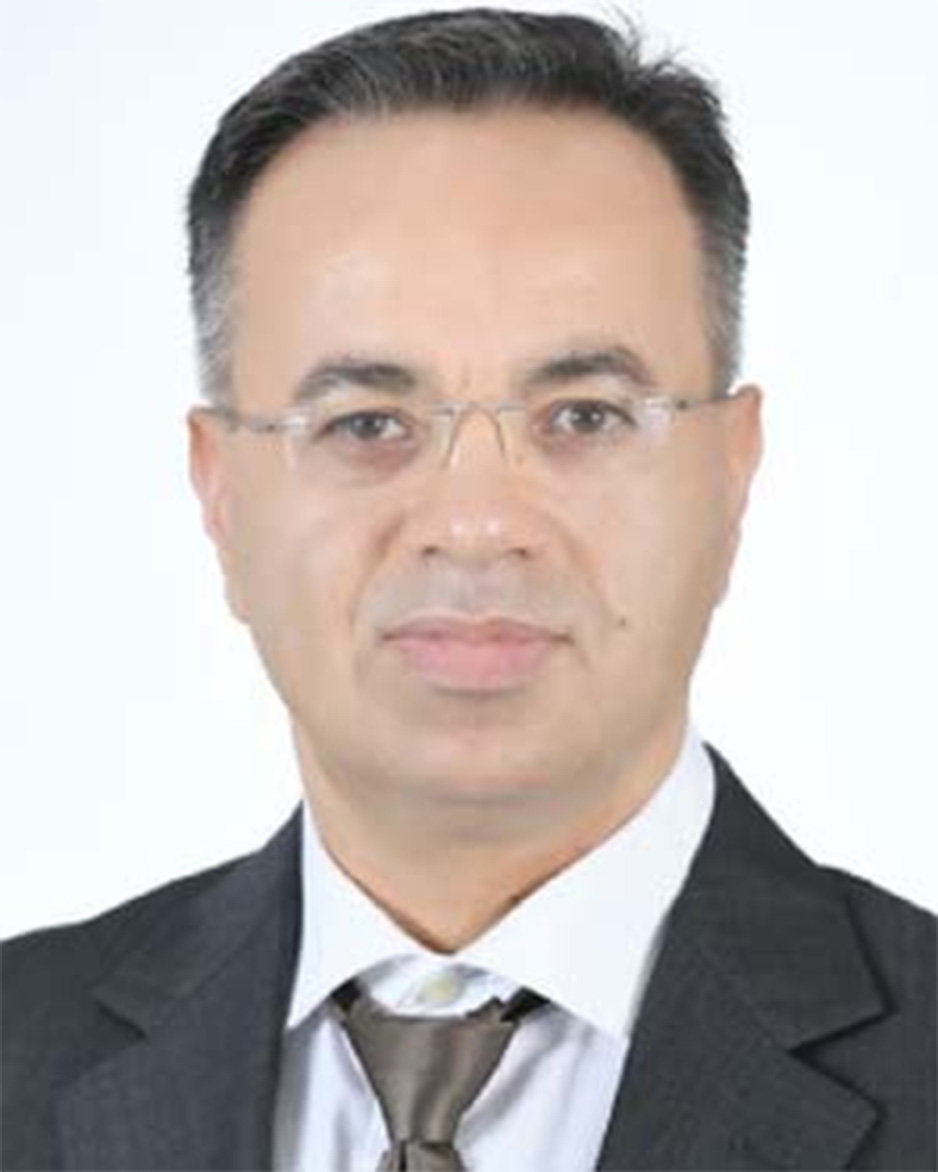}}] {Sami Muhaidat} (Senior Member, IEEE) received his Ph.D. in Electrical and Computer Engineering from the University of Waterloo, Ontario, in 2006. From 2007 to 2008, he was an NSERC Postdoctoral Fellow in the Department of Electrical and Computer Engineering at the University of Toronto, Canada. From 2008 to 2012, he served as an Assistant Professor in the School of Engineering Science at Simon Fraser University, British Columbia, Canada. Currently, he is a Professor and the Associate Dean for Research in the College of Computing and Mathematical Sciences at Khalifa University. He is also an Adjunct Professor at Carleton University, Ontario, Canada. Sami’s research interests include advanced digital signal processing techniques for wireless communications, intelligent surfaces, machine learning for communications, optical communications, and multiple-access techniques. He has served in various editorial roles, including as Area Editor for the IEEE Transactions on Communications, Guest Editor for the IEEE Network special issue on "Native Artificial Intelligence in Integrated Terrestrial and Non-Terrestrial Networks in 6G," and Guest Editor for the IEEE Open Journal of Vehicular Technology (OJVT) special issue on "Recent Advances in Security and Privacy for 6G Networks." Additionally, he has held positions as Senior Editor and Editor for IEEE Communications Letters, Editor for the IEEE Transactions on Communications, and Associate Editor for the IEEE Transactions on Vehicular Technology.
\end{IEEEbiography}
\begin{IEEEbiography}[{\includegraphics[width=1in,height=1.25in,clip,keepaspectratio]{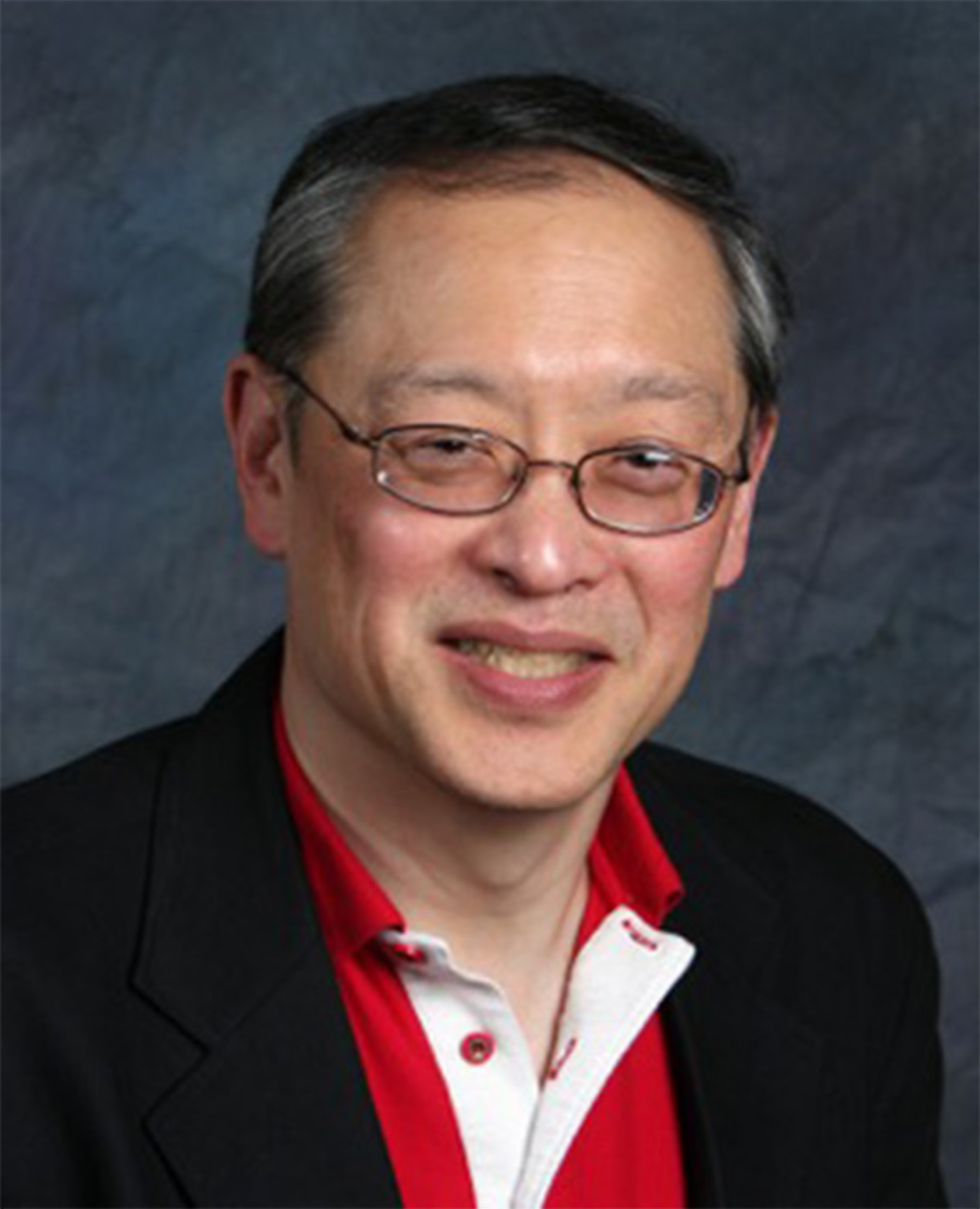}}] {Lyndon Ong} (Member, IEEE) was a Principal Architect in the Office of the CTO at Ciena Corporation and had headed numerous projects in network control and SDN, receiving the Ciena Technical Fellow award in 2014 in recognition of his industry leadership and innovation. He was co-chair of the Open RAN Alliance (O-RAN) Cloudification and Orchestration Working Group and Project Leader for the ONF’s Open Transport Configuration and Control (OTCC) Project. He also co-chaired the Systems Optimization Working Group of the IEEE International Network Generations Roadmap effort. Dr. Ong received his doctoral degree from Columbia University in 1991.
\end{IEEEbiography}
\begin{IEEEbiography}[{\includegraphics[width=1in,height=1.25in,clip,keepaspectratio]{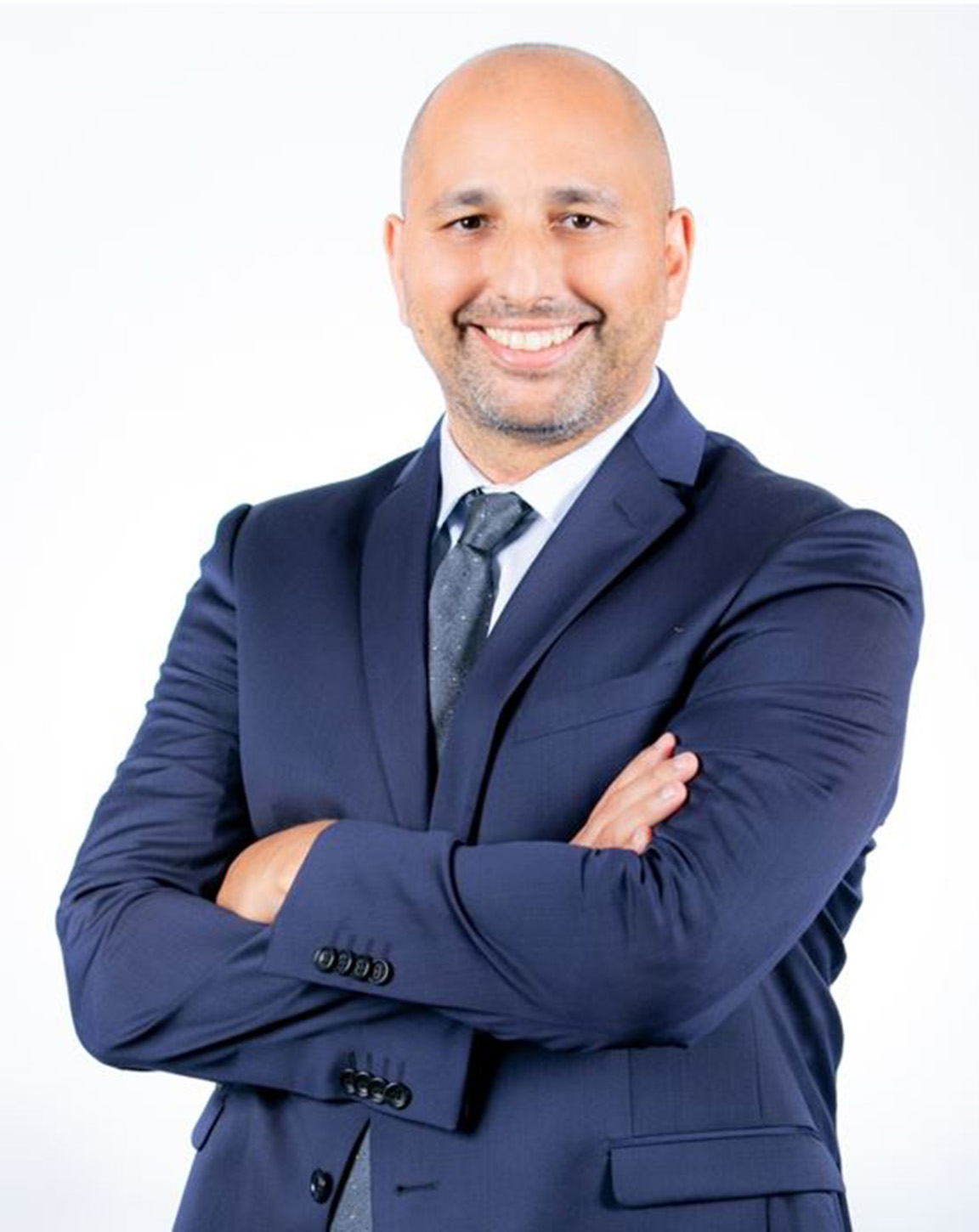}}] {M\'{e}rouane Debbah} (Fellow, IEEE) is a Professor at Khalifa University of Science and Technology in Abu Dhabi and founding Director of the KU 6G Research Center. He is a frequent keynote speaker at international events in the field of telecommunication and AI. His research has been lying at the interface of fundamental mathematics, algorithms, statistics, information and communication sciences with a special focus on random matrix theory and learning algorithms. In the Communication field, he has been at the heart of the development of small cells (4G), Massive MIMO (5G) and Large Intelligent Surfaces (6G) technologies. In the AI field, he is known for his work on Large Language Models, distributed AI systems for networks and semantic communications. He received multiple prestigious distinctions, prizes and best paper awards for his contributions to both fields. He is an IEEE Fellow, a WWRF Fellow, a Eurasip Fellow, an AAIA Fellow, an Institut Louis Bachelier Fellow, an AIIA Fellow  and a Membre émérite SEE. He is actually chair of  the IEEE Large Generative AI Models in Telecom (GenAINet) Emerging Technology Initiative and  a member of the Marconi Prize Selection Advisory Committee.
\end{IEEEbiography}


\end{document}